\documentclass[aps,pre,twocolumn,amsmath,amssymb,floatfix,superscriptaddress]{revtex4-1}

\usepackage{graphicx}
\usepackage{dcolumn}
\usepackage{bm}
\usepackage{latexsym}
\usepackage{color}
\usepackage[colorlinks=true,linkcolor=blue,citecolor=blue]{hyperref}
\usepackage{xcolor}

\newcommand*\Laplace{\mathop{}\!\mathbin{\triangle}}
\newcommand*\DAlembert{\mathop{}\!\mathbin{\Box}}

\begin{document}

\title{Physics of relativistic collisionless shocks:\\ The scattering center frame}

\author{Guy Pelletier} 
\affiliation{Universit\'e Grenoble Alpes, CNRS-INSU, Institut de Plan\'etologie et d'Astrophysique de Grenoble (IPAG), F-38041 Grenoble, France}
\author{Laurent Gremillet}
\affiliation{CEA, DAM, DIF, F-91297 Arpajon, France}
\author{Arno Vanthieghem}
\affiliation{Institut d'Astrophysique de Paris, CNRS -- Sorbonne Universit\'e, 98 bis boulevard Arago, F-75014 Paris}
\affiliation{Sorbonne Universit\'e, Institut Lagrange de Paris (ILP),
98 bis bvd Arago, F-75014 Paris, France}
\author{Martin Lemoine}
\affiliation{Institut d'Astrophysique de Paris, CNRS -- Sorbonne Universit\'e, 98 bis boulevard Arago, F-75014 Paris}

\date{\today}

\begin{abstract}
In this first paper of a series dedicated to the microphysics of unmagnetized, relativistic  collisionless pair shocks, we discuss the physics of the Weibel-type transverse current filamentation instability (CFI) that develops in the shock precursor, through the interaction of an ultrarelativistic suprathermal particle beam with the background plasma. We introduce in particular the notion of ``Weibel frame'', or scattering center frame, in which the microturbulence is of mostly magnetic  nature. We calculate the properties of this frame, using first a kinetic formulation of the linear phase of the instability, relying on Maxwell-J\"uttner distribution functions, then using a quasistatic model of the nonlinear stage of the instability. Both methods show that: (i) the ``Weibel frame'' moves at subrelativistic velocities relative to the background plasma, therefore at relativistic velocities relative to the shock front; (ii) the velocity of the ``Weibel frame'' relative to the background plasma scales with $\xi_{\rm b}$, {\it i.e.},  the pressure of the suprathermal particle beam in units of the momentum flux density incoming into the shock; and (iii), the ``Weibel frame'' moves slightly less fast than the background plasma relative to the shock front. Our theoretical results are found to be in satisfactory agreement with the measurements carried out in dedicated large-scale 2D3V PIC simulations.
\end{abstract}

\maketitle

\section{Introduction}\label{sec:introd}
\subsection{Motivations and objectives}
The Weibel-type current filamentation instability (CFI) that develops in anisotropic plasma flows~\cite{Weibel_1959, Fried_1959, Davidson_1972, Califano_1997, Achterberg_2007_I}, has gained a lot of attention in past decades because of its relevance to various fields of physics. Through the generation of skin-depth-scale current filaments surrounded by toroidal magnetic fields, which gradually isotropize the counterstreaming plasmas, this instability is a key process in the formation of astrophysical collisionless shock waves~\cite{Moiseev_1963}. Because it also arises in the precursor of weakly magnetized shock waves, \emph{i.e.}, the upstream region where the unshocked plasma flows against a beam of suprathermal particles, it further sustains the shock transition and likely generates the magnetized turbulence that is required for particle acceleration and radiation. Such microphysics may well underpin the outstanding phenomenon of gamma-ray burst afterglows, in which a substantial fraction of the $\sim10^{52}\,$ergs liberated by the cataclysmic event is radiated away by shock-accelerated electrons, on hours to months timescales~\cite{Medvedev_1999,1999ApJ...511..852G}.

In laser-driven high-energy-density physics and laboratory astrophysics, the CFI also plays a central role~\cite{Sentoku_2003, Adam_2006, Allen_2012, Debayle_2010, Masson-Laborde_2010, Mondal_2012, Quinn_2012, Fiuza_2012, Ruyer_2015b}. In particular, it has been recently observed to grow in the interpenetration of laser-ablated plasmas~\cite{Fox_2013, Huntington_2015}, marking a first step towards the generation of collisionless shocks in the laboratory \cite{Drake_2012, Chen_PRL_114_215001_2015, Lobet_2015, Ruyer_2016, *Ruyer_2017}.

The development of the CFI in symmetric colliding plasmas has been studied theoretically in both the linear ~\cite{Medvedev_1999, Wiersma_2004, Lyubarsky_2006a, Achterberg_2007_I} and nonlinear \cite{Medvedev_2005, Milosavljevic_2006a, Achterberg_2007_II, Bret_2013, Ruyer_2015a, Vanthieghem_2018} regimes, and numerically using particle-in-cell (PIC) simulations, \cite{Davidson_1972, Lee_1973, Silva_2003, Frederiksen_2004, Jaroschek_2005, Nishikawa_2009, Shvets_2009, Bret_2010a}. In such a symmetric configuration, the lab frame appears as a privileged frame in which to discuss the physics of the CFI and it is found, indeed, that the CFI generates an essentially magnetic structure in that frame.

However, in the precursor of relativistic collisionless shocks, the CFI develops in a highly asymmetric configuration: in the reference frame in which the shock front lies at rest (the shock rest frame $\mathcal R_{\rm s}$), the background plasma appears cold, and streams at relativistic velocities through a quasi-isotropic, ultrarelativistically hot gas of suprathermal particles. Conversely, in the background plasma rest frame $\mathcal R_{\rm p}$, the suprathermal particles form a dense beam of angular dispersion $ \ll 1$, carrying high inertia, and moving through a tenuous gas of sub- or mildly- relativistic temperature. The growth of the CFI in the precursor of relativistic collisionless shocks has been discussed in a number of numerical studies~\cite{Kato_2007,Spitkovsky_2008a, Martins_2009,Keshet_2009,Sironi_2009b, Sironi_2013,Haugbolle_2011,Kumar_2015}, but only a few theoretical studies have discussed its properties in such conditions~\cite{Lemoine_2010, Rabinak_2011, Lemoine_2011, Shaisultanov_2012}. In particular, the notion of a frame in which this CFI is essentially of magnetic nature has not, to the best of our knowledge, received attention so far; the only exception being Ref.~\cite{Ruyer_2016, *Ruyer_2017} where this frame was introduced to simplify the calculation of the CFI in the precursor of a nonrelativistic electron-ion shock. 
 
The present paper is the first of a series in which we discuss the microphysics of unmagnetized, relativistic collisionless pair shocks. Here, we address in detail this notion of a ``Weibel frame'', in which the electromagnetic configuration is essentially magnetic in nature. As shown in accompanying papers of this series, and in particular~\cite{L1}, this reference frame plays a fundamental role in the physics of collisionless shock waves. In Paper~II~\cite{pap2}, it is shown how the noninertial character of the ``Weibel frame'' controls the heating and slowdown of the background plasma.
In Paper~III~\cite{pap3}, this reference frame is used to calculate the scattering rate of suprathermal particles; the latter quantity controls the residence time of suprathermal particles in the upstream, hence the acceleration timescale, thus the maximum energy etc. Therefore, this notion of a ``Weibel frame'' has direct phenomenological consequences and potential observable radiative signatures. Finally, in Paper~IV~\cite{pap4}, we discuss the growth of the filamentary turbulence in the precursor of a collisionless shock. In the following, as in all accompanying papers, we support our present theoretical findings through detailed comparisons to dedicated large-scale 2D3V PIC simulations of relativistic collisionless shocks.

The present paper is laid out as follows. Section~\ref{sec:linwf} introduces the notion of the ``Weibel frame'' in a fluid model of the linear phase of the current filamentation instability. Section~\ref{sec:kinwf} then extends those results to a fully kinetic model of the CFI; the details of these calculations are provided in Apps.~\ref{app:simp_exp_eps}, \ref{app:low-temp_expansions} and \ref{app:series_expansion}. Section~\ref{sec:nlinwf} introduces the notion of the ``Weibel frame'' in a nonlinear model of the filamentation phase, in which the filaments are modeled in quasistatic equilibrium at all points in the precursor. Finally, Sec.~\ref{sec:linPIC} and \ref{sec:nlinPIC} provide detailed comparisons of those various models to dedicated PIC simulations. A summary and conclusions are provided in Sec.~\ref{sec:conc}. We use units such that $k_{\rm B} = c = 1$. The metric signature is $(-,+,+,+)$.  

\subsection{Setup}\label{sec:setup}
In this work, we consider an unmagnetized, relativistic collisionless shock wave propagating through an electron-positron plasma. In the shock rest frame $\mathcal R_{\rm s}$, the background plasma is initially incoming at relativistic velocity $\beta_\infty \simeq -1$ (Lorentz factor $\gamma_\infty \gg 1$) along the $x-$axis from $+\infty$. Inside the precursor, its mean velocity is written $\beta_{\rm p}$ (Lorentz factor $\gamma_{\rm p}$). The precursor is defined as the region permeated by a beam of suprathermal particles, which were reflected off the shock surface or accelerated through a Fermi-like process, possibly up to high energies. 

Our conventions are as follows: the index $_{\rm p}$ refers to the background plasma, while the index $_{\rm b}$ refers to the suprathermal particle population. For species $\alpha$, $n_\alpha$ represents the density, $w_\alpha$ the enthalpy density, $p_\alpha$ the pressure, and $T_\alpha$ the temperature, all measured in the species rest frame, while ${u_\alpha}^\mu = \gamma_\alpha\left(1,\bm{\beta_\alpha} \right)$
represents its four-velocity. All throughout, $\alpha$ will denote a species (`p' or `b') and not a space-time index. Both the background plasma and the suprathermal beam are composed of electrons and positrons, so that there are actually four species. However, unless otherwise noted, we do not make any particular distinction between electrons and positrons in a given population, and therefore drop any reference to the charge. The above proper hydrodynamic quantities nevertheless correspond to one charged species (electrons or positrons) of one population (either the background plasma or the suprathermal beam). 
As in other papers of this series, we rely on the following frames: the shock front rest frame $\mathcal R_{\rm s}$, the background plasma rest frame $\mathcal R_{\rm p}$, the downstream rest frame $\mathcal R_{\rm d}$ and the ``Weibel frame'' $\mathcal R_{\rm w}$, to be determined hereafter. Quantities evaluated in one or the other reference frame are indicated with the respective subscripts $_{\vert\rm s}$, $_{\vert\rm p}$, $_{\vert\rm d}$ and $_{\vert\rm w}$. By default, however, frame-dependent quantities that lack subscripts are defined in the lab frame, which coincides with the shock rest frame $\mathcal R_{\rm s}$. Finally, thermodynamic moments $n_\alpha$, $p_\alpha$, $T_\alpha$ and $w_\alpha$ are always defined as proper, unless otherwise stated; they are thus defined in the rest frame of species $\alpha$.

\begin{figure*}
  \centering
  \includegraphics[width=0.95\textwidth]{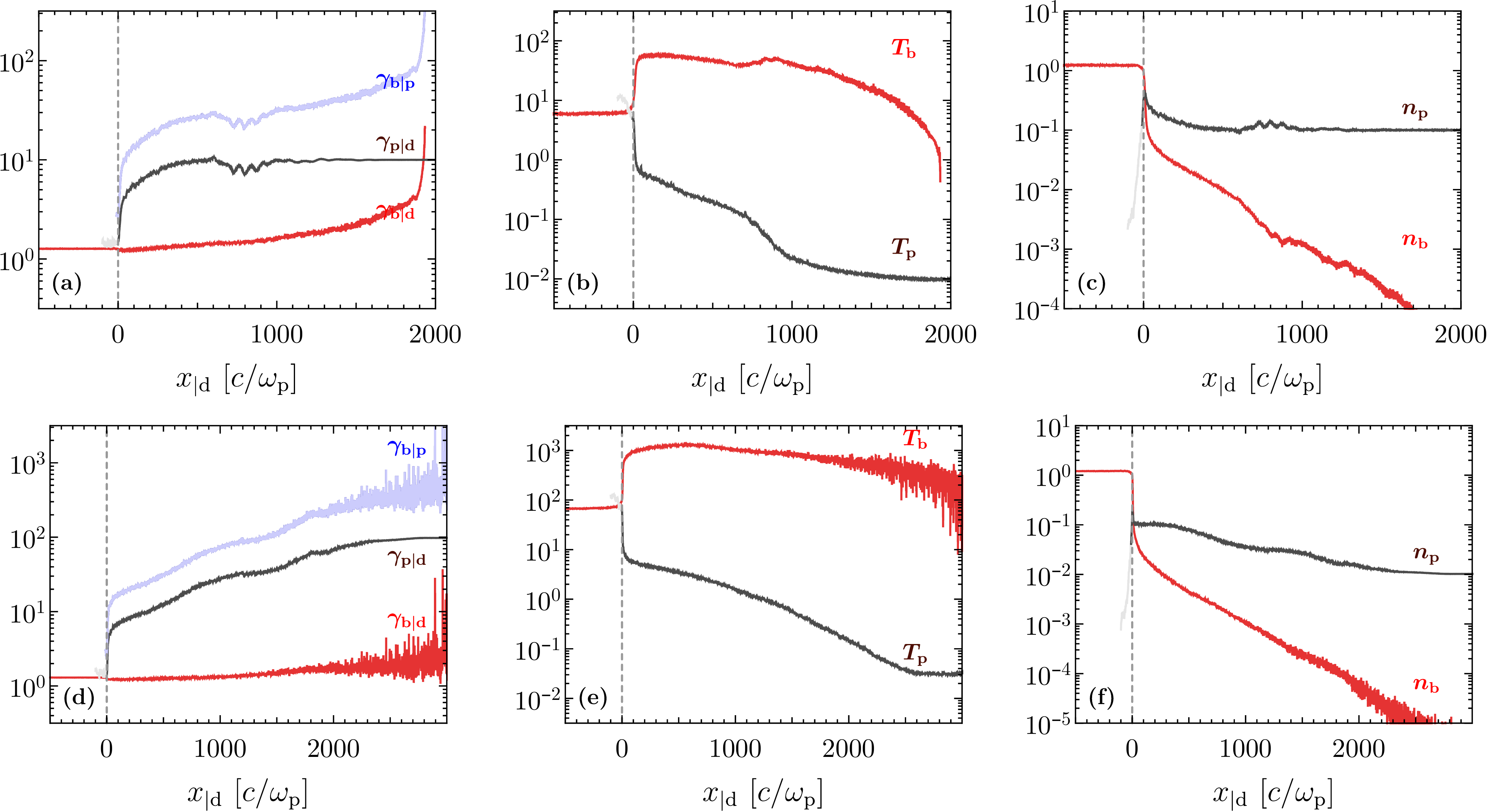}
  \caption{Spatial profiles of the main hydrodynamic quantities characterizing the background plasma and the beam of suprathermal particles in the shock precursor, as a function of distance to the shock front $x_{\rm \vert d}$ (simulation frame). Upper row, panels (a), (b) and (c): profiles extracted from a PIC simulation of  shock Lorentz factor $\gamma_\infty = 17$ (corresponding to a relative upstream-downstream Lorentz factor $\gamma_{\infty\vert\rm d} = 10$) at time $t = 3600\,\omega_{\rm p}^{-1}$; lower row, panels (d), (e) and (f): from a PIC simulation of shock Lorentz factor $\gamma_\infty = 173$ ($\gamma_{\infty\vert\rm d} = 100$) at time $t = 6900\,\omega_{\rm p}^{-1}$. Panels (a) and (d): Lorentz factor of the background plasma in the simulation (downstream) rest frame (dark gray), of the beam in the simulation frame (light red) and the relative Lorentz factor between the beam and the background plasma (light blue); panels (b) and (e): proper temperature of the background plasma (dark gray) and of the beam (light red); panels (c) and (f): proper density of the background plasma (dark gray) and of the beam (light red). The asymmetry of the beam-plasma configuration is manifest in those figures.
  \label{fig:profs}}
\end{figure*}

According to the fluid shock jump conditions~\cite{1976PhFl...19.1130B}, the typical temperature of shocked particles is $T_{\rm b} = \kappa_{T_{\rm b}}\gamma_\infty m_e$, where $\kappa_{T_{\rm b}} = (\widehat\Gamma_{\rm b}-1)(2-\widehat\Gamma_{\rm b})^{1/2}\widehat\Gamma_{\rm b}^{-1/2}$, in terms of the adiabatic index $\widehat\Gamma_{\rm b}$ of the shocked gas; the latter is relativistically hot, so $\widehat\Gamma_{\rm b} = 4/3$ in 3D, but $\widehat\Gamma_{\rm b} = 3/2$ in 2D, which must be considered when drawing comparison with a 2D3V (2D in configuration space, 3D in momentum space) PIC simulation; thus $\kappa_{T_{\rm b}} = 1/(3\sqrt{2})$ in 3D and $\kappa_{T_{\rm b}} = 1/(2\sqrt{3})$ in 2D.
Strictly speaking, the temperature of suprathermal particles is by definition larger than that of the shocked plasma, defined above. Yet, as this temperature of suprathermal particles always scales with $\gamma_\infty$, we retain the above definition, emphasizing that $\kappa_{T_{\rm b}}$ is generically larger than the above, by about an order of magnitude, so that $\kappa_{T_{\rm b}}\sim\,$a~few. This is illustrated in Fig.~\ref{fig:profs}, which shows the profiles of the main hydrodynamic properties of the background plasma and suprathermal beam for two reference PIC simulations, with respective Lorentz factors $\gamma_\infty = 17$ and $\gamma_\infty = 173$.

Our PIC simulation are described further on in Sec.~\ref{sec:linPIC}. Let us note here, however, that the reference frame of these 2D3V simulations (2D in configuration space, 3D in momentum space) coincide with the downstream rest frame $\mathcal R_{\rm d}$. Hence, a Lorentz factor $\gamma_\infty = 17$ (resp. $\gamma_\infty = 173$) in the $\mathcal R_{\rm s}$ frame corresponds to a simulation frame Lorentz factor 
$\gamma_{\infty\vert\rm d} = 10$ (resp. $\gamma_{\infty\vert\rm d} = 100$).  Simulations are conducted in the $x-y$ plane, with $\widehat{\boldsymbol{x}}$ oriented along the shock normal, and $\widehat{\boldsymbol{y}}$ defining the transverse dimension.

The ultrarelativistic suprathermal particle gas is nearly isotropic in the shock rest frame, meaning a drift Lorentz factor $\gamma_{\rm b} \sim 1$ in the shock rest frame (see Fig.~\ref{fig:profs}). This population is further characterized by its pressure $p_{\rm b}$, written $\xi_{\rm b}$ in units of the incoming momentum flux at infinity $F_\infty = \gamma_\infty^2\beta_\infty^2n_\infty m_e$:
\begin{equation}
p_{\rm b}\,=\,\xi_{\rm b} F_\infty\ .
\label{eq:xib}
\end{equation}

The leading instability that develops in the precursor of weakly magnetized ultrarelativistic shocks is the Weibel-type CFI described above. In principle, this instability is defined in all momentum space $(k_\parallel,\,\mathbf{k_\perp})$ where $k_\parallel=\mathbf{k}\cdot \widehat{\boldsymbol{x}}$ and $\mathbf{k_\perp} = k_\perp \widehat{\boldsymbol{y}}$ in our PIC simulations. The purely transverse modes correspond to $k_\parallel \ll k_\perp$, while the so-called oblique modes correspond to the limit $k_\parallel \simeq k_\perp \gg \omega_{\rm p}$, with $\omega_{\rm p} = \left(4\pi n_\infty e^2/m_e\right)^{1/2}$ the plasma frequency of (one charged species of) the unperturbed far-upstream background plasma. Oblique modes are likely relevant far in the precursor, but most likely Landau damped once the background plasma heats up \cite{Bret_2010a,Lemoine_2011,Shaisultanov_2012}. We thus assume that the transverse modes dominate in most of the precursor and restrict ourself to the limit $k_\parallel \rightarrow 0$ for simplicity.

\begin{figure}
\centering
\includegraphics[width=0.45\textwidth]{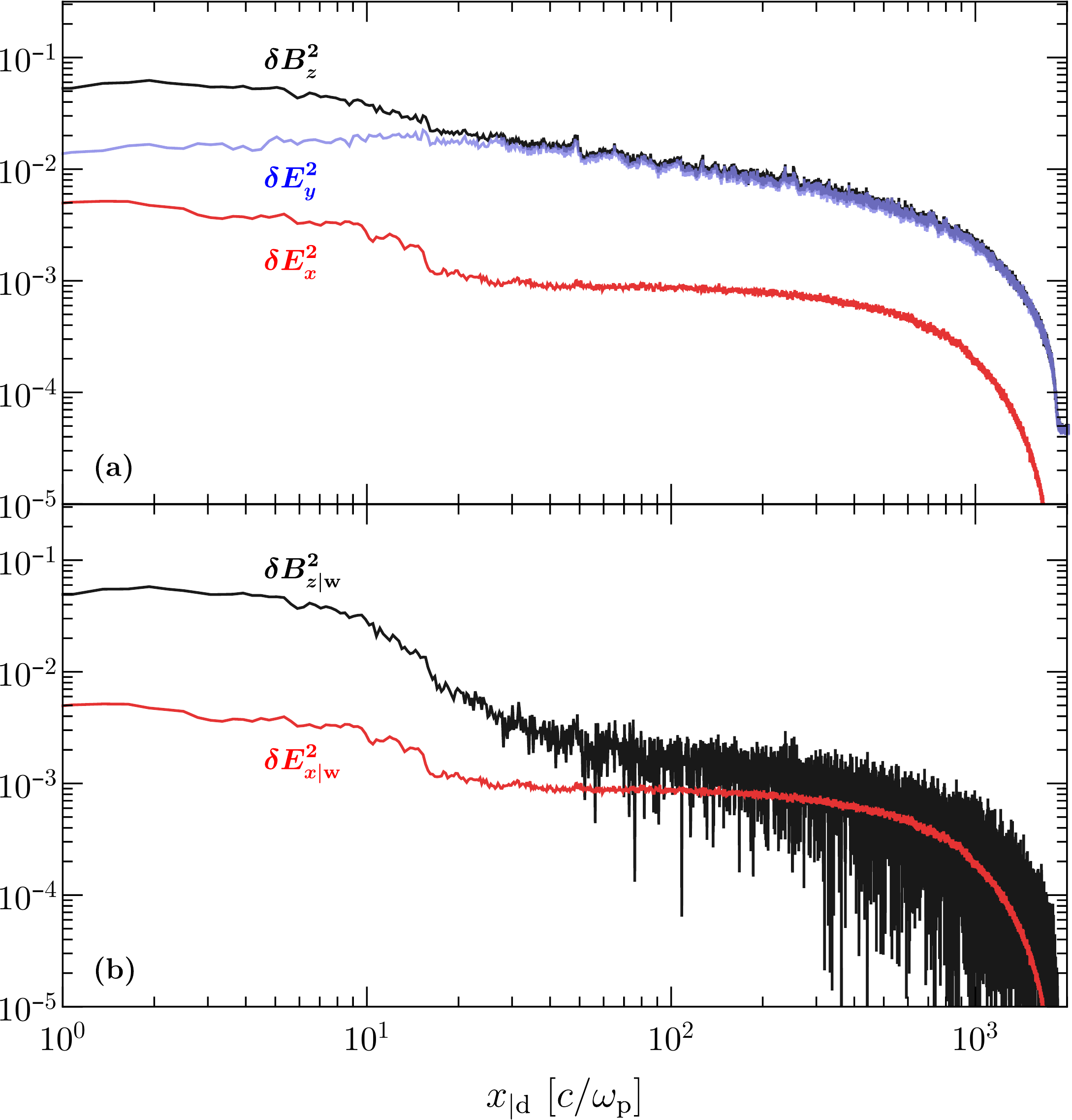}\smallskip\\
\includegraphics[width=0.45\textwidth]{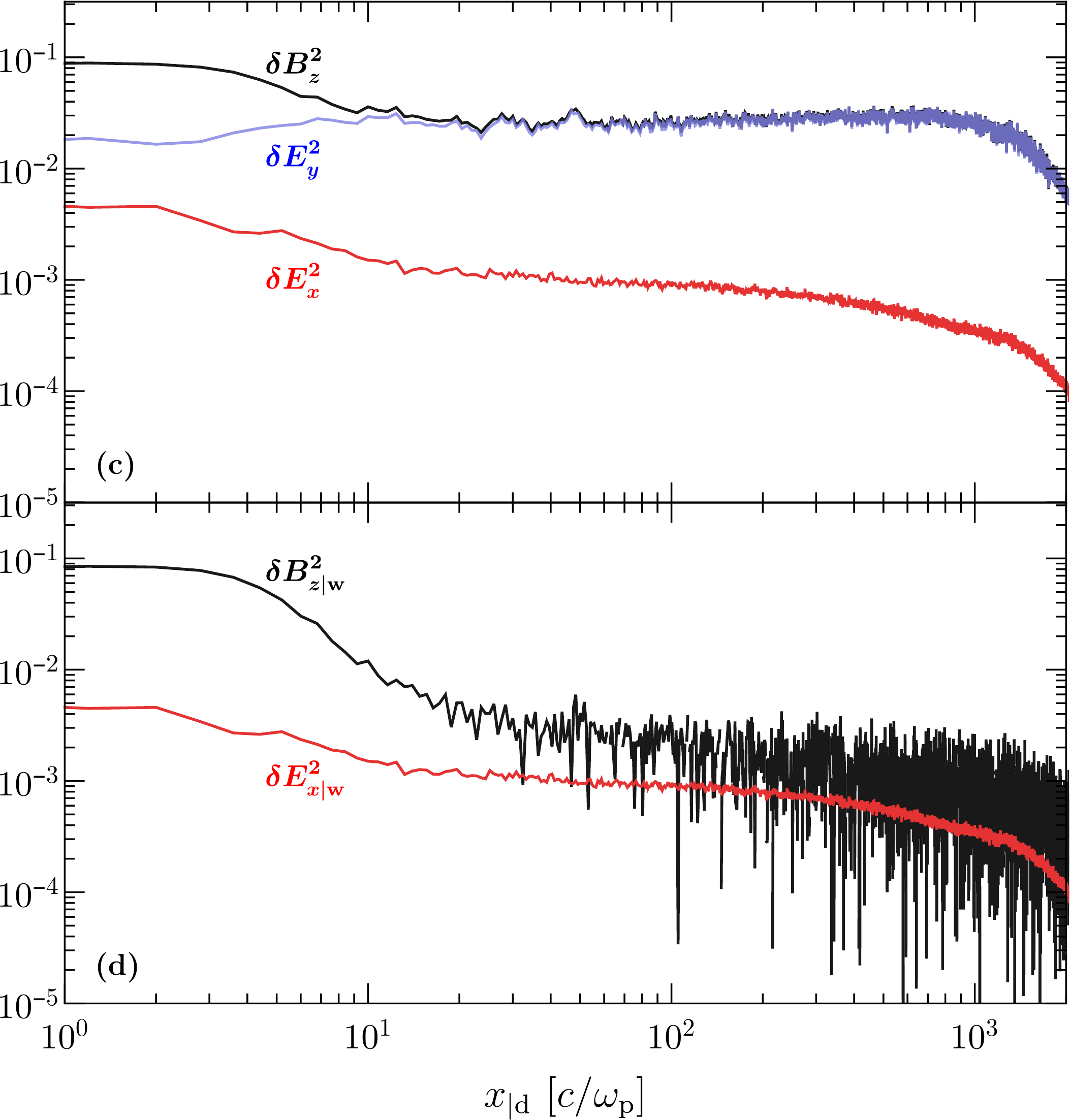}
\caption{Spatial profiles of normalized electromagnetic field energy densities in transverse magnetic field $\delta B_z$ (black), transverse electric field $\delta E_y$ (blue) and longitudinal electric field $\delta E_x$ (red), as a function of distance to the shock. Panels (a) and (b): extracted from simulation $\gamma_\infty = 17$ (simulation frame $\gamma_{\infty\vert\rm d} = 10$). Panels (c) and (d): simulation $\gamma_\infty = 173$ (simulation frame $\gamma_{\infty\vert\rm d} = 100$). Panels (a) and (c) show the energy densities measured in the simulation frame; panels (b) and (d) show the corresponding energy densities deboosted to the ``Weibel frame'' $\mathcal R_{\rm w}$ (where $\delta E_{y\vert\rm w}$ vanishes by definition). The velocity of $\mathcal R_{\rm w}$ is measured in the simulation as $\beta_{\rm w\vert d} = \langle\delta E_y^2\rangle^{1/2}/\langle\delta B_z^2\rangle^{1/2}$, where the average is taken over the transverse dimension of the simulation box. See text for details.
\label{fig:emdens}
}
\end{figure}

Our 2D3V PIC simulations indicate that this is a rea- sonable approximation. Consider indeed Fig.~\ref{fig:emdens}, which plots the energy densities in electromagnetic components $\delta E_x$, $\delta E_y$ and $\delta B_z$ in our two reference PIC simulations, normalized to the incoming momentum flux at infinity $F_\infty$. The purely transverse CFI generates a transverse $\delta B_z$ and its associated electrostatic $\delta E_y$ component; by definition of the ``Weibel frame'' $\mathcal R_{\rm w}$ (see further on), $\delta E_{y\vert\rm w} =,0$, and hence $\delta E_y = \beta_{\rm w\vert\rm d}\delta B_z$ in the $\mathcal R_{\rm d}$ simulation frame. The growth of this instability also generates an inductive longitudinal electric field $\delta E_x$. Nonstrictly transverse CFI modes (with $k_\parallel \neq 0$) further generate an electrostatic longitudinal $\delta E_x$ component. Figure~\ref{fig:emdens} reveals that the energy density in the transverse field components dominates that in the longitudinal electric field everywhere in the precursor in the $\mathcal R_{\rm d}$ frame, and that the electric field is smaller than the magnetic field in the near precursor where the ratio can be measured accurately (up to $x$ of the order of a few hundred to a thousand $c/\omega_{\rm p}$). Finally, the transverse magnetic field energy remains larger than the longitudinal electric field component when deboosted back to the $\mathcal R_{\rm w}$ frame. Therefore, this ``Weibel frame'' appears well defined in the 2D3V simulations.

Let us emphasize that the limit $k_\parallel \neq 0$ certainly remains of interest; as a matter of fact, a nontrivial structure along the streaming axis turns out to be a mandatory requirement to achieve pitch-angle diffusion of suprathermal particles~\cite{Achterberg_2007_I,pap3}. Yet, we expect that the main features of the (already nontrivial) calculations that follow will remain valid in a limit $0<k_\parallel \ll k_\perp$.  We thus consider purely transverse electromagnetic perturbations, \emph{i.e.}, $\delta A^\mu = \left(\delta \Phi,\delta A^x,0,0\right)$, and $\partial_x = 0$ for all quantities. 
 
Figure~\ref{fig:emdens} suggests that the ratio $\left\langle\delta E_y^2\right\rangle^{1/2}/\left\langle\delta B_z^2\right\rangle^{1/2}$, namely $\beta_{\rm w\vert d}$, depends on $x$, which implies that the ``Weibel frame'' is not globally inertial. This frame actually decelerates from the far to the near precursor, as discussed in \cite{L1}. The noninertial nature of $\mathcal R_{\rm w}$ has important consequences for the physics of the shock, most notably the deceleration and heating of the background plasma, which form the focus of a subsequent paper in this series~\cite{pap2}. 

In the present paper, we characterize the velocity $\beta_{\rm w\vert d}$ (more specifically, $\beta_{\rm w\vert p}$) at each point $x$ of the precursor, given the physical conditions at this point. The discussion that follows thus relies on an implicit WKB-like approximation, which stipulates that the `Weibel frame'' has time to adjust at each point to the local physical conditions on a deceleration length scale $\left\vert u_{\rm w}\,{\rm d}x/{\rm d} u_{\rm w}\right\vert$. Improving on this assumption would necessitate a proper inclusion of noninertial effects, which are characterized by the velocity profile ${\rm d}u_{\rm w}/{\rm d}x$, in the calculations that follow. However, this velocity profile is itself determined by the response of the background plasma and the suprathermal beam to the microturbulence. Such an endeavor thus represents a rather formidable task, well beyond the scope of any current study.

\section{The ``Weibel frame'' in a fluid model}\label{sec:linwf}

Here we present a fluid derivation of the CFI and of its associated ``Weibel frame'', in the context of the precursor of an unmagnetized, relativistic electron-positron shock. Although Sec.~\ref{sec:linPIC} will demonstrate that, in actual relativistic shock precursors, kinetic corrections are mandatory to describe the physics of the CFI, the following fluid model retains the advantage of simplicity as well as a pedagogical virtue. A fully kinetic description of the CFI and its ``Weibel frame'' in the case of Maxwell-J\"uttner plasma distribution functions will be provided in the forthcoming Sec.~\ref{sec:kinwf}.

To preserve covariance, we use a relativistic fluid formalism. The conservation of the total energy-momentum tensor can be written in the compact way~\cite{Achterberg_2007_I}:
\begin{equation}
  w_\alpha {u_\alpha}^\mu \partial_\mu {u_\alpha}^\nu + {h_{\alpha}}^{\mu\nu} \partial_\mu p_\alpha
   = q_\alpha n_\alpha {u_\alpha}_\mu F^{\nu\mu} \,,
\end{equation}
introducing ${h_{\alpha}}^{\mu\nu} = \eta^{\mu\nu} + {u_\alpha}^\mu {u_\alpha}^\nu$, which projects orthogonally to ${u_\alpha}^\mu$ (since ${u_\alpha}^\mu {u_\alpha}_\mu=-1$). The dynamical equation thus becomes, to first order in the perturbations,
\begin{equation}
  w_\alpha {u_\alpha}^\mu \partial_\mu {\delta u_\alpha}^\nu + {h_{\alpha}}^{\mu\nu} \partial_\mu \delta p_\alpha
  = q_\alpha n_\alpha {u_\alpha}_\mu \delta F^{\nu\mu}\ .
  \end{equation}
In the following, the four-velocity perturbation is decomposed as
${\delta u_\alpha}^\mu = \left(\delta \gamma_\alpha,\gamma_\alpha^3{\delta \beta_{\alpha}}^x,\gamma_\alpha {\delta \beta_{\alpha}}^y, \gamma_\alpha{\delta \beta_{\alpha}}^z\right)$,
with $\delta \gamma_\alpha = \gamma_\alpha^3 \beta_\alpha {\delta \beta_\alpha}^ x$. Using the short-hand notation
$\delta \boldsymbol{\beta_{\perp\alpha}} = \left(0,{\delta \beta_\alpha}^y,\,{\delta\beta_\alpha}^z\right)$ and
$\boldsymbol{\nabla_\perp} = \left(0,\partial_y,\,\partial_z\right)$, the system can be rewritten explicitly as
\begin{align}
  \gamma_\alpha^2 w_\alpha\partial_t{\delta \beta_\alpha}^x + \beta_\alpha \partial_t \delta p_\alpha
  &= - \frac{q_\alpha n_\alpha}{\gamma_\alpha} \partial_t \delta A^x \label{eq:dbx} \,, \\
  \gamma_\alpha^2 w_\alpha \partial_t \delta\boldsymbol{\beta_{\perp\alpha }} + \boldsymbol{\nabla_\perp}\delta p_\alpha
  &= - q_\alpha n_\alpha \gamma_\alpha \boldsymbol{\nabla_\perp} \left(\delta\Phi - \beta_\alpha \delta A^x\right) \,.
  \label{eq:dbp}
\end{align}
Current conservation written to first order also yields
\begin{equation}
  \gamma_\alpha \partial_t \delta n_\alpha + n_\alpha \partial_t \delta\gamma_\alpha +
  \gamma_\alpha n_\alpha\boldsymbol{\nabla_\perp} \cdot \delta \boldsymbol{\beta_{\perp\alpha}} = 0 \,.
  \label{eq:dn}
  \end{equation}
The pressure perturbation can be related to the density perturbation through the adiabatic index $\widehat{\Gamma}_\alpha$:
\begin{equation}
  \delta p_\alpha = \widehat{\Gamma}_\alpha \frac{p_\alpha}{n_\alpha}\delta n_\alpha \,.
  \label{eq:pn}
\end{equation}
In the following, we use the isentropic sound speed squared
\begin{equation}
  c_\alpha^2 \equiv \widehat\Gamma_\alpha \frac{p_\alpha}{w_\alpha} \,,
  \label{eq:ca}
\end{equation}
so that $\delta p_\alpha = c_\alpha^2(w_\alpha/n_\alpha)\delta n_\alpha$. Given the relation between $\delta \gamma_\alpha$ and ${\delta \beta_{\alpha}}^x$, Eq.~\eqref{eq:dbx} can then be used to express the pressure perturbation in terms of the perturbed
\emph{apparent} density $\delta N_\alpha \equiv \delta (\gamma_\alpha n_\alpha)$:
\begin{equation}
  \partial_t \delta p_\alpha = \frac{c_\alpha^2}{\gamma_\alpha^2\left(1-c_\alpha^2\beta_\alpha^2 \right)} \left[q_\alpha \gamma_\alpha n_\alpha \beta_\alpha \partial_t \delta A^x + \frac{\gamma_\alpha  w_\alpha}{n_\alpha} \partial_t
  \delta N_\alpha \right]  \,.
    \label{eq:dpdn}
\end{equation}
This equation involves the effective sound speed squared of the streaming plasma species,
\begin{equation}
  c_{\rm eff\,\alpha}^2 = \frac{c_\alpha^2}{\gamma_\alpha^2 \left(1-c_\alpha^2 \beta_\alpha^2 \right)} \, .
  \label{eq:ceff}
\end{equation}
We have $c_{\rm eff\,\alpha}^2 \simeq 5T_\alpha/(3\gamma_\alpha^2)$ and $c_{\rm eff\,\alpha}^2 \simeq 1/(2\gamma_\alpha^2)$ in the nonrelativistic ($T_\alpha \ll m_e$, $\widehat\Gamma_\alpha \simeq 5/3$) and ultrarelativistic ($T_\alpha \gg m_e$, $\widehat\Gamma_\alpha \simeq 4/3$)
thermal limits of a 3D gas, respectively. After integration, Eq.~\eqref{eq:dpdn} provides a direct relationship between $\delta p_\alpha$ and $\delta A_x$ and $\delta N_\alpha$, which can be inserted in Eq.~\eqref{eq:dbp} to yield
\begin{align}
  \gamma_\alpha^2 w_\alpha \partial_t \delta\boldsymbol{\beta_{\perp\alpha}}
  + & c_{{\rm eff}\,\alpha}^2 \frac{\gamma_\alpha w_\alpha}{n_\alpha} \boldsymbol{\nabla_\perp} \delta N_\alpha = \nonumber\\
&\quad
 -\,q_\alpha \gamma_\alpha n_\alpha \left(\boldsymbol{\nabla_\perp}\delta\Phi -\beta_\alpha \boldsymbol{\nabla_\perp} \delta A^x \right) \,.
 \label{eq:dbp2}
\end{align}
Finally, combining this equation with Eq~\eqref{eq:dn}, one obtains
\begin{align}
  \partial_t^2 \delta N_\alpha - & c_{\rm eff\,\alpha}^2 \Laplace \delta N_\alpha =  \nonumber\\
&\quad\quad  \frac{q_\alpha n_\alpha^2}{w_\alpha}\left[ \Laplace \delta\Phi - \beta_\alpha \left( 1-c_{\rm eff\,\alpha}^2 \right) \Laplace \delta A^x \right] \,,\nonumber\\ 
&
  \label{eq:dn2}
\end{align}
which determines the response of the {\rm apparent} charge density to the electromagnetic perturbation: in Fourier space, with $\delta \rho_\alpha = q_\alpha \delta N_\alpha$, one finds
\begin{equation}
  \delta \rho_\alpha = \frac{\Omega_{\rm p\alpha}^2}{4\pi}\frac{k^2}{\omega^2-c_{\rm eff\,\alpha}^2k^2}
  \left[\delta \Phi - \beta_{\alpha} \left(1-c_{\rm eff\,\alpha}^2 \right) \delta A^x \right] \, ,
  \label{eq:drho}
\end{equation}
where the relativistic (proper-frame) plasma frequency, $\Omega_{\rm p\alpha}$, is defined by
\begin{equation}  \label{eq:dw}
  \Omega_{\rm p\alpha}^2 \equiv \frac{4\pi n_\alpha q_\alpha^2}{w_\alpha/n_\alpha} \,.
\end{equation}
Note that the far-upstream plasma frequency $\Omega_{\rm pp}$ coincides with $\omega_{\rm p}$ defined earlier.

The above also allows us to determine the response of the current density, ${\delta j_\alpha}^x$, which comprises both a perturbed conduction current density as well as a perturbed advection current density:
\begin{equation}
  {\delta j_\alpha}^x = \rho_\alpha {\delta \beta_\alpha}^x + \beta_\alpha\delta \rho_\alpha \,. \label{eq:djx}
\end{equation}
Equations~\eqref{eq:dbx}, \eqref{eq:dpdn} and \eqref{eq:drho} can then be combined to derive the response in Fourier space:
\begin{align}  \label{eq:rdjx}
  {\delta j_\alpha}^x = \frac{\Omega_{\rm p\alpha}^2}{4\pi}\biggl\{& 
   \beta_\alpha \left( 1-c_{\rm eff\,\alpha}^2 \right) \frac{k^2}{\omega^2-c_{\rm eff\,\alpha}^2k^2} \delta \Phi   \nonumber\\
 & -  \left[ 1 - \beta_\alpha^2 \left( 1-c_{\rm eff\,\alpha}^2 \right) \frac{\omega^2-k^2}{\omega^2-c_{\rm eff\,\alpha}^2k^2} \right]
   \delta A^x  \biggr\} \, .
\end{align}

We define the ``Weibel frame'' as that in which the linear instability becomes purely magnetic, \emph{i.e.}, the electrostatic potential and the total electric charge density vanish. From the response of the charge density, one finds that in this frame the following relation must be
fulfilled:
\begin{equation}   \label{eq:Wf}
  \sum_\alpha \Omega_{\rm p\alpha}^2 \beta_{\alpha \vert \rm w}  \frac{ 1-c_{\rm eff\, \alpha}^2}{\zeta^2-c_{\rm eff\, \alpha}^2}= 0 \,,
\end{equation}
where $\zeta=\omega/k$ is the (complex) phase velocity .
In the shock precursor, the background plasma, of proper density $n_{\rm p}$ and nonrelativistic temperature $T_{\rm p} \lesssim 1$, flows at a velocity $\beta_{\rm p \vert s} \simeq -1$ against the suprathermal beam, of proper density $n_{\rm b}$ and relativistic temperature $T_{\rm b} \gg 1$. To leading thermal corrections, the above equation can be recast in the form
\begin{equation}
  n_{\rm p} \beta_{\rm p \vert w} \left( 1 + \frac{\widehat\Gamma_{\rm p} T_{\rm p}}{\gamma_{\rm p \vert w}^2 \zeta^2} \right)
  + \frac{n_{\rm b} \beta_{\rm b \vert w}}{T_{\rm b}} \left( 1- \frac{1}{\widehat\Gamma_{\rm b}}\right)  \simeq 0 \,,
\label{eq:wf_hydro0}
\end{equation}
where we have assumed $\vert \zeta \vert ^2 \gg c_{\rm eff\,\alpha}^2$ for both populations, as is consistent within a hydrodynamic model (see Sec.~\ref{sec:kinwf}). This relation indicates that the velocity of $\mathcal R_{\rm w}$ is, in principle, mode-dependent. Yet the response of the beam proves to be more sensitive to thermal effects than that of the plasma since the inequality $\zeta^2 \gg c_{\rm eff\,\alpha}^2$ implies $T_{\rm p} / \gamma_{\rm p \vert w}^2 \zeta^2 \ll 1$. As a consequence,
\begin{equation}
  \beta_{\rm w \vert p } \equiv  - \beta_{\rm p \vert w} \simeq \frac{n_{\rm b} \beta_{\rm b \vert w}}{n_{\rm p} T_{\rm b}}
  \left( 1- \frac{1}{\widehat\Gamma_{\rm b}} \right)
\label{eq:wf_hydro1} 
\end{equation}
is a good (mode-independent) approximation of the velocity of $\mathcal R_{\rm w}$ relative to the background plasma frame.

The beam is more conveniently characterized by its normalized pressure $\xi_{\rm b}$ and temperature $T_{\rm b}\,=\,\kappa_{T_{\rm b}}\gamma_\infty m_e$, so that $n_{\rm b}\,=\,\kappa_{T_{\rm b}}^{-1}\gamma_\infty\xi_{\rm b}n_\infty$, see Sec.~\ref{sec:setup}. Current conservation further implies $n_\infty/n_{\rm p} \,=\,\gamma_{\rm p}\beta_{\rm p}/(\gamma_\infty\beta_\infty)$, with $\beta_{\rm p}\,\simeq\,\beta_\infty\,\simeq\,-1$, so that
\begin{equation}
  \beta_{\rm w\vert p} \simeq \beta_{\rm b \vert w}\, \xi_{\rm b}\, \frac{\gamma_{\rm p}}{\gamma_\infty} \,\frac{\widehat\Gamma_{\rm b}- 1}{\kappa_{T_{\rm b}}^2 \widehat\Gamma_{\rm b}}\,.
\label{eq:wff2}
\end{equation}
The beam moves relativistically with respect to the background plasma, therefore either $\beta_{\rm b \vert w} \sim 1$ or $\beta_{\rm p \vert w} \sim -1$. As $\xi_{\rm b} < 1$, however, the former must hold, which provides the final result:
\begin{equation} \label{eq:Wfc3}
  \beta_{\rm w \vert p} \simeq \xi_{\rm b}\, \frac{\gamma_{\rm p}}{\gamma_\infty} \,\frac{\widehat\Gamma_{\rm b}- 1}{\kappa_{T_{\rm b}}^2 \widehat\Gamma_{\rm b}} \,.
\end{equation}
In the case of negligible deceleration of the incoming plasma ($\gamma_{\rm p} \simeq \gamma_\infty$) and for a 3D adiabatic index $\widehat\Gamma_{\rm b} = 4/3$, 
\begin{equation}
  \beta_{\rm w \vert p} \simeq \frac{1}{4\kappa_{T_{\rm b}}^2}\xi_{\rm b} \, .
\end{equation}
One can also directly calculate $\beta_{\rm w}$:
\begin{equation} \label{eq:Wvelf1}
\beta_{\rm w} = \frac{\beta_{\rm w\vert p} + \beta_{\rm p}}{1+\beta_{\rm w\vert p} \beta_{\rm p}}   
 \simeq \beta_{\rm p}\left(1 - \frac{1}{4\kappa_{T_{\rm b}}^2}\frac{\xi_{\rm b}}{\gamma_{\rm p}^2}\right) \,.
\end{equation}

The above indicates that: (i) the ``Weibel frame'' $\mathcal R_{\rm w}$ moves at a sub-relativistic velocity relative to the background plasma [\emph{i.e.}, $u_{\rm w\vert p} =\beta_{\rm w\vert p}(1-\beta_{\rm w\vert p}^2)^{-1/2} < 1$], and therefore at a relativistic velocity $\simeq \beta_{\rm p}$ relative to the shock front; (ii) $\beta_{\rm w\vert p}$ is of opposite sign to $\beta_{\rm p}$, which implies that, in magnitude, the ``Weibel frame'' moves slightly less fast than the background plasma relative to the shock front; (iii) the relative velocity $\beta_{\rm w\vert p}$ scales with $\xi_{\rm b}$.

Finally, because $\xi_{\rm b}$ is a function of distance $x$ to the shock, Eq.~(\ref{eq:Wvelf1}) indicates that $\beta_{\rm w}$ itself depends on $x$, {\it i.e.}, the ``Weibel frame'' is not globally inertial, as already observed in Fig.~\ref{fig:emdens}. This observation has important implications with respect to the physics of the shock, which are addressed in detail in the follow-up paper~\cite{pap2}.

In Sec.~\ref{sec:kinwf}, we extend the above calculations to a kinetic description. Although the expression for $\beta_{\rm w\vert p}$ will be found to take somewhat different values, the above three features will remain valid.

Using the above equations, it becomes straightforward to derive the dispersion relation of the CFI in this warm fluid description. Consider the CFI in $\mathcal R_{\rm w}$, where $\delta \Phi_{\rm\vert w} = 0$. The response
${\delta j_{\rm p}}_{\rm\vert w}^x$ of the background plasma can be written
\begin{equation}
  {\delta j_{\rm p}}_{\rm\vert w}^x \simeq \frac{\omega_{\rm p}^2}{4\pi}\, \delta A_{\rm\vert w}^x 
\end{equation}
to first order in $\xi_{\rm b}$, since the term that has been neglected here with respect to Eq.~\eqref{eq:rdjx} is of the order of
$\beta_{\rm p\vert w}^2 \sim \xi_{\rm b}^2$. Note also that we have assumed that the background plasma temperature remains sub-relativistic over most of the precursor, as discussed in Paper~II~\cite{pap2}, so that $\Omega_{\rm pp} \simeq \omega_{\rm p}$.
From the Maxwell equations written in the Lorentz gauge ($k^\mu \delta A_\mu = 0$),
\emph{i.e.}, $\DAlembert \delta A^\mu = -4\pi\sum_\alpha {\delta j_\alpha}^\mu$, one derives the dispersion relation in the ``Weibel frame'' (subscript $_{\rm w}$ omitted for clarity):
\begin{align}
  \left(\omega^2-k^2-\omega_{\rm p}^2\right) &= \Omega_{\rm pb}^2 \left[ 1-\beta_{\rm b \vert w}^2 \left( 1-c_{\rm eff\,b}^2 \right)
  \frac{\omega^2-k^2}{\omega^2-k^2c_{\rm eff\,b}^2} \right] \,,\nonumber\\
&
\end{align}
which, assuming $\vert\omega\vert^2 \ll k^2$ and $c_{\rm eff\,b}^2 \ll 1$, can be approximated by
\begin{equation}
 \left( \omega^2 - k^2c_{\rm eff\,b}^2 \right) \left(\omega^2 - k^2-\omega_{\rm p}^2 \right) \simeq \Omega_{\rm pb}^2 \beta_{\rm b \vert w}^2k^2 \,,
\end{equation}
with solution
\begin{equation}
  \omega^2 \simeq k^2 c_{\rm eff\,b}^2 - \Omega_{\rm pb}^2 \beta_{\rm b\vert w}^2 \frac{k^2}{k^2 + \omega_{\rm p}^2} \,.
\label{eq:Wsol1}
\end{equation}
The stabilizing effect of the beam dispersion is manifest in this equation; this effect has been noted before in~\cite{Rabinak_2011,Lemoine_2011}.

\section{Relativistic kinetic model} \label{sec:kinwf}

In this Section, we evaluate the ``Weibel frame'' velocity and the local instability growth rate within a rigorous kinetic formalism. As discussed further below, kinetic effects must indeed be taken into account when considering the development of the CFI in the shock precursor. The derivation of the kinetic dielectric tensor involves rather heavy calculations, which are relegated to Apps.~\ref{app:simp_exp_eps}, \ref{app:low-temp_expansions} and \ref{app:series_expansion}. Below, we describe the general method, summarize the approximate expressions of the relevant dielectric tensor elements, and provide the estimates of the growth rate of the CFI and the associated estimate of $\beta_{\rm w}$ in various limits. The latter results, in particular, are given in Sec.~\ref{subsec:Weibel_rates_velocities}. 

For the time being, the reference frame in which we work is left unspecified. We first recall the linear dispersion relation fulfilled by the CFI modes \cite{Silva_2002}:
\begin{equation} \label{eq:disp}
  \varepsilon_{yy}(\varepsilon_{xx}-1/\zeta^2)=\varepsilon_{xy}^2 \,,
\end{equation}
with $\zeta = \omega /k$ as the complex phase velocity. In a fully relativistic framework, the dielectric tensor elements read ($i,j\,=\,1,\,2,\,3$)
\cite{Ichimaru_1973}
\begin{align} \label{eq:eps}
  \varepsilon_{ij}(\omega,k) = &\delta_{ij} +\sum_\alpha \frac{\gamma_\alpha \omega_{\rm p\alpha}^2}
  {\zeta^2 k^2} \int \frac{u_i}{\gamma}\frac{\partial f_\alpha^{(0)}} {\partial u^j}\,{\rm d}^3 u  
  \nonumber\\
&\,+\, \sum_\alpha \frac{\gamma_\alpha \omega_{\rm p\alpha}^2}{\zeta^2 k^3}  
  \int \frac{u_i u_j}{\gamma^2 m_e} \frac{k\partial  f_\alpha^{(0)}/\partial u_y}{\zeta  -v_y} \, {\rm d}^3 u \,,
\end{align}
where $v_y = u_y/(\gamma m_e)$, $\omega_{\rm p\alpha}^2=4\pi n_\alpha e^2/m_e$ represents the nonrelativistic plasma frequency squared of species $\alpha$ ($n_\alpha$ represents as before the proper density) and $f_\alpha^{(0)}(\mathbf{u})$ the corresponding unperturbed momentum distribution function, normalized such that $\int {\rm d}^3u\, f_\alpha^{(0)} = 1$. If the non-diagonal tensor element $\varepsilon_{xy}$ happens to vanish, Eq.~\eqref{eq:disp} implies either $\varepsilon_{yy}=0$ or $\zeta^2 \varepsilon_{xx}-1=0$. These two dispersion
relations describe, respectively, purely electrostatic modes (with $\mathbf{\delta E} \parallel \widehat{\mathbf{y}}$) and purely electromagnetic (or inductive) modes (with $\mathbf{\delta E}\parallel \widehat{\mathbf{x}}$). Assuming that $f_\alpha^{(0)}(\mathbf{u})$ is even
in $p_y$, $\varepsilon_{xy}$ reduces to
\begin{equation}\label{eq:epsxy_init}
  \varepsilon_{xy}(\omega,k) =  \sum_\alpha \frac{\gamma_\alpha \omega_{\rm p\alpha}^2}{\omega^2}
  \int  \frac{u_x u_y}{m_e \gamma^2} \frac{\partial f_\alpha^{(0)}/\partial u_y} {\zeta -v_y} \, {\rm d}^3u \,.
\end{equation}
This expression is generally nonzero, and hence the CFI excites mixed electromagnetic/electrostatic fluctuations.
This feature has been often overlooked in the past, $\delta E_y = 0$ being assumed from the outset in a number of calculations
\cite{Molvig_1975,Cary_1981,Okada_1980a,Hill_2005}. The electromagnetic ($\delta E_x$) and electrostatic ($\delta E_y$)
components of the solutions to Eq.~\eqref{eq:disp} verify
\cite{Bret_2004,Bret_2007}:
\begin{equation}\label{eq:EysEx}
  \frac{\delta E_y}{\delta E_x} = -\frac{\varepsilon_{xy}}{\varepsilon_{yy}} \,.
\end{equation}

In Sect.~\ref{sec:linwf}, $\beta_{\rm p \vert w}$ was determined by solving $\varepsilon_{xy}=0$ in the fluid limit, exploiting the fact that, to leading order, this equation is independent of the complex frequency $\zeta$. In the general kinetic case, however, $\varepsilon_{xy}$ depends on $\zeta$, whose knowledge involves solving Eq.~\eqref{eq:disp}. In practice, we are interested in the fastest-growing mode ($\zeta_{\rm max}$), which should be calculated in the (unknown) ``Weibel frame''. Instead of solving simultaneously Eq.~\eqref{eq:disp} and $\varepsilon_{xy \vert \rm w}=0$ for $\zeta_{\rm max \vert w}$ and $\beta_{\rm p \vert w}$, we follow a different approach, noting that the $(\delta E_y, \delta B_z)$ fluctuations pertaining
to a given mode in the plasma and ``Weibel frames'' are related through
\begin{align}
  \delta E_{y \vert \rm p} &=  \gamma_{\rm w \vert p}\beta_{\rm w \vert p} \delta B_{z \vert \rm w} \, ,\\
  \delta B_{z \vert \rm p} &= \gamma_{\rm w \vert p} \delta B_{z\vert \rm w} \,,
\end{align}
since $\delta E_{y \vert \rm w}=0$. The velocity of the ``Weibel frame'' relative to the plasma rest frame is therefore given by
\begin{equation}
  \beta_{\rm w\vert p} = \frac{\delta E_{y \vert \rm p} }{\delta B_{z \vert \rm p}} \,.
\end{equation}
Making use of Eq.~$\eqref{eq:EysEx}$ and of $\delta B_{z \vert \rm p} = - \delta E_{x \vert \rm p} / \zeta_{\rm max \vert p}$
, one obtains
\begin{equation}
  \beta_{\rm w \vert p} = \zeta_{\rm max \vert p} \frac{\varepsilon_{xy \vert \rm p}}{\varepsilon_{yy \vert \rm p}} \,.
\label{eq:beta_w_ex}
\end{equation}
This formula has the advantage of involving only quantities measured in the plasma frame. 

We apply the above formalism to the case of Maxwell-J\"uttner momentum distribution functions \cite{Juttner_1911}:
\begin{equation}\label{eq:edf}
  f_\alpha^{(0)}(\mathbf{u}) = \frac{\mu_\alpha}{4\pi m_e^3\gamma_\alpha K_2 (\mu_\alpha)}        
  \exp \left[-\gamma_\alpha \mu_\alpha (\gamma - \beta_\alpha u_x/m_e) \right] \,,
\end{equation}
where $\beta_\alpha \equiv \langle u^x /\gamma m_e \rangle$ is the normalized mean drift velocity of species $\alpha$ (corresponding drift Lorentz factor $\gamma_\alpha$), $\mu_\alpha \equiv m_e /T_\alpha$ and $K_2$ denotes a modified Bessel function of the second kind.

Compact expressions of the tensor elements $\varepsilon_{lm}$ can be obtained in terms of one-dimensional integrals over the velocity parallel to the wave vector ($v_\parallel$) \cite{Wright_1975,Bret_2007}. These calculations are detailed in Appendix~\ref{app:simp_exp_eps}.
In Sec.~\ref{sec:linPIC}, such expressions will be used for the numerical resolution of Eq.~\eqref{eq:beta_w_ex} along with Eq.~\eqref{eq:disp}. The parameters of the background plasma and suprathermal beam will be then extracted from PIC simulations of relativistic collisionless shocks. In the remainder of this section, we will derive analytic approximations of $\zeta_{\rm max \vert p}$ and $\beta_{\rm w \vert p}$, valid in distinct instability regimes for the plasma and beam particles.

The starting points of these calculations are the alternative expressions \eqref{eq:epsxx_alt}, \eqref{eq:epsyy_alt} and \eqref{eq:epsxy_alt} of the
dielectric tensor. For instance, $\varepsilon_{xx}$ can be rewritten as \eqref{eq:epsxx_alt}
\begin{equation}
  \varepsilon_{xx} = 1 + \sum_\alpha\frac{\omega_{\rm p\alpha}^2}{k^2 \zeta^2}   \mu_\alpha \gamma_\alpha ^2 \beta_\alpha^2
  - \frac{\omega_{\rm p\alpha}^2}{4 k^2 \zeta}  \frac{\mu_\alpha^2 \gamma_\alpha}{K_2(\mu_\alpha)} \mathcal{A}_{xx}^\alpha \,, 
  \label{eq:epsxx_alt_rep}
\end{equation}
where 
\begin{align}
  \mathcal{A}_{xx}^\alpha &= \frac{2}{\mu_\alpha \sqrt{1-\zeta^2}}  \int _{-\infty}^\infty {\rm d}s \,\frac{1}{\chi_\alpha -s} \nonumber\\
&\times
  \biggl\{ \frac{\gamma_\alpha^2 \beta_\alpha^2}{(s^2+1)^{3/2}}
  + \frac{1}{\mu_\alpha} \left[ \frac{1}{s^2+1}
  + \frac{3\beta_\alpha^2\gamma_\alpha^2}{(s^2+1)^2} \right]  \nonumber\\
 &\quad\,\,  +  \frac{1}{\mu_\alpha^2} \left[\frac{1}{(s^2+1)^{3/2}}
  +\frac{3\beta_\alpha^2\gamma_\alpha^2}{(s^2+1)^{5/2}} \right] \biggr\} e^{-\mu_\alpha \sqrt{s^2+1}} \,,
  \label{eq:Axx_alt_rep}
\end{align}
and $\chi_\alpha = \gamma_\alpha \zeta /\sqrt{1-\zeta^2}$. The integrals involved in $\varepsilon_{yy}$ and $\varepsilon_{xy}$ can be put in a similar form,
see \eqref{eq:int_Ayy_alt} and \eqref{eq:int_Axy_alt}. Introducing $\Delta s_\alpha$ the characteristic width of the integrand in Eq.~\eqref{eq:Axx_alt_rep}
[except for the denominator $(\chi_\alpha -s)^{-1}$], two limiting cases can be considered for each plasma species:
\begin{itemize}
\item the `hydrodynamic' limit: $\vert \chi_\alpha \vert \gg \Delta s_\alpha$;
\item the `kinetic' limit: $\vert \chi_\alpha \vert \ll \Delta s_\alpha $.
\end{itemize}
The dimensionless variable $s$ is introduced immediately before Eq.~(\ref{eq:epsxx_alt}); it corresponds to $\gamma_\alpha\beta_\parallel\gamma_\parallel$, with $\beta_\parallel$ the component of the particle velocity along the wavevector. Since the wavevector is transverse to the streaming direction, the typical extent of $s$ in the above integral is, up to the resonant factor, controlled by the proper temperature $1/\mu_\alpha$; the parameter $\chi_\alpha$ itself corresponds to the apparent phase four-velocity of the mode. Therefore, the meaning of the hydrodynamic limit is that the apparent typical transverse momentum (normalized to $m_e$) exceeds the apparent phase four-velocity, while the kinetic limit corresponds to the opposite case. In the following, approximate formulas of the dielectric tensor will be derived in these two limits.

\subsection{Evaluation of the dielectric tensor for the background plasma}

\subsubsection{Hydrodynamic limit}

In the outermost part of the precursor, the background plasma is characterized by a nonrelativistic proper temperature, $\mu_{\rm p} \gg 1$~\cite{pap2}.
In this limit, $\Delta s_{\rm p} \simeq \sqrt{2/\mu_{\rm p}}$, and hence the hydrodynamic response of the background plasma implies
$\sqrt{\mu_{\rm p}/2} \vert \chi_{\rm p} \vert \gg 1$. This condition coincides with the large-argument limit ($\tilde{\chi}_{\rm p}\simeq\vert \chi_{\rm p} \vert \sqrt{\mu_{\rm p}/2}  \gg 1$)
of the $\mathcal{Z}$ and $\mathcal{Z}'$ functions involved in the low-temperature expressions derived in Appendix~\ref{app:low-temp_expansions}.
These formulas can be further expanded to first order in $1/\mu_{\rm p}$ by making use of the asymptotic series
$\mathcal{Z}(\eta) \simeq -1/\eta - 1/2\eta^3 - 3/4\eta^5\cdots$ \cite{NRL_2013}:
\begin{align}
  \varepsilon_{xx}^{\rm p} &\simeq 1 - \frac{\omega_{\rm p}^2}{k^2 \zeta^2}  \left\{ 1 + \left[\beta_{\rm p}^2
  + \frac{1}{\mu_{\rm p}}\left( 1-\frac{5}{2}\beta_{\rm p}^2 \right) \left( \frac{1}{\zeta^2} -1 \right) \right] \right\}  \,,
  \label{eq:exx_hyd1} \\
  \varepsilon_{yy}^{\rm p} &\simeq 1 - \frac{\omega_{\rm p}^2}{k^2\zeta^2} \left[ 1+ \frac{3\gamma_{\rm p}}{\mu_{\rm p}} \left( \frac{1}{\zeta^2} -1 \right) \right] \,,
  \label{eq:eyy_hyd1} \\
  \varepsilon_{xy}^{\rm p} &\simeq -\frac{\omega_{\rm p}^2 \beta_{\rm p}}{k^2 \zeta^3} \left[ 1-\frac{3}{2\mu_{\rm p}} \left( 1-\frac{2}{\gamma_{\rm p}^2} \right)
  \left(1-\zeta^2 \right) \right] \, .
  \label{eq:exy_hyd1}
\end{align}
In the rest frame of the background plasma, $\beta_{\rm p \vert p} = 0$; further assuming the weak-growth limit, $\vert\zeta^2\vert \ll  1$, the above relations simplify to
\begin{align}
  \varepsilon_{xx}^{\rm p} &\simeq 1 - \frac{\omega_{\rm p}^2}{k^2 \zeta^2} \left(1 + \frac{1}{\mu_{\rm p} \zeta^2} \right) \,,
  \label{eq:exx_plas_hyd} \\
  \varepsilon_{yy}^{\rm p} &\simeq 1 - \frac{\omega_{\rm p}^2}{k^2 \zeta^2} \left(1 + \frac{3}{\mu_{\rm p} \zeta^2} \right) \,,
  \label{eq:eyy_plas_hyd} \\
  \varepsilon_{xy}^{\rm p} &= 0 \,.
  \label{eq:exy_plas_hyd}
\end{align}

\subsubsection{Kinetic limit}

We now consider the limit $\tilde{\chi}_{\rm p} \ll 1$ of Eqs.~\eqref{eq:epsxx_warm}, \eqref{eq:epsyy_warm} and \eqref{eq:epsxy_warm}.
Using the power series $\mathcal{Z}(\eta) \simeq i\sqrt{\pi}\exp(-\eta^2)-2\eta \cdots$ \cite{NRL_2013} and assuming $\vert\zeta^2\vert \ll 1$, this yields, in the background plasma rest frame
\begin{align}
  \varepsilon_{xx}^{\rm p}  &\simeq 1 + i \sqrt{\frac{\pi \mu_{\rm p}}{2}} \frac{\omega_{\rm p}^2}{k^2 \zeta} \,, 
  \label{eq:exx_plas_kin} \\
  \varepsilon_{yy}^{\rm p}  &\simeq 1 + \frac{\omega_{\rm p}^2\mu_{\rm p}}{k^2} \left(1+ i \sqrt{\frac{\pi \mu_{\rm p}}{2}}\,  \zeta \right) \,,
  \label{eq:eyy_plas_kin} \\
  \varepsilon_{xy}^{\rm p} &= 0 \,.
  \label{eq:exy_plas_kin}
\end{align}

\subsection{Evaluation of the dielectric tensor for the suprathermal particles}

\subsubsection{Hydrodynamic limit}

In contrast to the background plasma, the beam particles are characterized by an ultrarelativistic drift velocity in the background plasma rest frame ($\gamma_{\rm b} \equiv \gamma_{\rm b\vert p}\gg 1$) and a relativistic proper temperature ($\mu_{\rm b} \ll 1$).
As a result, the integrand of  Eq.~\eqref{eq:Axx_alt_rep} presents the approximate width $\Delta s_{\rm b} \simeq 1$, so that
the hydrodynamic response of the suprathermal particles requires $\vert \chi_{\rm b} \vert \gg 1$. The corresponding dielectric
tensor, $\varepsilon_{lm,\rm b}$, is obtained by expanding  $(\chi_{\rm b}-s)^{-1} \simeq \chi_{\rm b}^{-1} \left[1+(s/\chi_{\rm b})^2 \right]$
in Eqs.~\eqref{eq:int_Axx_alt}-\eqref{eq:int_Axy_alt}, and evaluating  the various resulting integrals. For $\varepsilon_{xx,\rm b}$,
this gives
\begin{align}
  \mathcal{A}_{xx}^{\rm b} \simeq & \frac{4}{\mu_{\rm b}^2 \gamma_{\rm b}\zeta}
  \left[2 \gamma_{\rm b}^2 \beta_{\rm b}^2 \frac{\partial^2}{\partial b^2}J(0,\mu_{\rm b},1)
  -  \frac{\partial}{\partial b}J(0,\mu_{\rm b},1) \right] \nonumber\\
& 
  + \frac{4 (1-\zeta^2)}{\mu_{\rm b}^3 \gamma_{\rm b}^3 \zeta^3} 
  \left[2 \gamma_{\rm b}^2 \beta_{\rm b}^2 \frac{\partial^2}{\partial b^2}I(0,\mu_{\rm b},1)
  - \frac{\partial}{\partial b}I(0,\mu_{\rm b},1) \right] \,,
\end{align} 
where the functions $I(t,\lambda,b)$ and $J(t,\lambda,b)$ are defined by Eqs.~\eqref{eq:int_I} and \eqref{eq:int_J}, respectively.
Working out the derivatives, we find
\begin{align}
  \mathcal{A}_{xx}^{\rm b} \simeq & \frac{4}{\mu_{\rm b}^2 \gamma_{\rm b}\zeta}
  \left[\gamma_{\rm b}^2 \beta_{\rm b}^2 \mu_{\rm b} K_2(\mu_{\rm b}) + K_1(\mu_{\rm b}) \right] \nonumber\\
& + \frac{4 (1-\zeta^2)}{\mu_{\rm b}^3 \gamma_{\rm b}^3 \zeta^3} 
  \left[2 \gamma_{\rm b}^2 \beta_{\rm b}^2 \mu_{\rm b} K_1(\mu_{\rm b}) +  K_0(\mu_{\rm b}) \right] \,.
\end{align}
Inserting this expression into Eq.~\eqref{eq:epsxx_alt_rep} yields, to leading order,
\begin{equation}
  \varepsilon_{xx}^{\rm b}  \simeq 1 - \frac{\omega_{\rm pb}^2 \mu_{\rm b} \beta_{\rm b}^2}{2k^2 \zeta^4} \,,
\label{eq:epsxx_beam_hyd}
\end{equation}
assuming $\vert\zeta^2\vert \ll 1$ and $\gamma_{\rm b}^{-2} \ll \mu_{\rm b} \ll 1$, and expanding the Bessel functions accordingly.

Applying the same procedure to Eqs.~\eqref{eq:int_Ayy_alt} and \eqref{eq:int_Axy_alt} leads to the hydrodynamic expressions
\begin{align}
  \varepsilon_{yy}^{\rm b}  &\simeq 1 -\frac{\omega_{\rm pb}^2 \mu_{\rm b}}{2 k^2 \zeta^2}
  \left\{ 1 + \frac{1}{\mu_{\rm b}^2 \gamma_{\rm b}^4 \zeta^2}
  \left[12 - \frac{5}{2}\gamma_{\rm b}^2 \mu_{\rm b}^2 \ln \mu_{\rm b} \right]\right\} \,,
  \label{eq:epsyy_beam_hyd} \\
  \varepsilon_{xy}^{\rm b}  &\simeq - \frac{\omega_{\rm pb}^2\mu_{\rm b} \beta_{\rm b}}{2 k^2 \zeta^3} \,.
  \label{eq:epsxy_beam_hyd}
\end{align}

\subsubsection{Kinetic limit}

The kinetic response of the beam particles can be readily obtained, to leading order in $\vert \chi_{\rm b} \vert$, from the expansion
$(\chi_{\rm b}-s)^{-1}\simeq -i\pi \delta(s) - P(1/s)$ in Eq.~\eqref{eq:epsxx_alt}, where $\delta(s)$ is the Dirac delta function and
$P$ denotes the Cauchy principal value, which here vanishes. In general, however, the beam particles appear to be only marginally kinetic
in PIC shock simulations, so it could be useful to go to the next order. The series expansions derived in Appendix~\ref{app:series_expansion}
are well suited to this purpose. In the limit $\vert \chi_{\rm b} \vert \ll 1$, Eqs.~\eqref{eq:int_Axx_series}, \eqref{eq:int_Ayy_series}
and \eqref{eq:int_Axy_series} reduce to 
\begin{align}
  \mathcal{A}_{xx}^{\rm b} \simeq& - 4 \gamma_{\rm b}
  \biggl\{ i \sqrt{\frac{\pi}{2\mu_{\rm b}}}\left[\frac{1}{\mu_{\rm b}}K_{3/2}(\mu_{\rm b})
  + \beta_{\rm b}^2 \gamma_{\rm b}^2 K_{5/2}(\mu_{\rm b}) \right] \nonumber\\
&\quad - \gamma_{\rm b} \zeta \left[ \frac{1}{\mu_{\rm b}} K_2(\mu_{\rm b}) + \beta_{\rm b}^2 \gamma_{\rm b}^2 K_{3}(\mu_{\rm b}) \right] \biggr\} \,, \\
  \mathcal{A}_{yy}^{\rm b} \simeq& -\frac{2\gamma_{\rm b} \zeta}{\mu_{\rm b}} 
  \biggl\{ K_0(\mu_{\rm b}) + \frac{2}{\mu_{\rm b}}K_2(\mu_{\rm b})\nonumber\\
&\quad  + i \gamma_{\rm b} \sqrt{\frac{\pi \mu_{\rm b}}{2}} \zeta
  \left[K_{1/2}(\mu_{\rm b}) + \frac{2+\beta_{\rm b}^2}{\mu_{\rm b}} K_{3/2}(\mu_{\rm b}) \right] \biggr\}  \,, \\
  \mathcal{A}_{xy}^{\rm b} \simeq&  -2 \gamma_{\rm b} \beta_{\rm b} \biggl\{ i \sqrt{\frac{\pi}{2 \mu_{\rm b}}} K_{5/2} (\mu_{\rm b})\nonumber\\
&\quad  - \zeta \left[ \gamma_{\rm b}^2 K_3(\mu_{\rm b}) - \frac{1}{\mu_{\rm b}} K_2(\mu_{\rm b}) \right] \biggr\} \,.
\end{align}
Combining these approximate expressions with Eqs.~\eqref{eq:epsxx_alt}-\eqref{eq:epsxy_alt} and expanding the Bessel functions in the small-argument limit gives
\begin{align}
  \varepsilon_{xx}^{\rm b}  &\simeq 1 + \frac{\omega_{\rm pb}^2 \mu_{\rm b} \gamma_{\rm b}^2 \beta_{\rm b}^2}{k^2 \zeta^2} 
  + \frac{\omega_{\rm pb}^2 \mu_{\rm b} \gamma_{\rm b}^3}{4 k^2 \zeta} \left[ 3 i \pi - 16 \gamma_{\rm b} \zeta \right]  \,,
  \label{eq:epsxx_beam_kin} \\
  \varepsilon_{yy}^{\rm b}  &\simeq 1 + \frac{\omega_{\rm pb}^2 \mu_{\rm b} \gamma_{\rm b}^2}{k^2}
  \left[ 1 + \frac{3i \pi }{4} \gamma_{\rm b} \zeta \right]  \,,
  \label{eq:epsyy_beam_kin} \\
  \varepsilon_{xy}^{\rm b}  &\simeq \frac{\omega_{\rm pb}^2 \mu_{\rm b} \beta_{\rm b}  \gamma_{\rm b}^2}{k^2 \zeta}
  \left[ 1 + \frac{3 i \pi}{4} \gamma_{\rm b} \zeta - 4 \gamma_{\rm b}^2 \zeta^2 \right] \,.
  \label{eq:epsxy_beam_kin}
\end{align}
where we have further assumed  $\gamma_{\rm b} \ll 1$.

\subsection{CFI growth rates and frame velocities in various plasma response regimes}
\label{subsec:Weibel_rates_velocities}
The previous formulas may now be applied to the case of the precursor of a relativistic shock to derive the growth rate of the purely transverse CFI and the velocity of the ``Weibel frame'' $\mathcal R_{\rm w}$. We consider the various limits in which the plasma and/or the beam can be described in a fluid-like or kinetic approximation, keeping in mind that the most relevant limit for the precursor is that of both kinetic beam and background plasma.

For reference, let us recall that in the limit $\vert\zeta\vert = \vert\omega/k\vert \ll 1$, which is applicable here, the plasma can be described in the hydrodynamic regime iff $\gamma_{\rm p}\vert\zeta \vert \ll \sqrt{2/\mu_{\rm p}}$ (with $\mu_{\rm p} = m/T_{\rm p}$ assumed greater than unity). As for the beam, it can be described in the hydrodynamic regime iff $\gamma_{\rm b}\vert\zeta\vert \ll 1$. 

In the following, we solve for the dispersion relation in the background plasma rest frame, in order to derive $\beta_{\rm w\vert p}$ according to Eq.~\eqref{eq:beta_w_ex}. We also assume that the plasma frame is close to the ``Weibel frame'', so that the off-diagonal term $\epsilon_{xy}^2$ can be neglected in the dispersion relation as written in the background plasma frame. The dispersion relation may then be approximated as
\begin{equation}
  \zeta^2 \varepsilon_{xx} -1 \simeq 0 \,.
\end{equation}

Finally, in order to make contact with our previous notations, we will repeatedly use the substitution: $\omega_{\rm pb}^2\mu_{\rm b}/\omega_{\rm p}^2 = \xi_{\rm b} (n_\infty/n_{\rm p})/\kappa_{T_{\rm b}}^2$. This notably implies $\omega_{\rm pb}^2\mu_{\rm b}/\omega_{\rm p}^2 \ll 1$. We also have $\gamma_{\rm b\vert p} \simeq \gamma_{\rm p\vert s}$ and $\beta_{\rm b\vert p} \simeq 1$.

\subsubsection{Hydrodynamic plasma and beam}

In the hydrodynamic regime (and in the background plasma rest frame), $\vert\zeta\vert \ll \mu_{\rm p}^{-1/2}$ and $\vert\zeta\vert \ll 1/\gamma_{\rm b\vert p}$. Hence, the dispersion relation gives to leading order:
\begin{equation}
  \zeta^2 \simeq - \frac{\omega_{\rm pb}^2 \mu_{\rm b} \beta_{\rm b\vert p}^2}{k^2 + 2\omega_{\rm p}^2} \,,
\label{eq:disp_rel_hyd}
\end{equation}
and so the growth rate saturates at $\Gamma_{\rm max} \simeq \omega_{\rm pb}^2 \mu_{\rm b} $ for $k \gg \sqrt{2}\omega_{\rm p}$.

Adding up the hydrodynamic plasma and beam contributions into Eq.~\eqref{eq:beta_w_ex} and retaining only leading order terms yields the ``Weibel frame'' velocity
\begin{equation}
  \beta_{\rm w \vert p} \simeq
  \frac{\omega_{\rm pb}^2 \mu_{\rm b} \beta_{\rm b\vert p}}{-k^2 \zeta^2 + 2\omega_{\rm p}^2 
  + \omega_{\rm pb}^2 \mu_{\rm b}
   } \,.
\label{eq:beta_w_hyd_int1}
\end{equation}
As $\vert\zeta\vert \ll \omega_{\rm p}/k$ according to Eq.~\eqref{eq:disp_rel_hyd}, the expression for $\beta_{\rm w\vert p}$ boils down to
\begin{equation}
\beta_{\rm w\vert p} \simeq \frac{\omega_{\rm pb}^2 \mu_{\rm b} \beta_{\rm b\vert p}}{2 \omega_{\rm p}^2}  \simeq \frac{1}{2\kappa_{T_{\rm b}}^2}\xi_{\rm b}\frac{n_\infty}{n_{\rm p}}\,.
\label{eq:beta_w_ind}
\end{equation}
As $\beta_{\rm b \vert p}\simeq \beta_{\rm b \vert w} \simeq 1$, we recover the formula derived within a fluid approach, Eq.~\eqref{eq:wff2}, provided one sets in the latter the adiabatic index $\widehat\Gamma_{\rm b}= 2$. This is the value expected for a gas with one degree of freedom in the relativistic limit; the reduced effective dimensionality for the beam response results from the assumption of a purely 1D transverse fluctuation spectrum.

Finally, the fully hydrodynamic regime holds as long as $\min (\sqrt{\mu_{\rm p}/2},\gamma_{\rm b\vert p}) \vert \zeta_{\rm max} \vert  \gg  1$.  Now, expressing Eq.~\eqref{eq:disp_rel_hyd} at $k_{\rm max} \equiv \sqrt{2}\omega_{\rm p}$ gives
\begin{equation}
  \zeta_{\rm max} \simeq i\left(\frac{\omega_{\rm pb}^2 \mu_{\rm b} }{4 \omega_{\rm p}^2}\right)^{1/2} \,,
\label{eq:zeta_max_kin_hyd}
\end{equation}
so that another way of expressing the validity of the hydrodynamic regime is:
\begin{equation}
  \min \left( \frac{\mu_{\rm p}}{2}, \gamma_{\rm b\vert p}^2 \right) > \frac{4 \omega_{\rm p}^2}{\omega_{\rm pb}^2\mu_{\rm b} } \simeq \frac{4\kappa_{T_{\rm b}}^2}{\xi_{\rm b}} \frac{n_{\rm p}}{n_\infty} \,. 
\label{eq:thr_plasma_beam_hydro_kin}
\end{equation}

\subsubsection{Kinetic plasma and hydrodynamic beam}

In this limit, $1/\gamma_{\rm b\vert p} \gg \vert\zeta\vert \gg \mu_{\rm p}^{-1/2}$; to leading order, the dispersion relation takes the form
\begin{equation}
  i \sqrt{2 \pi \mu_{\rm p}} \omega_{\rm pb}^2 \zeta - \frac{\omega_{\rm pb}^2\mu_{\rm b} \beta_{\rm b\vert p}^2}{\zeta^2} - k^2 \simeq 0 \,.  
\end{equation}
The growth rate reaches its maximum value $\Gamma_{\rm max} = \sqrt{\omega_{\rm pb}^2\mu_{\rm b} }$ for
$k \lesssim  k_{\rm max} \simeq (2 \pi \omega_{\rm pb}^2 \mu_{\rm p} \mu_{\rm b} )^{1/6} \omega_{\rm p}^{2/3}$.
We define $\vert \zeta_{\rm max} \vert = \Gamma_{\rm max}/k_{\rm max} \simeq \left(\omega_{\rm pb}^2 \mu_{\rm b} /\sqrt{2 \pi \mu_{\rm p}} \omega_{\rm p}^2\right)^{1/3}$.

Making use of Eqs.~\eqref{eq:eyy_plas_kin}, \eqref{eq:exy_plas_kin}, \eqref{eq:epsyy_beam_hyd}, and \eqref{eq:epsxy_beam_hyd}, we express Eq.~\eqref{eq:beta_w_ex} as
\begin{equation}
  \beta_{\rm w \vert p} \simeq - \frac{\omega_{\rm pb}^2 \mu_{\rm b} \beta_{\rm b}}{2 \omega_{\rm p}^2\mu_{\rm p} \zeta_{\rm max}^2} \,,
\label{eq:eq:beta_w_kin_hyd_int1}
\end{equation}
which gives
\begin{equation}
 \beta_{\rm w \vert p} \simeq \left( \frac{\pi}{4} \frac{\omega_{\rm pb}^2 \mu_{\rm b}}{\omega_{\rm p}^2 \mu_{\rm p}^2 } \right)^{1/3}
  \simeq \left(\frac{\pi }{4\kappa_{T_{\rm b}}^2}\right)^{1/3} \,\xi_{\rm b}^{1/3}\,\left(\frac{n_\infty}{n_{\rm p}} \frac{T_{\rm p}^2}{m^2} \right)^{1/3} \,.
\end{equation}

\subsubsection{Hydrodynamic plasma and kinetic beam}

In this limit, we now have $1/\gamma_{\rm b\vert p} \ll \vert\zeta\vert \ll \mu_{\rm p}^{-1/2}$.
Using Eqs.~\eqref{eq:exx_plas_hyd} and \eqref{eq:epsxx_beam_kin}, the dispersion relation reduces to
\begin{equation}
  2\omega_{\rm pb}^2 \mu_{\rm b} \gamma_{\rm b\vert p}^2 \left[1 + i \frac{3\pi}{4}\gamma_{\rm b\vert p} \zeta \right]
  - 2\omega_{\rm p}^2\left[1 + \frac{1}{\mu_{\rm p} \zeta^2} \right] - k^2 \simeq 0 \,.
\label{eq:rel_disp_hyd_kin}
\end{equation}

Let us first assume that $\omega_{\rm p}^2/(\mu_{\rm p} \zeta^2)$ can be neglected in front of 
$2\omega_{\rm pb}^2 \mu_{\rm b} \gamma_{\rm b\vert p}^3 \zeta$. An unstable solution then exists provided $\gamma_{\rm b\vert p}^2 > \omega_{\rm p}^2/\left(\omega_{\rm pb}^2 \mu_{\rm b}\right) \sim 1/\xi_{\rm b}$. This is the same condition as encountered in the $4-$fluid system of Sec.~\ref{sec:linwf}, see Eq.~\eqref{eq:Wsol1} where $c_{\rm eff,b}^2 \sim 1/\gamma_{\rm b\vert p}^2$ and $\Omega_{\rm pb}^2 \sim  \omega_{\rm pb}^2 \mu_{\rm b}$.

If this condition is fulfilled, the fastest-growing solution corresponds to $\Gamma_{\rm max} \simeq \frac{8}{9\pi}\sqrt{\frac{2}{3}} \frac{[\omega_{\rm pb}^2\mu_{\rm b}\gamma_{\rm b\vert p}^2 -\omega_{\rm p}^2]^{3/2}}{\omega_{\rm pb}^2\mu_{\rm b} \gamma_{\rm b\vert p}^3}$
and $\zeta_{\rm max} \simeq \frac{8 i}{9\pi \gamma_{\rm b\vert p}} \left(1-\frac{\omega_{\rm p}^2}{\omega_{\rm pb}^2\mu_{\rm b} \gamma_{\rm b\vert p}^2} \right)$. Note that the conditions for
a hydrodynamic plasma ($\sqrt{\frac{\mu_{\rm p}}{2}} \vert \zeta_{\rm max}\vert > 1$) and a kinetic beam ($\gamma_{\rm b\vert p} \vert \zeta_{\rm max}\vert <1$) are then  verified, albeit  marginally.

Moreover, combining Eqs.~\eqref{eq:eyy_plas_hyd}, \eqref{eq:exy_plas_hyd}, \eqref{eq:epsyy_beam_kin}, and \eqref{eq:epsxy_beam_kin} gives the ``Weibel frame''
velocity
\begin{equation}
  \beta_{\rm w \vert p} \simeq \frac{2 \omega_{\rm pb}^2 \mu_{\rm b} \gamma_{\rm b\vert p}^2 \zeta_{\rm max}^2}{k^2 \zeta_{\rm max}^2 - 2\omega_{\rm p}^2+2\omega_{\rm pb}^2 \mu_{\rm b} \gamma_{\rm b\vert p}^2 \zeta_{\rm max}^2} \,.
\label{eq:beta_w_ hyd_kin}
\end{equation}
Inserting the above expressions of $\Gamma_{\rm max}$ and $\zeta_{\rm max}$, there follows
\begin{align}
  \beta_{\rm w \vert p} &\simeq \left(\frac{8}{9\pi} \right)^2 \frac{\omega_{\rm pb}^2\mu_{\rm b}\beta_{\rm b\vert p}}{\omega_{\rm p}^2} \left(1-\frac{\omega_{\rm p}^2}{\omega_{\rm pb}^2\mu_{\rm b}\gamma_{\rm b\vert p}^2}\right)^2 \\
  &\simeq \left(\frac{8}{9\pi\kappa_{T_{\rm b}}} \right)^2 \xi_{\rm b}\frac{n_\infty}{n_{\rm p}} \left( 1-\frac{n_{\rm p}}{n_\infty}\frac{\kappa_{T_{\rm b}}^2}{\gamma_{\rm b\vert p}^2 \xi_{\rm b}} \right)^2 \,.
\end{align}

In the opposite limit, in which $\omega_{\rm p}^2/(\mu_{\rm p} \zeta^2)$ dominates over 
$2\omega_{\rm pb}^2 \mu_{\rm b} \gamma_{\rm b\vert p}^3 \zeta$ in Eq.~\eqref{eq:rel_disp_hyd_kin}, the dominant mode satisfies $\Gamma_{\rm max} \simeq  \sqrt{2\omega_{\rm p}^2/\mu_{\rm p}}$ and $\zeta_{\rm max} \simeq i[2\mu_{\rm p} (1-\omega_{\rm pb}^2\mu_{\rm b}\gamma_{\rm b\vert p}^2/\omega_{\rm p}^2)]^{-1/2}$
for $k_{\rm max}  \equiv  \sqrt{2\omega_{\rm p}^2 - 2\omega_{\rm pb}^2\mu_{\rm b}\gamma_{\rm b\vert p}^2}$. To leading order, we thus derive
\begin{align}
  \beta_{\rm w \vert p} &\simeq \frac{\omega_{\rm pb}^2\mu_{\rm b}\gamma_{\rm b\vert p}^2}{2\omega_{\rm p}^2 \mu_{\rm p} \left(1- \omega_{\rm pb}^2\mu_{\rm b}\gamma_{\rm b\vert p}^2/\omega_{\rm p}^2\right)} \\
  &\simeq \frac{1}{2\kappa_{T_{\rm b}}^2}\gamma_{\rm b\vert p}^2 \,\xi_{\rm b}\,\frac{T_{\rm p}}{m}\frac{n_\infty}{n_{\rm p}}\left( 1- \frac{1}{\kappa_{T_{\rm b}}^2}\gamma_{\rm b\vert p}^2\xi_{\rm b} \frac{n_\infty}{n_{\rm p}} \right)^{-1}
\end{align}

\subsubsection{Kinetic plasma and beam}

Finally, we consider the case of a fully kinetic beam-plasma system. This regime is of particular importance since it is found to hold in most of the precursor region in long-time shock simulations (see Sec.~\ref{sec:linPIC}). Using the expressions \eqref{eq:exx_plas_kin}
and \eqref{eq:epsxx_beam_kin}, the dispersion relation writes
\begin{equation}
  i \sqrt{2 \pi \mu_{\rm p}} \omega_{\rm p}^2 \zeta + 2 \omega_{\rm pb}^2 \mu_{\rm b} \gamma_{\rm b\vert p}^2 \left(1 + i \frac{3\pi}{4} \gamma_{\rm b\vert p} \zeta \right) - k^2 \simeq 0 \,.
\end{equation}
The dominant CFI mode is then defined by
\begin{align}
  \Gamma_{\rm max} &\simeq \frac{4}{3}\sqrt{\frac{2}{3}}
  \frac{(\omega_{\rm pb}^2\mu_{\rm b})^{3/2} \gamma_{\rm b\vert p}^3\beta_{\rm b\vert p}^3}{\sqrt{2\pi \mu_{\rm p}} \omega_{\rm p}^2 + \frac{3\pi}{2}\omega_{\rm pb}^2 \mu_{\rm b} \gamma_{\rm b\vert p}^3} \,,
  \label{eq:wi_kin_kin} \\
  \zeta_{\rm max} &\simeq i\frac{4}{3} \frac{\omega_{\rm pb}^2 \mu_{\rm b}\gamma_{\rm b\vert p}^2}{\sqrt{2\pi \mu_{\rm p}} \omega_{\rm p}^2+ \frac{3\pi}{2}\omega_{\rm pb}^2 \mu_{\rm b} \gamma_{\rm b\vert p}^3} \,,
  \label{eq:zeta_kin_kin}
\end{align} 
and $k_{\rm max} \simeq \sqrt{\frac{2}{3} \mu_{\rm b}}  \omega_{\rm pb}\gamma_{\rm b\vert p}$.

The corresponding expression for the ``Weibel frame'' velocity is obtained by combining Eqs.~\eqref{eq:eyy_plas_kin}, \eqref{eq:exy_plas_kin}, \eqref{eq:epsyy_beam_kin},
and \eqref{eq:epsxy_beam_kin}. After some algebra, one finds
\begin{align}
  \beta_{\rm w \vert p} &\simeq \omega_{\rm pb}^2 \mu_{\rm b} \gamma_{\rm b\vert p} \beta_{\rm b\vert p}
  \frac{1+ i\frac{3\pi}{4}\gamma_{\rm b\vert p} \zeta_{\rm max} - 4\gamma_{\rm b\vert p}^2 \zeta_{\rm max}^2}{\omega_{\rm p}^2 \mu_{\rm p} + \frac{2}{3} \omega_{\rm pb}^2 \mu_{\rm b} \gamma_{\rm b\vert p}^2} 
\end{align}
or, to leading order and in terms of our usual parameters,
\begin{align}
 \beta_{\rm w \vert p} &\simeq \frac{1}{\kappa_{T_{\rm b}}^2}\,\xi_{\rm b}\,\frac{n_\infty}{n_{\rm p}} \gamma_{\rm b\vert p}^2 \frac{T_{\rm p}}{m} 
\label{eq:kinpkinbbw}
\end{align}

\subsection{Comparison to PIC simulations} \label{sec:linPIC}

In this Section, we compare the relative velocity $\beta_{\rm w\vert p}$ estimated using the kinetic model of the CFI developed in Sec.~\ref{subsec:Weibel_rates_velocities} with that extracted from 2D3V PIC simulations.

These simulations have been performed using the massively parallel, relativistic PIC code \textsc{calder} \cite{Lefebvre_2003}. The shock is generated by means of the standard mirror technique \cite{Spitkovsky_2008a}. The background pair plasma is continuously injected into the domain from the right-hand boundary, and is made to reflect specularly off the left-hand boundary ($x_{\vert\rm d}=0$). Electrons and positrons are injected with a Maxwell-J\"uttner momentum distribution
[Eq.~\eqref{eq:edf}] of proper temperature $T_\infty= 0.01m_e$ and mean drift velocity $\beta_{\infty\vert\rm d}=-(1-1/\gamma_{\infty\vert\rm d}^{2})^{1/2}$ in the simulation frame. As mentioned earlier, this simulation reference frame coincides with the downstream plasma rest frame (as a consequence of the use of the mirror technique). Our two reference PIC simulations use $\gamma_{\infty\vert\rm d} = 10$ and $\gamma_{\infty\vert\rm d} = 100$, which respectively correspond to shock Lorentz factors (with respect to the upstream) of $\gamma_\infty = 17$ and $\gamma_\infty = 173$.

As the simulation proceeds, the right-hand boundary (injector) is progressively displaced towards the right so as to keep the reflected ballistic particles inside the (increasingly large) domain, while speeding up the calculation at early times \cite{Spitkovsky_2008a}. In order to quench the numerical Cherenkov instability, which notoriously hampers simulations of relativistically drifting plasmas, we make use of the Godfrey-Vay filtering scheme, combined with the Cole-Karkkainnen finite difference field solver \cite{Godfrey_2014}. We use a mesh size  $\Delta x = \Delta y = 0.1\, c/\omega_{\rm p}$ and a time step $\Delta t = 0.99 \Delta x/c$. Periodic boundary conditions are employed in the tranverse direction for both particles and fields. The initial domain size is $L_x\times L_y = 2700 \times 340\, (c/\omega_{\rm p})^2$ for $\gamma_{\infty\vert\rm d} = 10$ and $L_x\times L_y = 2000 \times 200\, (c/\omega_{\rm p})^2$ for $\gamma_{\infty\vert\rm d} = 100$. Each cell is initially filled with 10 macro-particles per species (electrons or positrons). The simulation is run until $t_{\rm max} = 3600\, \omega_{\rm p}^{-1}$ (resp. $t_{\rm max} = 6900\, \omega_{\rm p}^{-1}$) for our simulation with $\gamma_{\infty\vert\rm d}=10$ (resp. $\gamma_{\infty\vert\rm d}=100$).

In order to test our model of the CFI developing in the precursor through the interpenetration of suprathermal and background plasma populations, we need to carefully distinguish these two in the simulations. In order to do so, we track the particles according to the sign of their $x-$momentum and how many times this sign has changed, due only to interactions with the electromagnetic turbulence. We then define the background plasma as those particles that move toward negative values of $x$ and that do not have undergone turnarounds, {\it i.e.}, any change of sign of $u^x$.  We define the beam particles as those that move towards positive values of $x$, independently of their number of turnarounds. This definition leaves a minority of particles: those that move with $u^x < 0$ and have undergone at least one turnaround. In this population, however, it becomes difficult to distinguish particles that originate from the right boundary of the simulation box from those that originate from the left boundary; these populations have different temperatures, so that a single-fluid description of this compound population would introduce errors.

Downstream of the shock, the left-moving and right-moving particle populations are identical, because the plasma has isotropized in this simulation frame. There, our definition of suprathermal particles only counts half of the particles, therefore our $\xi_{\rm b} \simeq 1/4$ in this region: in 2D3V simulations, the downstream pressure represents $1/2$ of the energy density, which itself amounts to the incoming energy flux into the shock rest frame $F_\infty$.

We extract hydrodynamic moments $n_\alpha$, $T_\alpha$, $u_\alpha$ for each of the beam and background plasma population, assuming that they obey Maxwell-J\"uttner
momentum distributions. The spatial profiles of these various hydrodynamic quantities have already been presented in Fig.~\ref{fig:profs}, as extracted from the simulations with $\gamma_{\infty\vert\rm d} = 10$ and $\gamma_{\infty\vert\rm d} = 100$ at respective times $t \simeq 3600\, \omega_{\rm p}^{-1}$ and $t \simeq 6900\, \omega_{\rm p}^{-1}$. One can see that $n_{\rm p}$ and $\gamma_{\rm p \vert\rm d}$ vary weakly across the precursor region, except near the shock front where the incoming plasma slows down significantly and experiences compression. By contrast, the plasma temperature steadily increases from its far-upstream value ($T_{\rm p}=0.01 m_e$) to unity and beyond when approaching the shock front. This heating results from the interaction with the beam particles, whose density rises by $\sim 4-5$ orders of magnitude across the precursor~\cite{pap2}. The beam Lorentz factor in the simulation frame is close to unity, confirming that the beam drifts at a weakly relativistic velocity in the shock frame. 

\begin{figure}
  \includegraphics[width=0.45\textwidth]{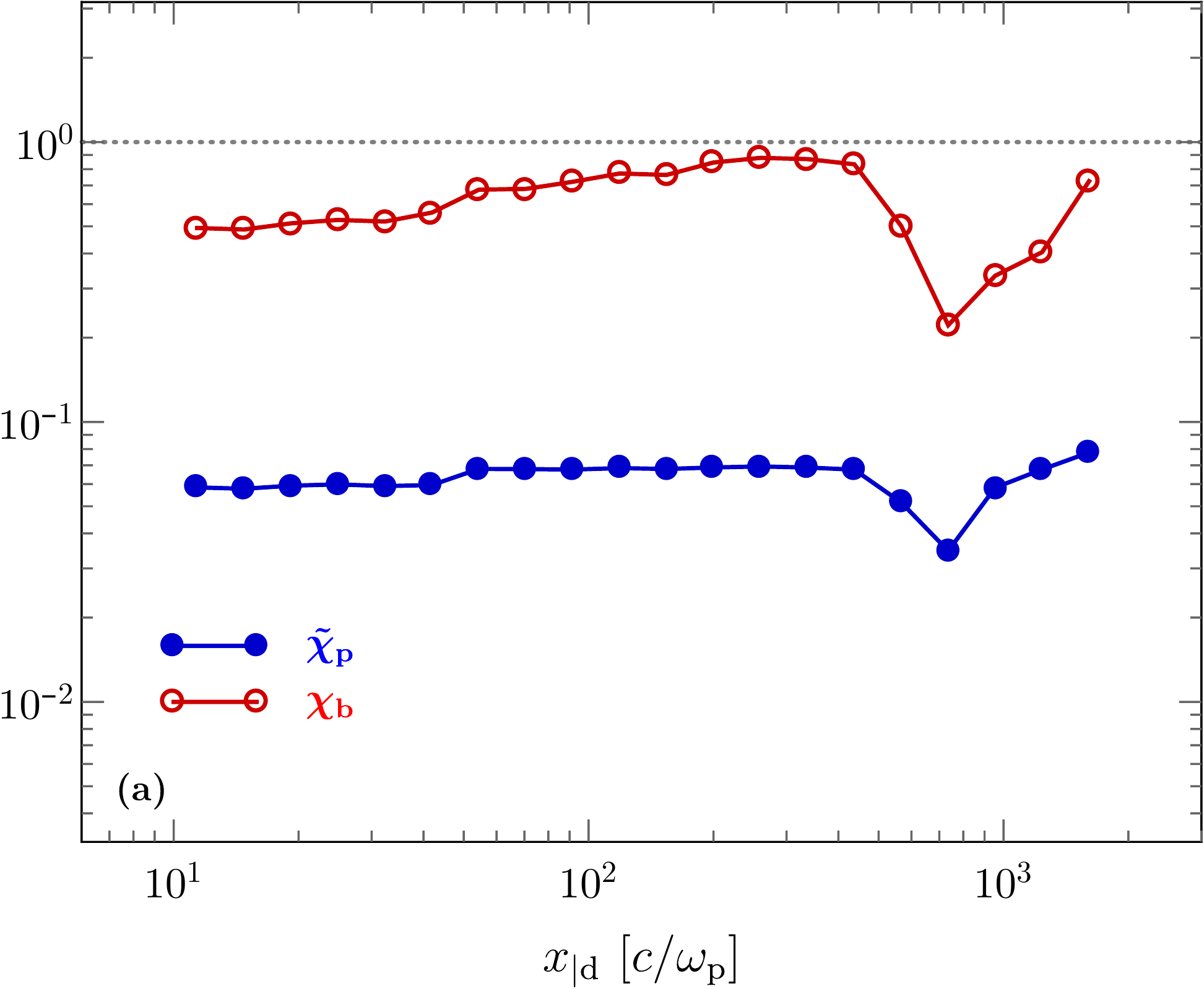}\smallskip\\
  \includegraphics[width=0.45\textwidth]{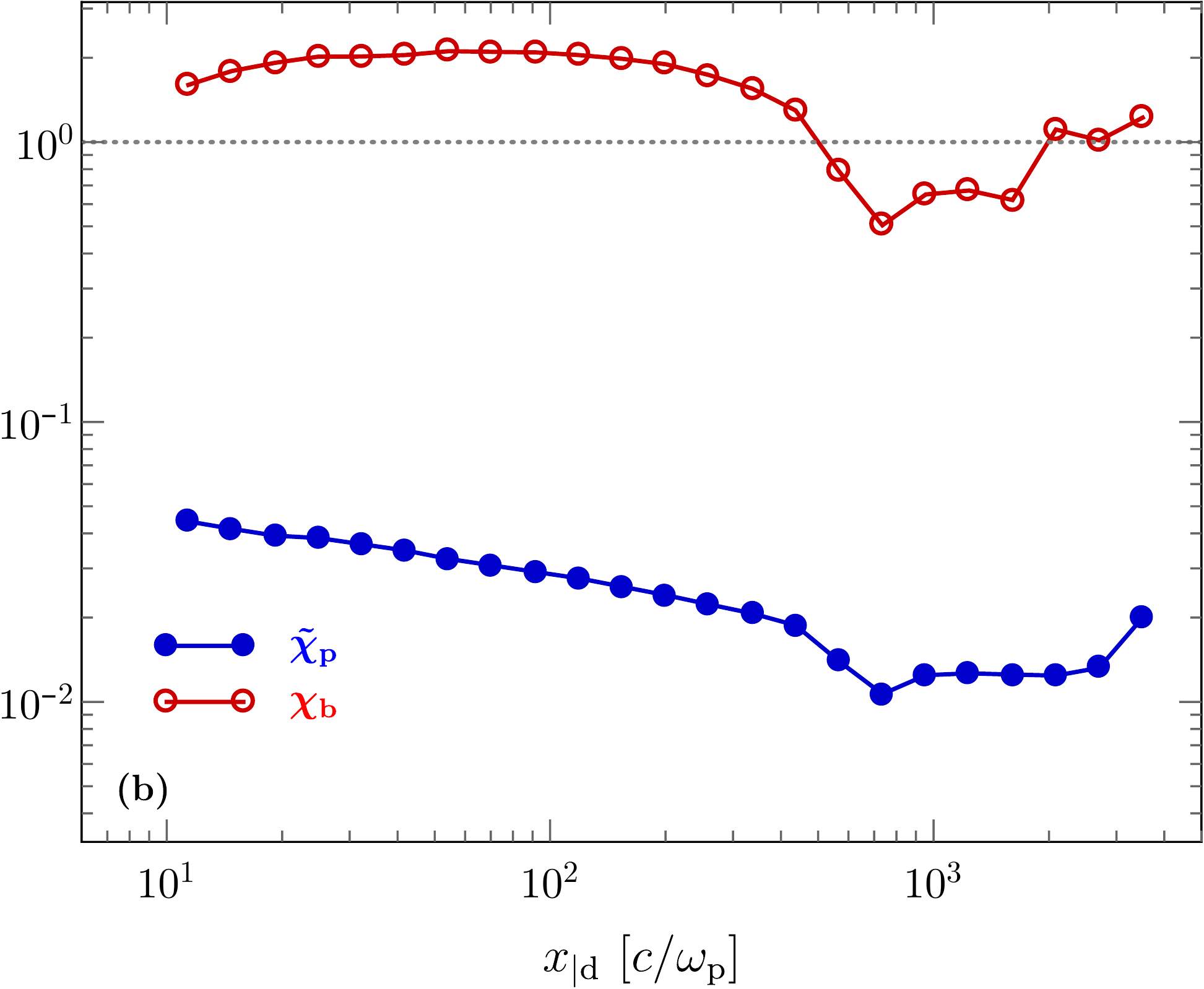}
 \caption{Parameters $\chi_{\rm b}$ and $\tilde\chi_{\rm p}$ of the beam and the background plasma as a function of distance to the shock front, extracted from PIC simulations with $\gamma_\infty = 17$ [panel (a), top] and $\gamma_\infty = 173$ [panel (b), bottom]. We recall that the hydrodynamic (kinetic) regime for the beam and/or the plasma corresponds to $\chi_{\rm b} \gg 1$ ($\chi_{\rm b} \ll 1$) and/or $\tilde\chi_{\rm p} \gg 1$ ($\tilde\chi_{\rm p} \ll 1$), respectively. This figure thus indicates that the plasma should be described in the kinetic regime, and that the beam regime is intermediate.
  \label{fig:comp_PIC_chi}}
\end{figure}

The general dispersion relation \eqref{eq:disp} is numerically solved using the reduced forms \eqref{eq:epsxx}, \eqref{eq:epsyy} and \eqref{eq:epsxy}
of the dielectric tensor elements, as detailed in Ref.~\cite{Bret_2010b}. At each sampled location, we look for the growth rate ($\Gamma_{\rm max}$),
wave number ($k_{\rm max}$) and wave phase velocity ($\xi_{\rm max}$) of the fastest-growing mode, and then use Eq.~\eqref{eq:beta_w_ex} to evaluate
the corresponding value of the ``Weibel frame'' velocity ($\beta_{\rm w \vert p}$).

Figures~\ref{fig:comp_PIC_chi} display the spatial profiles of the $\chi_{\rm b}$ and $\tilde{\chi}_{\rm p}$ parameters defined by Eqs.~\eqref{eq:chi} and \eqref{eq:chi_tilde}, respectively. Interestingly, both the background plasma and beam populations appear to lie in the kinetic CFI regime ($\tilde{\chi}_{\rm p} <1$ and $\chi_{\rm b}<1$) throughout the precursor region. However, whereas $\tilde{\chi}_{\rm p}$ shows weak variations around relatively low values ($\tilde{\chi}_{\rm p} \sim 0.05$), so that taking the kinetic plasma limit is well justified, the $\chi_{\rm b}$ values are larger by about an order
of magnitude and present stronger variations. Where $\xi_{\rm b}$ becomes of the order of unity, the beam response is then only marginally kinetic; consequently, analytical approximations present an error of about a factor 2 with respect to the full numerical calculation of $\beta_{\rm w\vert p}$. 

\begin{figure}
\includegraphics[width=0.45\textwidth]{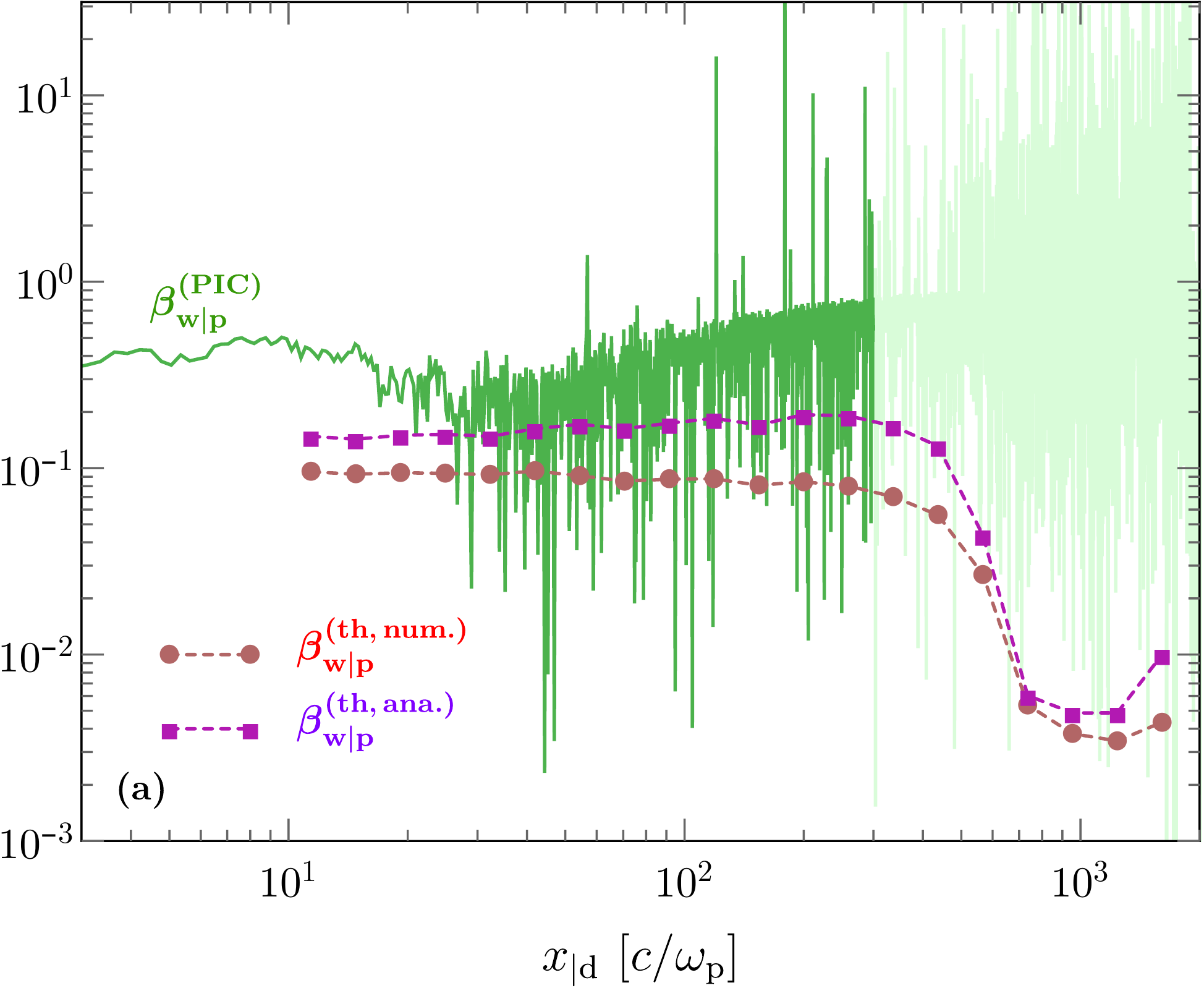}\smallskip\\
\includegraphics[width=0.45\textwidth]{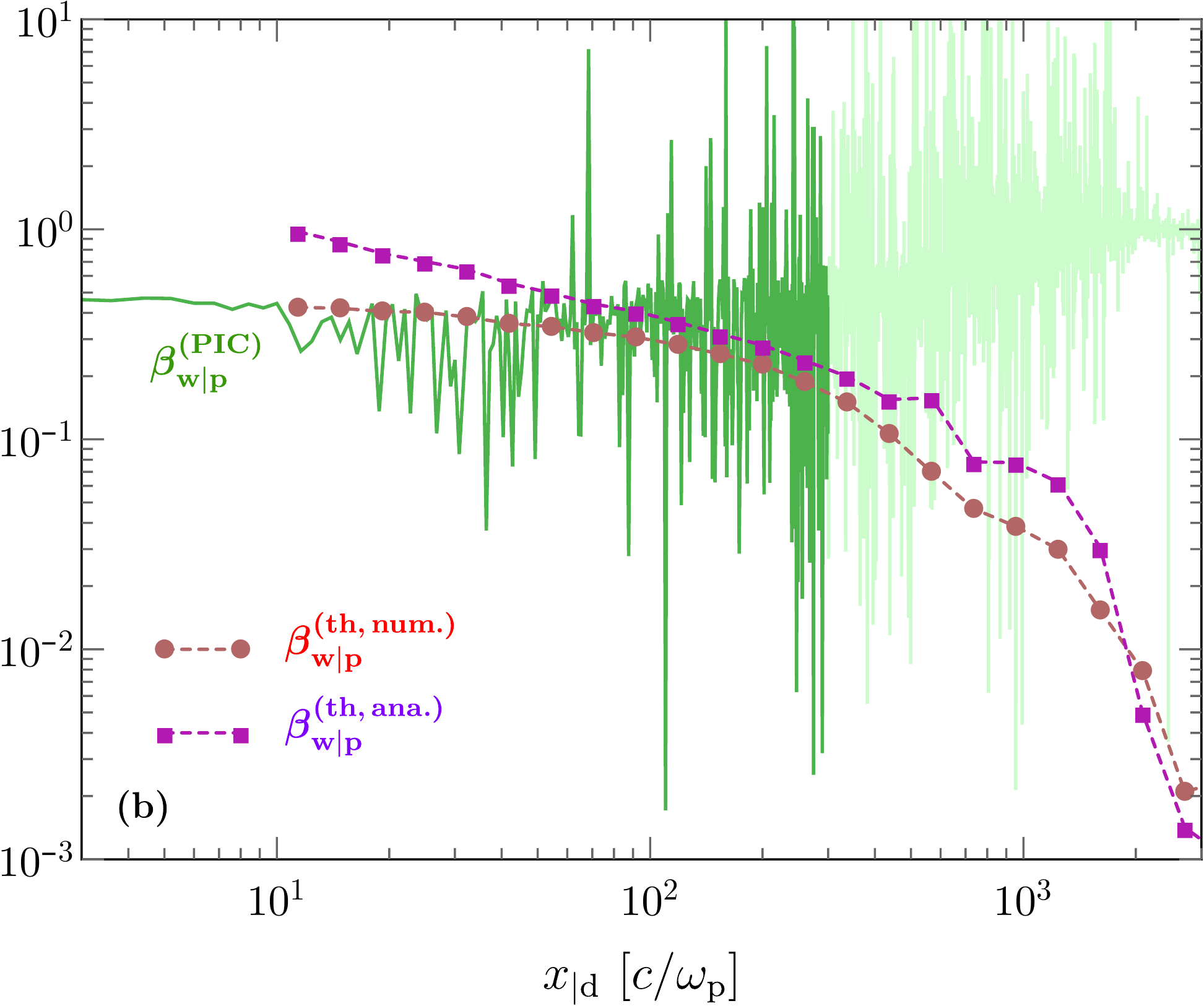}
  \caption{Theoretical estimates of the relative velocity between the ``Weibel frame'' $\mathcal R_{\rm w}$ and the background plasma, $\beta_{\rm w\vert p}$, as a function of distance to the shock front, compared with the velocity extracted from our reference PIC simulations through the ratio $\left\langle\delta E_y^2\right\rangle^{1/2}/\left\langle\delta B_z^2\right\rangle^{1/2}$ (solid green); where this value cannot be measured accurately from the simulation, the data are colored in light green (see text for details). The red circle symbols/dashed curve uses the numerical solution to the general kinetic dispersion relation \eqref{eq:disp} to derive $\zeta_{\rm max}$, while the purple square/dashed curve plots the analytic approximations derived in Sec.~\ref{subsec:Weibel_rates_velocities}. Panel (a), top: $\gamma_\infty = 17$ ($\gamma_{\infty\vert\rm d} = 10$); panel (b), bottom: $\gamma_\infty = 173$ ($\gamma_{\infty\vert\rm d} = 100$);
  \label{fig:comp_PIC_bw}}
\end{figure}

As discussed in Sec.~\ref{sec:setup}, we estimate the 3-velocity of the ``Weibel frame'' in PIC simulations as the ratio $\beta_{\rm w\vert d}=\left\langle\delta E_y^2\right\rangle^{1/2}/\left\langle\delta B_z^2\right\rangle^{1/2}$, where averaging is performed over the transverse dimension. Given the locally measured value of the background plasma velocity $\beta_{\rm p\vert d}$, we convert it to the instantaneous local plasma rest frame through the standard transform $\beta_{\rm w\vert p}=\left(\beta_{\rm w\vert d}-\beta_{\rm p\vert d}\right)/\left(1-\beta_{\rm w\vert d}\beta_{\rm p\vert d}\right)$. We then compare in Fig.~\ref{fig:comp_PIC_bw} this measurement with our theoretical estimates of the relative velocity, $\beta_{\rm w \vert p}$, between the ``Weibel frame'' and the background plasma, 
using both the numerical calculation (red circle/dashed line) and the analytical approximation (purple square/dashed). Both for $\gamma_\infty=17$ and $\gamma_\infty = 173$, it is seen that, as predicted, $\beta_{\rm w\vert p}$ remains subrelativistic  throughout the precursor, increasing from $\beta_{\rm w\vert p}\sim 10^{-3}$ at the far end of the precursor up to $\beta_{\rm w\vert p}\simeq 0.1$ near the shock front. The theoretical estimates appear to provide reasonable match to the PIC data in the region where $\beta_{\rm w\vert p}$ can be measured accurately.

For both reference simulations, Fig.~\ref{fig:comp_PIC_bw} reveals significant fluctuations in the measured values of $\beta_{\rm w\vert p}$, with increasing amplitude at large $x$, which deserves some discussion. Far from the shock front, the magnitude of our observable $\left\langle\delta E_y^2\right\rangle^{1/2}/\left\langle\delta B_z^2\right\rangle^{1/2}$ is close to unity (see {\it e.g.}, Fig.~\ref{fig:emdens}), which implies that any amount of numerical noise can artificially push $\vert\beta_{\rm w\vert d}\vert$ to values larger than unity, even though its true value might be $<1$. Furthermore, when transforming values to the background plasma rest frame, any error in $\beta_{\rm w\vert d}$ is amplified by $\sim\gamma_{\rm p\vert d}^2$, which is large far from the shock.

In a given portion of the precursor, the ``Weibel frame'' can be considered well defined where $\vert\beta_{\rm w\vert d}\vert<1$ for most data points. In Fig.~\ref{fig:comp_PIC_bw}, values $\vert\beta_{\rm w\vert p}\vert<1$ correspond to values $\vert\beta_{\rm w\vert d}\vert<1$, but the use of raw numerical data, binned linearly, and plotted on a log-scale,  precludes a clear identification of the region where this frame is well-defined, at least by eye. By rebinning the data, however, we infer that the ``Weibel frame'' is well defined in the range $x\lesssim10^3\,c/\omega_{\rm p}$, for both PIC simulations, as claimed in Sec.~\ref{sec:setup}. At larger distances from the shock, $\vert\beta_{\rm w\vert p}\vert$ fluctuates too often on both sides of unity, either because of numerical noise, or because of the contribution of electrostatic modes.
Given the magnitude of the fluctuations, we have chosen to plot the data in full color only in the region $x\leq 300\,c/\omega_{\rm p}$, for the sake of being conservative.

\section{The ``Weibel frame'' in the nonlinear regime}
\label{sec:nlinwf}
\subsection{Theoretical model}

We adopt here another perspective on the ``Weibel frame''. Specifically, we  consider the nonlinear evolution of the CFI, once the current filaments have formed, borrowing on the work of Ref.~\cite{Vanthieghem_2018}. We assume that, at each point in the shock precursor, the CFI has developed a quasistatic, transversally periodic system of current filaments. By quasistatic, it is meant that an equilibrium approximately holds between magnetic and thermal pressures in the filaments, according to the physical conditions at the point considered, and that these conditions evolve slowly enough that this equilibrium has time to adapt from one point to another. We also consider a 2D configuration, without loss of generality, so that the plasma is periodic along the $y$-axis and the species drift along the $x$-axis; the magnetic field is then transverse to the $(x,y)$-plane. The dominant components of the four-potential now explicitly depend on $y$ and are given by $A^\mu = \left[\phi(y),A_x(y),0,0\right]$. We assume a drifting Maxwell-J\"uttner distribution for each of the four species (corresponding to two counterstreaming pair distributions), so that
\begin{equation}\label{eq:density}
n_{\alpha} = n_{\alpha0}  \exp \left[ -\frac{\gamma_{\alpha} q_\alpha} {T_\alpha} \left(\phi - \beta_{\alpha} A_{x}\right) \right] \,,
\end{equation}
where $q_\alpha$ denotes the charge of the species and $n_{\alpha0}$ represents a normalization prefactor. We can inject these density profiles in the potential formulation of the Amp\`ere-Maxwell and Gauss-Maxwell equations, leading to
\begin{flalign}
\partial_y^2 A_{x} & =  
\sum\limits_\alpha \frac{\omega_\alpha^2}{c^2} \gamma_{\alpha} \beta_{\alpha } \sinh\left[\frac{\gamma_{\alpha} e }{T_\alpha}
\left(\phi - \beta_{\alpha } A_{x}\right)\right]\frac{m_e c^2}{e} \,, \label{eq:A_stat} \\
\partial_y^2 \phi &= \sum\limits_\alpha \frac{\omega_\alpha^2}{c^2} \gamma_{\alpha} \sinh \left[\frac{\gamma_{\alpha}e}{T_\alpha}
\left(\phi - \beta_{\alpha } A_{x}\right) \right]\frac{m_e c^2}{e} \label{eq:phi_stat} \,,
\end{flalign}
where $\omega_\alpha =\sqrt{4 \pi n_{\alpha0} e^2/m_e}$ scales to the plasma frequency of species $\alpha$. It is worth noting that this system can also be obtained in a 4-fluid framework with isothermal closure condition~\cite{Vanthieghem_2018}. 

Following~\cite{Vanthieghem_2018}, we introduce the plasma nonlinearity parameter:
\begin{equation}
 \Xi_{\rm p} = \left\vert \frac{\gamma_{\rm p} \beta_{\rm p}e}{T_{\rm p}}\,\max_y A_x(y) \right\vert \,.
\label{eq:xi}
\end{equation}
In the weakly nonlinear limit, $\Xi_{\rm p} \ll 1$, the sinh functions in the above equations can be approximated to unity, so that the vanishing of the electrostatic component entails:
\begin{equation}\label{eq:weibel_cond}
\frac{n_{\rm b} \gamma_{\rm b|w}^2 \beta_{\rm b|w}}{T_{\rm b}} + \frac{n_{\rm p} \gamma_{\rm p|w}^2 \beta_{\rm p|w}}{T_{\rm p}} = 0
\end{equation}
where the subscript $_{\rm w}$ has been introduced, because the above quantities are now defined in the ``Weibel frame'', $\mathcal{R}_{\rm w}$, in which $\phi_{\rm\vert w} = 0$.

In the weakly nonlinear limit, the velocity of $\mathcal{R}_{\rm w}$ can be computed exactly using relation~\eqref{eq:weibel_cond}. Writing $\beta_{\rm b\vert w}\,=\,(\beta_{\rm b}-\beta_{\rm w})/(1-\beta_{\rm b}\beta_{\rm w})$, $\gamma_{\rm b\vert w}\,=\,\gamma_{\rm b}\gamma_{\rm w}(1-\beta_{\rm b}\beta_{\rm w})$ etc., in any given frame, one finds that $\beta_{\rm w}$ is solution to the equation:
\begin{equation}
\beta_{\rm w}^2 - Q_{\rm w}\beta_{\rm w}  +1 = 0
\end{equation}
with 
\begin{equation}
Q_{\rm w} = \frac{ \frac{n_{\rm b}}{T_{\rm b}} \gamma_{\rm b}^2  \left( 1 + \beta^2_{\rm b} \right) + \frac{n_{\rm p}}{T_{\rm p}} \gamma_{\rm p}^2  \left( 1 + \beta^2_{\rm p} \right) }{\frac{n_{\rm b}}{T_{\rm b}} \gamma_{\rm b}^2 \beta_{\rm b} + \frac{n_{\rm p}}{T_{\rm p}} \gamma_{\rm p}^2 \beta_{\rm p}}  .
\label{eq:Q}
\end{equation}
Writing $n_{\rm b}$ in terms of $\xi_{\rm b}$ as before, we expand the above solution to first order in $\xi_{\rm b}$ to obtain the relative velocity $\beta_{\rm w\vert p}$:
\begin{equation}
\beta_{\rm w\vert p} \simeq +\gamma_{\rm b\vert p}^2\xi_{\rm b}\frac{T_{\rm p}}{m_e}\frac{n_\infty}{n_{\rm p}}\frac{\beta_\infty^2\beta_{\rm b\vert p}}{\kappa_{T_{\rm b}}^2}
\label{eq:bwpnlin}
\end{equation}
where $\gamma_{\rm b\vert p} \sim \gamma_{\rm p}$ represents the relative Lorentz factor between the beam and the background plasma. Interestingly, Eq.~(\ref{eq:bwpnlin}) corresponds to our earlier expression Eq.~\eqref{eq:kinpkinbbw} obtained from the linear dispersion relation of the CFI in the kinetic plasma -- kinetic beam limit.

 In Paper~II~\cite{pap2}, it is argued that $\gamma_{\rm p}^2\xi_{\rm b}$ is much smaller than unity in the far precursor, where the incoming background plasma maintains its initial velocity, {\it i.e.}, $\gamma_{\rm p} \simeq \gamma_\infty$, and of the order of unity in the near precursor, where $\gamma_{\rm p} < \gamma_\infty$ due to deceleration. The above thus implies that the ``Weibel frame'', in this nonlinear description, moves at subrelativistic velocities relative to the background plasma, as in the linear limit studied earlier. That $\beta_{\rm w\vert p}$ is positive means that the ``Weibel frame'' moves at slightly smaller absolute velocity towards the shock front than the background plasma.

 In the near precursor, as the background plasma is heated up to relativistic temperatures, $\beta_{\rm w\vert p}$ increases in magnitude; this implies that the background plasma decouples from the ``Weibel frame'', hence increasing the heating rate and leading to the shock transition.

 One can also compute the first nonlinear correction in $\Xi_{\rm p}$ to the above velocity. To this effect, we recast Eq.~\eqref{eq:phi_stat} in terms of the nonlinearity parameters of the beam ($\Xi_{\rm b}$) and the plasma ($\Xi_{\rm p}$) in relation~\eqref{eq:weibel_cond}:
\begin{equation}
 \frac{n_{b} \gamma_{\rm b|w}^2 \beta_{\rm b|w}}{T_{\rm b}} \frac{\sinh\Xi_{\rm b|w}}{\Xi_{\rm b|w}}\, + \,\frac{n_{p} \gamma_{\rm p|w}^2 \beta_{\rm p|w}}{T_{\rm p}}
   \frac{\sinh\Xi_{\rm p|w}}{\Xi_{\rm p|w}} = 0 \,.
\label{eq:weibel_cond_NL}
\end{equation}
As discussed in Paper~III~\cite{pap3}, the beam particles carry such inertia that they hardly participate in the filamentation, meaning $\Xi_{\rm b\vert w} \ll 1$. In $\mathcal{R}_{\rm w}$, $\Xi_{\rm b\vert w}$ indeed represents the ratio of the electromagnetic component $e A_{x\vert\rm w}$ to the typical momentum $T_{\rm b\vert w} = T_{\rm b}/\gamma_{\rm b\vert w}$ of the particles, and for suprathermal particles, this ratio is much smaller than unity. Assuming $\Xi_{\rm p\vert w} \lesssim 1$, we then obtain to lowest order
\begin{equation}\label{eq:weibel_cond_NL_2}
\frac{n_{b} \gamma_{\rm b|w}^2 \beta_{\rm b|w}}{T_{\rm b}} + \frac{n_{p} \gamma_{\rm p|w}^2 \beta_{\rm p|w}}{T_{\rm p}} \left(1 + \frac{ \Xi_{\rm p|w}^2}{6}  \right) \simeq 0 \,.
\end{equation}
In this configuration, the nonlinearity appears as a second-order correction to the ``Weibel frame'' speed. In detail, one obtains
\begin{equation}
\beta^{(\rm n-lin)}_{\rm w\vert p} \simeq \beta_{\rm w\vert p}\left(1-\frac{\Xi_{\rm p}^2}{6}\right) \,.
\end{equation} 

Note that the ``Weibel frame'' is not always well-defined in the strongly nonlinear limit: if $\Xi_{\rm p,b} \gg 1$, Eq.~\eqref{eq:phi_stat} decouples into a set of two relations between the temperatures and densities
\begin{flalign}
&\gamma_{\rm b|w} n_b \,=\, \gamma_{\rm p|w} n_p \,, \\
&\frac{\gamma_{\rm b|w} \beta_{\rm b|w}}{T_b} = -\frac{\gamma_{\rm p|w} \beta_{\rm p|w}}{T_p} \,, 
\end{flalign}
which overdetermine the system for a given set of parameters. From Eq.~\eqref{eq:phi_stat}, we see that the error on the electrostatic fields at leading order in $\Xi_{\rm p|w}$ if we evaluate the ``Weibel frame'' from relation~\eqref{eq:weibel_cond} evolves as $\delta \phi \propto \Xi_{\rm p|w}^3$. In the case of interest, however, PIC simulations indicate that the weakly nonlinear limit represents a good approximation in the precursor of relativistic shocks, see Sec.~\ref{sec:nlinPIC}.

\subsection{Comparison to PIC simulations}\label{sec:nlinPIC}

In this section, we confront the result of Sec.~\eqref{sec:nlinwf} with the PIC simulations presented in the Sec.~\ref{sec:linPIC}. A strong hypothesis underlying the above formulae is that of a marginally nonlinear plasma response $\Xi_{\rm p|w} < 1$. We wish to motivate this hypothesis by measuring this nonlinearity parameter. Assuming Eq.~\eqref{eq:density} holds, one expects 
\begin{align}
n_{\rm p}(y) &= n_{{\rm p}0}  \exp \left[ - \frac{\gamma_{\rm p|s}} {T_{\rm p}} \left(\phi - \beta_{\rm p|s} A_{x}\right) \right] \nonumber\\
&= n_{{\rm p}0} \exp \left[ \frac{\gamma_{\rm p|w} \beta_{\rm p|w}} {T_{\rm p}}   A_{x}\right] \,.
\end{align}
We can then estimate the nonlinearity parameter in the simulation through the following relation
\begin{equation}
\Xi_{\rm p\vert w} \,\simeq\,\sqrt{2}\left\langle \left[ \log\left(\frac{n_{\rm p}}{\langle n_{\rm p}\rangle}\right) - \left\langle \log \left(\frac{n_{\rm p}}{\langle  n_{\rm p}\rangle}\right) \right\rangle \right]^2\right\rangle^{1/2}\,,
\label{eq:xipw}
\end{equation}
where $\langle\cdot\rangle$ denotes the mean value along the direction transverse to the drift.

\begin{figure}
\centering
  \includegraphics[width=0.47\textwidth]{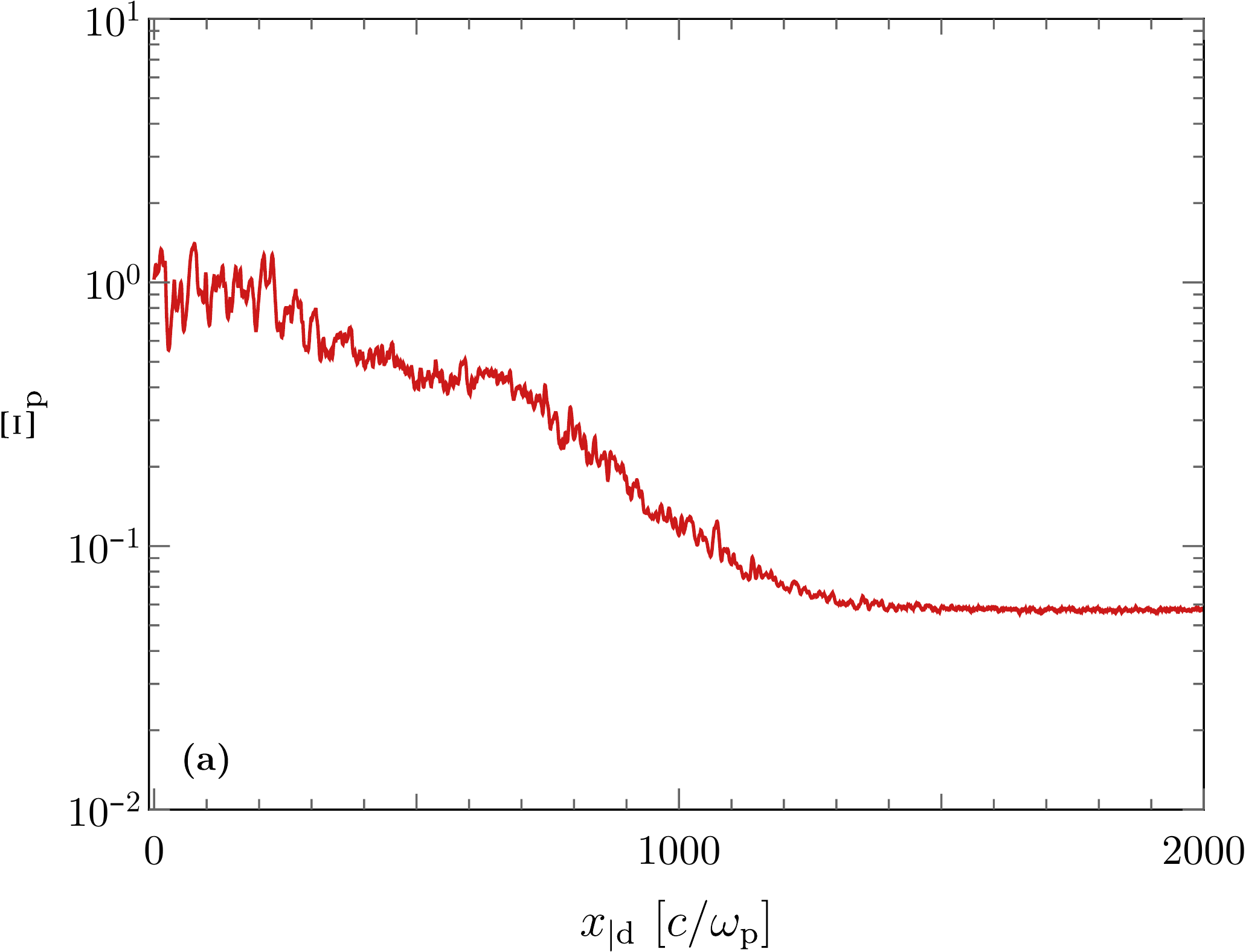}\smallskip\\
  \includegraphics[width=0.47\textwidth]{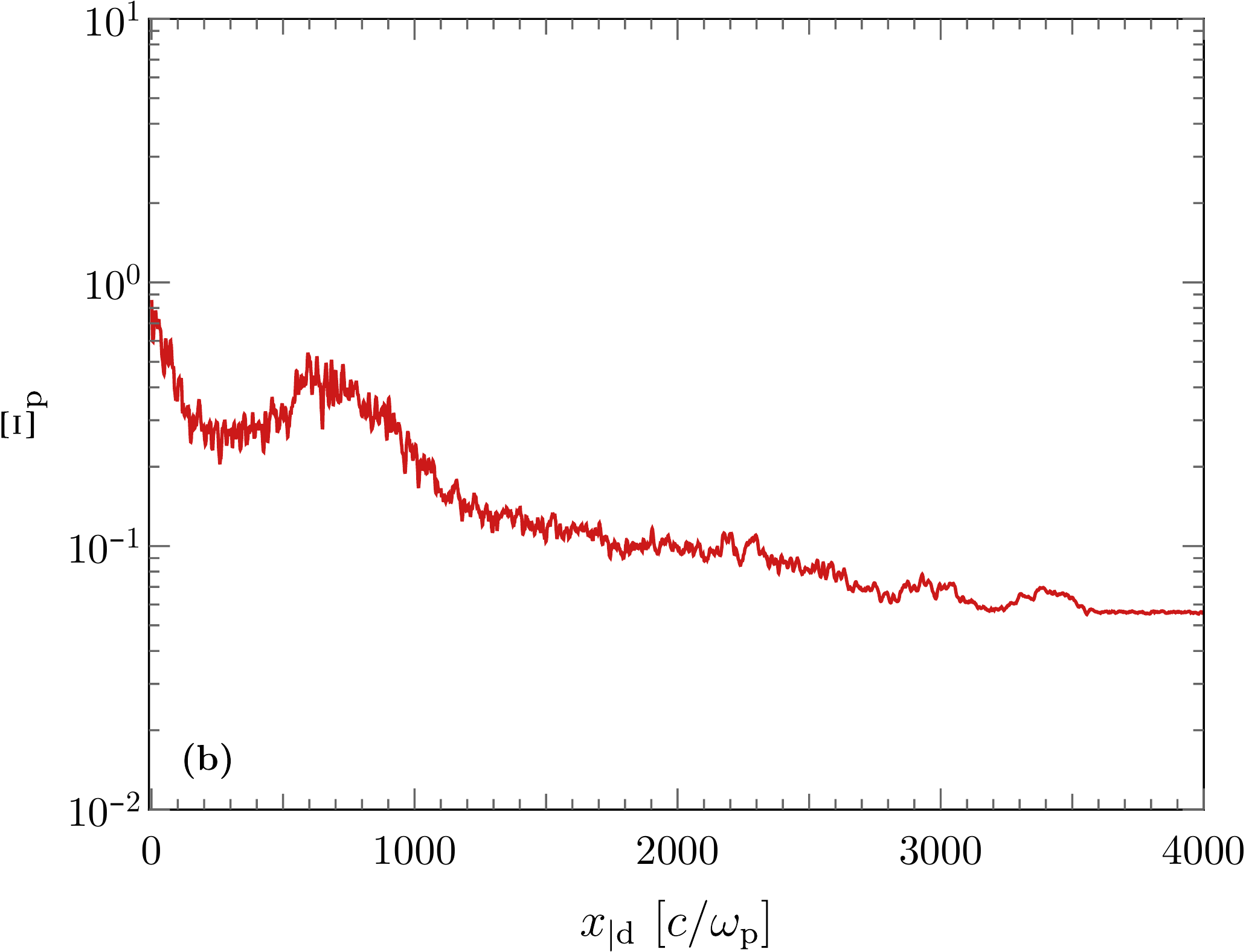}
\caption{Evolution of the nonlinearity parameter of the background plasma $\Xi_{\rm p\vert w}$. Panel (a), top: $\gamma_{\infty\vert\rm d} = 10$ at time $t = 3600\, \omega_{\rm p}^{-1}$. Panel (b), bottom: $\gamma_{\infty\vert\rm d} = 100$ at time $t = 6900\, \omega_{\rm p}^{-1}$.}
\label{fig:xipw}
\end{figure}

Using relation~\eqref{eq:xipw}, we present in Figs.~\ref{fig:xipw} our estimate of $\Xi_{\rm p\vert w}$ for the two reference simulations ($\gamma_\infty = 17$ in the top panel, $\gamma_\infty=173$ in the bottom panel). In both cases, the nonlinearity parameter tends to increase from values well below unity in the far precursor, up to near unity within a few hundred $c/\omega_{\rm p}$ to the shock front. 
This behavior suggests that one should use the present nonlinear equilibrium model to estimate $\beta_{\rm w\vert p}$ in the near precursor, say $x\lesssim300c/\omega_{\rm p}$, and the linear estimate obtained in Sec.~\ref{sec:kinwf} at larger distances, where $\Xi_{\rm p\vert w}$ falls to values small compared to unity. One must, however, keep in mind that the present approach is based on a fluid model, while the former is fully kinetic, and that the analysis of Sec.~\ref{sec:kinwf} indicates that, from the point of view of the instability, both the beam and the background plasma should be treated kinetically. Fortunately, both the present nonlinear estimate of $\beta_{\rm w\vert p}$, Eq.~(\ref{eq:bwpnlin}), and its linear kinetic counterpart, Eq.~\eqref{eq:kinpkinbbw}, match one another. Hence, our theoretical estimate of $\beta_{\rm w\vert p}$ can be considered as rather well established throughout the precursor.

\begin{figure}
\centering
  \includegraphics[width=0.45\textwidth]{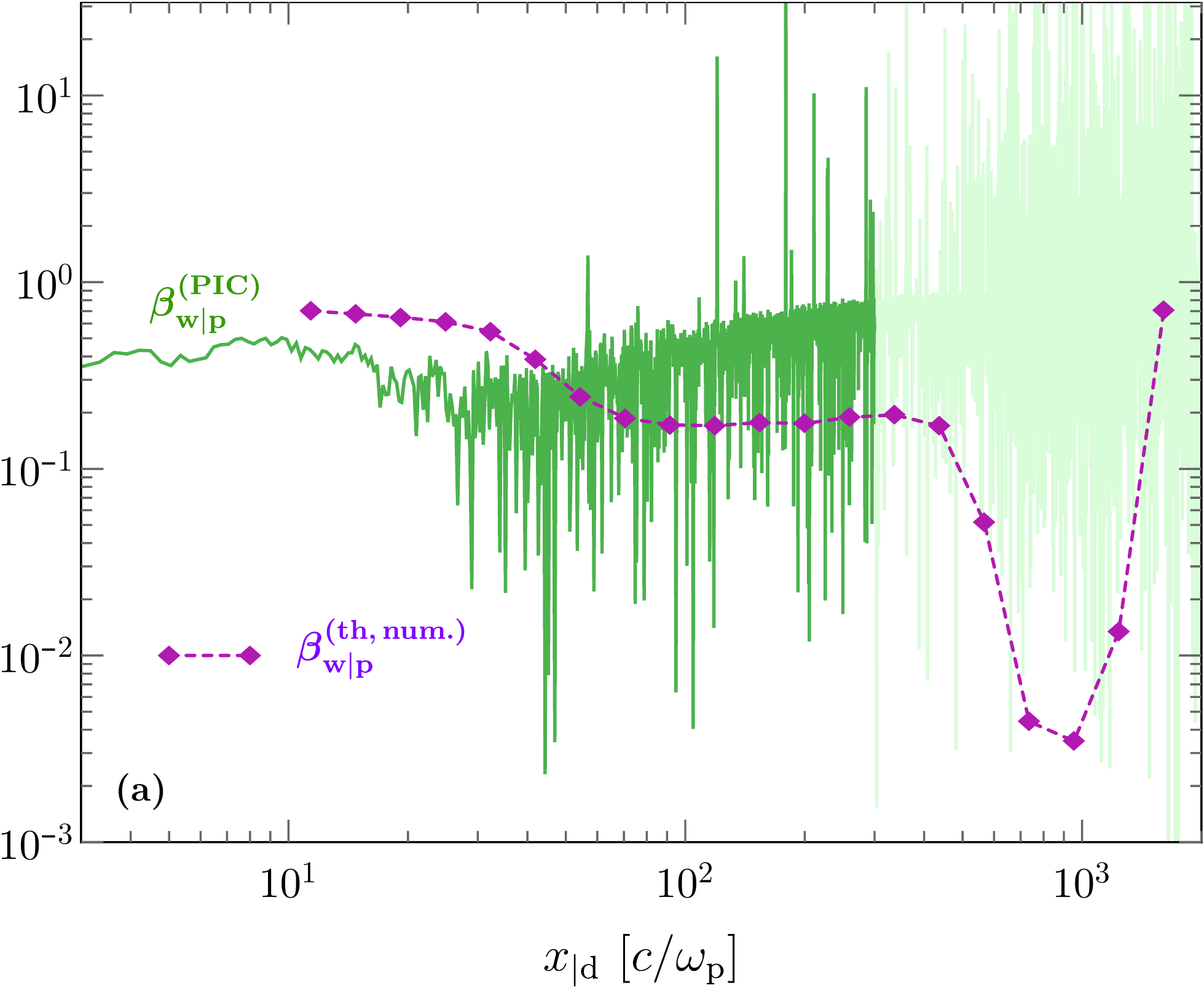}\smallskip\\
  \includegraphics[width=0.45\textwidth]{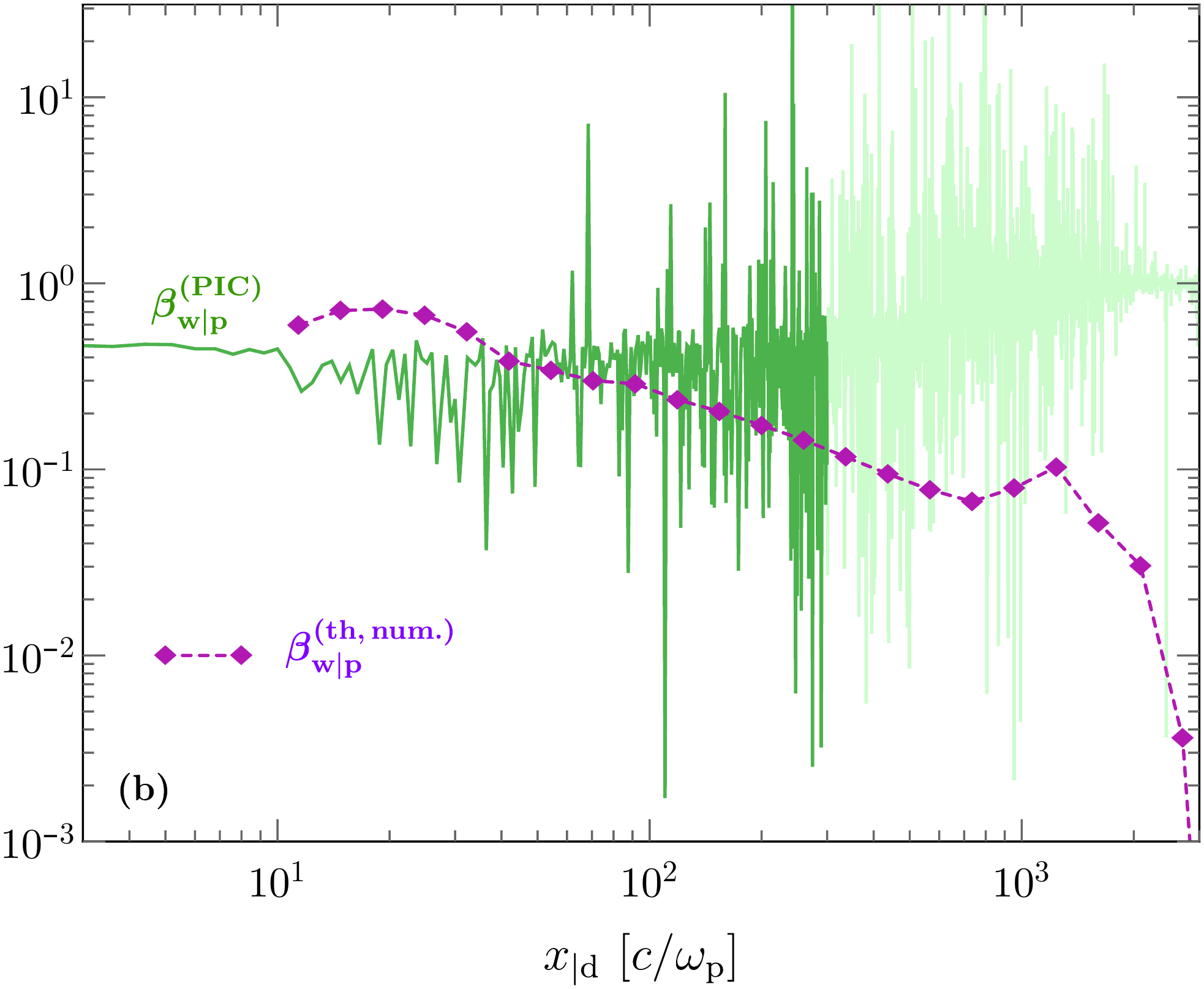}
\caption{Similar to Fig.~\ref{fig:comp_PIC_bw}, for our theoretical estimate of $\beta_{\rm w\vert p}$ obtained assuming nonlinear equilibrium filaments, as developed in Sec.~\ref{sec:nlinwf}. The purple diamond symbols/dashed curve uses \eqref{eq:bwpnlin} to compute the $\beta_{\rm w\vert p}$. Panel (a), top: $\gamma_\infty = 17$ ($\gamma_{\infty\vert\rm d} = 10$). Panel (b), bottom: $\gamma_\infty = 173$ ($\gamma_{\infty\vert\rm d} = 100$).}
\label{fig:bweibel}
\end{figure}

This is supported by Figs.~\ref{fig:bweibel}, which compare the values of $\beta_{\rm w\vert p}$ extracted from the simulations with the theoretical estimate \eqref{eq:bwpnlin}, assuming (weakly) nonlinear equilibrium filaments. As before, the PIC simulation data are light colored where they cannot be measured accurately. In both simulation cases, the theoretical estimates appear in reasonable agreement with the PIC results, especially in the near precursor $x \lesssim 10^3c/\omega_{\rm p}$, where the velocity of the ``Weibel frame'' is well defined.

\section{Conclusions} \label{sec:conc}
This paper belongs to a series of articles in which we build a theoretical model of unmagnetized, relativistic collisionless pair shocks, and compare it with dedicated PIC simulations. More specifically, we have discussed in the present work the physics of the purely transverse CFI that results from the interpenetration of the background plasma and the beam of suprathermal particles in the shock precursor. We have argued that there exists a frame $\mathcal R_{\rm w}$, referred to as the ``Weibel frame'' in which the instability is mostly magnetic in nature. We have derived the velocity of this frame at each point of the shock precursor, through a kinetic description of the linear stage of the instability, as well as through a quasistatic model of the nonlinear phase of the filamentation instability. This ``Weibel frame'' is of particular importance to the physics of the shock and of the acceleration process, because it represents the frame of the scattering centers. 

We emphasize the following properties of this ``Weibel frame'': (i) it is found to move at subrelativistic velocities relative to the background plasma, {(\it i.e.}, with 4-velocity $u_{\rm w\vert p} < 1$), and, therefore, at relativistic velocities toward the shock front. This result can be related to the strong asymmetry of the interation between the beam of suprathermal particles and the background plasma. We also find that, (ii), the 3-velocity $\beta_{\rm w\vert p}$ is of opposite sign to $\beta_{\rm p}$, the velocity of the background plasma in the shock rest frame, implying that the ``Weibel frame'' moves slightly less fast than the background plasma relative to the shock front. Furthermore, (iii), the relative velocity $\beta_{\rm w\vert p}$ typically scales as $\xi_{\rm b}$, which represents the beam pressure normalized to the incoming momentum flux at infinity. Finally, it is to be emphasized that the ``Weibel frame'' is not globally inertial, because its velocity $\beta_{\rm w}$ depends on the distance to the shock. As discussed in ~\cite{L1}, and more particularly \cite{pap2}, which addresses in detail the consequences of this noninertial nature, the ``Weibel frame'' decelerates from the far to the near precursor. In the present article, we have determined the velocity of this frame at each point in the precursor, given the local physical conditions, in a WKB-like approximation.

Our PIC simulations confirm these various features. In particular, they reveal that $\delta E_\perp < \delta B_\perp$ close to the shock front and over most of the precursor, and hence that the ``Weibel frame'' is well defined; relatively to the shock front, the ``Weibel frame'' velocity, measured through the ratio of $\delta E_\perp/\delta B_\perp$, is also found to be slightly below the background plasma velocity; finally, the relative velocity $\beta_{\rm w\vert p}$ generally decreases away from the shock front, like $\xi_{\rm b}$.

Our kinetic model relies on the use of Maxwell-J\"uttner distribution functions for the beam and the background plasma, and it provides (at the expense of rather complex calculations) simple approximations to the velocity of the ``Weibel frame'' and to the growth rate of the purely transverse CFI. Our quasistatic model of the nonlinear stage of the instability describes the current filaments as periodic magnetostatic structures in the ``Weibel frame'', in pressure equilibrium with the plasma. Our PIC simulations indicate that, over most of the precursor, those filaments are in a mildly nonlinear stage, with a nonlinearity parameter below unity. We expect that both models should capture the salient features of the instability and indeed, the resulting formulae turn out to bracket rather well the value $\beta_{\rm w\vert p}$ seen in PIC simulations.

In subsequent papers, it will be shown that the ``Weibel frame'' plays an essential role in shaping the microphysics of the shock transition, in particular the physics of heating and deceleration of the background plasma.

\begin{acknowledgments} We acknowledge financial support from the Programme National Hautes \'Energies (PNHE) of the C.N.R.S., the ANR-14-CE33-0019 MACH project and the ILP LABEX (under reference ANR-10-LABX-63) as part of the Idex SUPER, and which is financed by French state funds managed by the ANR within the "Investissements d'Avenir" program under reference ANR-11-IDEX-0004-02. This work was granted access to the HPC resources of TGCC/CCRT under the allocation 2018-A0030407666 made by GENCI. We also acknowledge PRACE for awarding us access to resource Joliot Curie-SKL at TGCC Center.
\end{acknowledgments}

\appendix

\begin{widetext}
\section{Reduced expressions of the kinetic dielectric tensor} \label{app:simp_exp_eps}

The triple integrals involved into Eq.~\eqref{eq:eps} after substitution of the model distribution \eqref{eq:edf} can be recast in the form of one-dimensional quadratures \cite{Bret_2007}. To do so, we mostly follow the lines sketched in Ref.~\cite{Wright_1975}, where it is shown that two of the momentum integrations can be carried out in closed form.

Let us first consider the dielectric tensor element $\varepsilon_{xx}$, which is involved in the purely electromagnetic dispersion relation of the CFI (\emph{i.e.}, in a frame close to the ``Weibel frame''). After straightforward algebra, it can be rewritten as
\begin{equation} \label{eq:epsxx1}
    \varepsilon_{xx} = 1  + \sum_\alpha\frac{\omega_{\rm p\alpha}^2}{\omega^2}  \mathcal{C}_\alpha \gamma_\alpha^2\mu_\alpha \beta_\alpha 
    \int {\rm d}^3u\, v_x e^{-\gamma_\alpha\mu_\alpha (\gamma-\beta_\alpha u_x)}
     - \frac{\omega_{\rm p\alpha}^2}{\zeta k^2}  \mathcal{C}_\alpha \gamma_\alpha^2\mu_\alpha  \int {\rm d}^3 u\,v_x^2 \frac{ e^{-\gamma_\alpha\mu_\alpha (\gamma-\beta_\alpha u_x)}}
    {\zeta - v_y} \,,
\end{equation}
where we have defined
\begin{equation}
  \mathcal{C}_\alpha = \frac{\mu_\alpha}{4\pi \gamma_\alpha K_2 (\mu_\alpha)} \,.
\end{equation}
The first integral in the right-hand side of \eqref{eq:epsxx1} can be exactly solved by
noting that
\begin{equation}
  \beta_\alpha = \int {\rm d}^3u\, v_x e^{- \gamma_\alpha\mu_\alpha (\gamma-\beta_\alpha u_x)} \,.
\end{equation}
We then proceed by changing to velocity variables in cylindrical coordinates along the wave vector:
$\mathbf{v}=(v_\parallel, v_\perp\cos \theta, v_\perp\sin \theta)$. Making use of ${\rm d}^3 u = \gamma^5{\rm d}^3 v$, Eq.~\eqref{eq:epsxx1} then
becomes
\begin{equation} \label{eq:epsxx2}
  \varepsilon_{xx}  = 1 + \sum_\alpha\frac{\omega_{\rm p\alpha}^2}{\omega^2}  \gamma_\alpha^2\beta_\alpha^2\mu_\alpha   
   - \frac{\omega_{\rm p\alpha}^2}{\zeta k^2} \mathcal{C}_\alpha \gamma_\alpha^2\mu_\alpha  \int  \frac{{\rm d}v_\parallel}{\zeta-v_\parallel}  \int_0^{1/\gamma_\parallel} {\rm d}v_\perp \,v_\perp^3 \gamma^5 e^{-\gamma_\alpha\mu_\alpha \gamma}
  \int_0^{2\pi} {\rm d}\theta\, \cos^2 \theta\, e^{\gamma_\alpha\beta_\alpha\mu_\alpha   \gamma v_\perp \cos \theta} \,, 
\end{equation}
with $\gamma_\parallel =(1-v_\parallel^2)^{-1/2}$.
Given the integral representation of the modified Bessel functions of the first kind \cite{Abramowitz_1972},
\begin{equation} \label{eq:In}
    I_n(z) = \frac{1}{\pi}\int _0^\pi {\rm d}t \, e^{z\cos t} \cos (nt) \,,
\end{equation}
we obtain 
\begin{equation} \label{eq:epsxx3}
    \varepsilon_{xx} = 1 +  \sum_\alpha\frac{\omega_{\rm p\alpha}^2}{\omega^2} \mathcal{C}_\alpha \mu_\alpha \gamma_\alpha^2\beta_\alpha^2
    - \frac{\pi\omega_{\rm p\alpha}^2}{k^2 \zeta}  \mathcal{C}_{\alpha} \gamma_\alpha^2\mu_\alpha  \int_{-1}^{1} {\rm d}v_\parallel \, \frac{\mathcal{S}_{30}^\alpha + \mathcal{S}_{32}^\alpha}
    {\zeta - v_\parallel} \,.
\end{equation}
Here we have introduced 
\begin{equation} \label{eq:Smn}
    \mathcal{S}_{mn}^\alpha (v_\parallel) = \int_0^{1/\gamma_\parallel} {\rm d}v_\perp \,\gamma^5 v_{\perp}^m I_n (\mu_\alpha \gamma_\alpha\beta_\alpha 
    v_\perp) e^{-\gamma_\alpha\mu_\alpha \gamma} \,.
\end{equation}  
From the definition $t = \gamma/\gamma_\parallel$ follow the relations $\gamma v_\perp=(t^2-1)^{1/2}$ and $v_\perp {\rm d}v_\perp = {\rm d}t/\gamma_\parallel^2 t^3$,
allowing us to express $\mathcal{S}_{30}^\alpha$ and $\mathcal{S}_{32}^\alpha$ as
\begin{align}
    &\mathcal{S}_{30}^\alpha(v_\parallel) = \gamma_\parallel \int_1^\infty {\rm d}t \, (t^2-1) I_0 \left( \mu_\alpha\gamma_\alpha \beta_\alpha \sqrt{t^2-1} \right) e^{-\gamma_\alpha\mu_\alpha \gamma_\parallel t} \,, \\
    &\mathcal{S}_{32}^\alpha(v_\parallel) = \gamma_\parallel \int_1^\infty {\rm d}t \, (t^2-1) I_2 \left( \mu_\alpha\gamma_\alpha \beta_\alpha \sqrt{t^2-1} \right) e^{-\gamma_\alpha\mu_\alpha \gamma_\parallel t} \,,
\end{align}
which can be further recast as
\begin{align}
\label{eq:s30-s32}
    &\mathcal{S}_{30}^\alpha (v_\parallel) =  \gamma_\parallel
    \left(\frac{1}{\gamma_\alpha^2\mu_\alpha^2} \frac{d^2}{d\gamma_\parallel^2} - 1 \right)  \int_1^\infty {\rm d}t \, I_0 \left( \mu_\alpha \gamma_\alpha\beta_\alpha \sqrt{t^2-1} \right) e^{-\gamma_\alpha\mu_\alpha \gamma_\parallel t} \,, \\
    &\mathcal{S}_{32}^\alpha (v_\parallel) =  \gamma_\parallel
    \left(\frac{1}{\gamma_\alpha^2\mu_\alpha^2} \frac{d^2}{d\gamma_\parallel^2} - \frac{2}{\mu_\alpha} \frac{d}{d\gamma_\parallel}  + 1 \right) \int_1^\infty {\rm d}t \, I_2 \left( \mu_\alpha \gamma_\alpha\beta_\alpha \sqrt{t^2-1} \right) e^{-\gamma_\alpha\mu_\alpha \gamma_\parallel t} \,,
\end{align}
We now take advantage of the following formula \cite{Gradshteyn_1980}
\begin{equation} \label{eq:grad}
    \int_0^\infty {\rm d}x\,\left( \frac{x-1}{x+1} \right)^{\nu/2} e^{-\delta x} I_\nu \left( \zeta \sqrt{x^2-1} \right)  
    = \frac{e^{-\sqrt{\delta^2 - \zeta^2}}} {\sqrt{\delta^2 - \zeta^2}} \left( \frac{\zeta} {\delta + \sqrt{\delta^2 - \zeta^2}} \right)^\nu \,,
\end{equation}
valid for $\Re \nu > -1$ and $\delta > \zeta$. There follows
\begin{align} 
    \mathcal{S}_{30}^\alpha &= \frac{\gamma_\parallel}{\gamma_\alpha\mu_\alpha}
     \left( \frac{1}{\gamma_\alpha^2\mu_\alpha^2} \frac{d^2}{d\gamma_\parallel^2} -1 \right)
     \left[ \frac{e^{-\gamma_\alpha\mu_\alpha \Pi_\alpha}} {\Pi_\alpha}  \right] \,, \\
     \mathcal{S}_{32}^\alpha &= \frac{\beta_\alpha^2 \gamma_\parallel}{\gamma_\alpha\mu_\alpha}
    \left( \frac{1}{\gamma_\alpha^2\mu_\alpha^2} \frac{d^2}{d\gamma_\parallel^2} - \frac{2}{\gamma_\alpha\mu_\alpha}\frac{d}{d\gamma_\parallel} + 1 \right)
    \left[ \frac{e^{-\gamma_\alpha\mu_\alpha \Pi_\alpha}} {\Pi_\alpha(\gamma_\parallel +\Pi_\alpha)^2} \right]  \,.
\end{align}
where $\Pi_\alpha = (\gamma_\parallel^{2} -\beta_\alpha^2)^{1/2}$. After evaluation of the first and second-order derivatives, one finds
\begin{equation} \label{eq:sxx}
    \mathcal{S}_{30}^\alpha + \mathcal{S}_{32}^\alpha
    =  \frac{2\gamma_\parallel}{\gamma_\alpha^2\mu_\alpha^2} 
     \left[
     \frac{\gamma_{\parallel}^2 + 2\beta_\alpha^2}{\Pi_\alpha^5}
    +\frac{\gamma_\alpha\mu_\alpha (\gamma_\parallel^2 + 2\beta_\alpha^2)}
    {\Pi_{\alpha}^4} + \frac{\gamma_\alpha^2\mu_\alpha^2 \beta_\alpha^2}{\Pi_\alpha^3}
    \right]
     e^{-\gamma_\alpha\mu_\alpha\Pi_\alpha} \,.
\end{equation}    
Substituting Eq.~\eqref{eq:sxx} into  Eq.~\eqref{eq:epsxx3} finally gives the following simplified expression for $\varepsilon_{xx}$:
\begin{equation} \label{eq:epsxx}
	\varepsilon_{xx} = 1 + \sum_\alpha\frac{\omega_{\rm p\alpha}^2}{\omega^2} \mu_\alpha \gamma_\alpha^2\beta_\alpha^2 
	-\frac{2\pi\omega_{\rm p\alpha}^2}{k^2 \zeta} \mathcal{C}_\alpha \gamma_\alpha 
      	\int _{-1}^{1} {\rm d}v_\parallel \, \frac{\gamma_\parallel}{\zeta - v_\parallel} 
	\left[ \frac{\gamma_\parallel^2 + 2\beta_\alpha^2} {\gamma_\alpha^2\mu_\alpha^2 \Pi_\alpha^5}
      	+\frac{\gamma_\parallel^2+2\beta_\alpha^2}{\gamma_\alpha\mu_\alpha \Pi_\alpha^4}
      	+\frac{\beta_\alpha^2}{\Pi_\alpha^3} \right]
	e^{-\gamma_\alpha\mu_\alpha \Pi_\alpha} \,.
\end{equation}

Let us now consider $\varepsilon_{yy}$, which writes
\begin{equation} \label{eq:epsyy1}
    \varepsilon_{yy} = 1 - \frac{\omega_{\rm p\alpha}^2}{\zeta k^2}\sum_\alpha \gamma_\alpha^2\mu_\alpha 
    \int {\rm d}^3 u\, v_y^2 \frac{ e^{-\gamma_\alpha\mu_\alpha (\gamma-\beta_\alpha u_x)}}
    {\zeta - v_y} \,.
\end{equation}
After changing to cylindrical velocity coordinates, one obtains
\begin{equation} \label{eq:epsyy2}
    \varepsilon_{yy} = 1-\sum_\alpha\frac{\omega_{\rm p\alpha}^2}{\zeta k^2}  \mathcal{C}_\alpha \gamma_\alpha^2\mu_\alpha 
    \int_{-1}^{1} {\rm d}v_\parallel\,\frac{v_\parallel^2}{\zeta-v_\parallel}  \int_0^{1/\gamma_\parallel} {\rm d}v_\perp \, \gamma^5 v_\perp e^{-\gamma_\alpha\mu_\alpha \gamma}
    \int_0^{2\pi} {\rm d}\theta\, e^{\gamma_\alpha\mu_\alpha \beta_\alpha \gamma v_\perp \cos \theta} \,,
\end{equation}
which, making use of Eq.~\eqref{eq:In}, reduces to
\begin{equation}\label{eq:epsyy3}
    \varepsilon_{yy} = 1 -\sum_\alpha\frac{2\pi\omega_{\rm p\alpha}^2 \zeta}{k^2}  \mathcal{C}_\alpha \gamma_\alpha^2\mu_\alpha 
   \int_{-1}^{1} {\rm d}v_\parallel \, v_\parallel^2 \frac{S_{10}^\alpha}{\zeta-v_\parallel} \,.
\end{equation}
Making use of Eq.~\eqref{eq:grad}, one obtains
\begin{equation}
\label{eq:s10}
    \mathcal{S}_{10}^\alpha (v_\parallel) =  \frac{\gamma_\parallel^3}{\gamma_\alpha^2\mu_\alpha^2} \frac{{\rm d}^2}{{\rm d}\gamma_\parallel^2} \int_0^\infty {\rm d}t \, 
    I_0 \left( \mu_\alpha \gamma_\alpha\beta_\alpha \sqrt{t^2-1} \right) e^{-\gamma_\alpha\mu_\alpha \gamma_\parallel t} = \left( \frac{\gamma_\parallel}{\gamma_\alpha\mu_\alpha} \right)^3 \frac{{\rm d}^2}{{\rm d}\gamma_\parallel^2} 
    \left[ \frac{e^{-\gamma_\alpha\mu_{\alpha} \Pi_{\alpha}}} {\Pi_{\alpha}} \right]  \,,
\end{equation}
and therefore
\begin{equation} \label{eq:syy}
    \mathcal{S}_{10}^\alpha (v_\parallel) = \left( \frac{\gamma_\parallel}{\gamma_\alpha\mu_\alpha} \right)^3    
    \left[ \frac{2\gamma_\parallel^2 +\beta_\alpha^2}{\Pi_\alpha^5} +\frac{\gamma_\alpha\mu_\alpha (2\gamma_\parallel^2 +\beta_\alpha^2)} {\Pi_\alpha^4}
    + \frac{\gamma_\alpha^2\mu_\alpha^2\gamma_\parallel^2}{\Pi_{\alpha}^3} \right] e^{-\gamma_\alpha\mu_\alpha \Pi_\alpha} \,.
\end{equation}    
Combining Eqs.~\eqref{eq:epsyy3} and \eqref{eq:syy} then readily yields
\begin{equation} \label{eq:epsyy}
      \varepsilon_{yy}  = 1 - \sum_\alpha\frac{2\pi\omega_{\rm p\alpha}^2}{\zeta k^2} \mathcal{C}_\alpha \gamma_\alpha
      \int _{-1}^1 {\rm d}v_\parallel \, \frac{v_\parallel^2 \gamma_\parallel^3} {\zeta - v_\parallel}
      \left[ \frac{2\gamma_\parallel^2+\beta_\alpha^2} {\gamma_\alpha^2\mu_\alpha^2 \Pi_\alpha^5}
      + \frac{2\gamma_\parallel^2+\beta_\alpha^2} {\gamma_\alpha\mu_\alpha \Pi_\alpha^4}
      + \frac{\gamma_\parallel^2}{\Pi_\alpha^3} \right]
      e^{-\gamma_\alpha\mu_\alpha \Pi_\alpha} \,.
\end{equation}

Finally, let us rewrite the non-diagonal dielectric tensor element $\varepsilon_{xy}$ as
\begin{equation} \label{eq:epsxy1}
    \varepsilon_{xy} =  \sum_\alpha\frac{\omega_{\rm p\alpha}^2}{k^2 \zeta} \mu_\alpha \gamma_\alpha^2\beta_\alpha 
    - \frac{\omega_{\rm p\alpha}^2}{k^2} \mathcal{C}_\alpha \gamma_\alpha^2\mu_\alpha  \int {\rm d}^3 u\,v_x \frac{e^{-\gamma_\alpha\mu_\alpha (\gamma-\beta_\alpha u_x)}}
    {\zeta - v_y} \,,
\end{equation}
Changing to cylindrical velocity variables leads to
\begin{equation} \label{eq:epsxy2}
    \varepsilon_{xy} =  \sum_\alpha\frac{\omega_{\rm p\alpha}^2}{k^2 \zeta} \mu_\alpha \gamma_\alpha^2\beta_\alpha 
    -\frac{2\pi\omega_{\rm p\alpha}^2}{k^2} \mathcal{C}_\alpha \gamma_\alpha^2\mu_\alpha \int_{-1}^{1} \frac{{\rm d}v_\parallel}{\zeta-v_\parallel}   \int_0^{1/\gamma_\parallel} {\rm d}v_\perp \, \gamma^5 v_\perp^2 e^{-\gamma_\alpha\mu_\alpha \gamma} 
    \int _0^{2\pi} {\rm d}\theta \, e^{\gamma_\alpha\mu_\alpha \beta_\alpha \gamma_\alpha
    v_\perp \cos \theta} \,.
\end{equation}
Using Eq.~\eqref{eq:In} gives
\begin{equation} \label{eq:epsxy3}
    \varepsilon_{xy} =  \sum_\alpha\frac{\omega_{\rm p\alpha}^2}{k^2 \zeta} \mu_\alpha \gamma_\alpha^2\beta_\alpha 
    - \frac{2\pi\omega_{\rm p\alpha}^2}{k^2} \mathcal{C}_\alpha \gamma_\alpha^2\mu_\alpha 
    \int_{-1}^{1} {\rm d}v_\parallel \, \frac{\mathcal{S}_{21}^\alpha}{\zeta-v_\parallel}  \,,
\end{equation}
where $\mathcal{S}_{21}^\alpha$ can be recast in the form
\begin{equation}
    \mathcal{S}_{21}^\alpha(v_\parallel) = \gamma_\parallel^2
    \int_1^\infty {\rm d}t \, \left(\frac{t-1}{t+1}\right)^{1/2} t(t+1)  I_1 \left( \gamma_\alpha\mu_\alpha \beta_\alpha \sqrt{t^2-1} \right) e^{-\gamma_\alpha\mu_\alpha \gamma_\parallel t} \,.
\end{equation}
Exploiting Eq.~\eqref{eq:grad} yields
\begin{equation}
\label{eq:s21b}
    \mathcal{S}_{21}^\alpha (v_\parallel) = \frac{\gamma_\parallel^2}{\beta_\alpha \gamma_\alpha^3\mu_\alpha^3}
    \left(\frac{{\rm d}^2}{d\gamma_\parallel^2} -\gamma_\alpha\mu_\alpha \frac{{\rm d}}{{\rm d}\gamma_\parallel} \right)
    \left[ \left(\frac{\gamma_\parallel}{\Pi_\alpha} -1 \right) e^{-\gamma_\alpha\mu_\alpha \Pi_\alpha} \right] \, ,
\end{equation}
After evaluation of the derivatives, one finds
\begin{equation} \label{eq:sxy}
    \mathcal{S}_{21}^\alpha (v_\parallel) = \beta_\alpha \left( \frac{\gamma_\parallel}{\gamma_\alpha\mu_\alpha}\right)^3 
    \left[\frac{\gamma_\alpha^2\mu_\alpha^2}{\Pi_\alpha^3} + \frac{3\gamma_\alpha\mu_\alpha}{\Pi_\alpha^4} + \frac{3}{\Pi_\alpha^5} \right]
    e^{-\gamma_\alpha\mu_\alpha \Pi_\alpha} \,,
\end{equation}    
Combining Eqs.~\eqref{eq:epsxy3} and \eqref{eq:sxy}, one obtains the simplified expression
\begin{equation} \label{eq:epsxy}
      \varepsilon_{xy}  =  \sum_\alpha\frac{\omega_{\rm p\alpha}^2}{k^2 \zeta} \mu_\alpha \gamma_\alpha^2\beta_\alpha 
      -\frac{2\pi\omega_{\rm p\alpha}^2}{k^2} \mathcal{C}_\alpha \gamma_\alpha\beta_\alpha 
      \int _{-1}^1 {\rm d}v_\parallel \, \frac{\gamma_\parallel^3} {\zeta - v_\parallel} 
      \left[ \frac{1}{\Pi_\alpha^3} + \frac{3}{\gamma_\alpha\mu_\alpha \Pi_\alpha^4} + \frac{3}{\gamma_\alpha^2\mu_\alpha^2 \Pi_\alpha^5} \right]
      e^{-\gamma_\alpha\mu_\alpha \Pi_\alpha} \,.
\end{equation}
 
It should be stressed that the compact expressions \eqref{eq:epsxx}, \eqref{eq:epsyy} and \eqref{eq:epsxy} are strictly valid for
$\Im \zeta >0$ only. They thus lend themselves readily to the numerical resolution of the dispersion relation (\ref{eq:disp}) when
searching for (purely growing) unstable modes only, as has been done in Refs.~\cite{Bret_2007,Bret_2008,Bret_2010b}. 
If damped modes are also examined, care must be taken for the analytic extension to the lower complex $\zeta$-plane of the integrals,
owing to the presence of branch points at $\zeta = \pm 1$ \cite{Mikhailovskii_1981,Bret_2010b}.

To obtain analytic approximations, it is convenient to make the change of integration variable $v_\parallel \to s$, such that 
$\gamma_\parallel=\sqrt{1+s^2/\gamma_\alpha^2}$. This gives the alternative expressions
\begin{align}
  \varepsilon_{xx} &= 1 +  \sum_\alpha\frac{\omega_{\rm p\alpha}^2}{k^2 \zeta^2} \mu_\alpha \gamma_\alpha ^2 \beta_\alpha^2
  - \frac{\omega_{\rm p\alpha}^2}{4 k^2 \zeta} \frac{\mu_\alpha^2 \gamma_\alpha}{K_2(\mu_\alpha)} \mathcal{A}_{xx}^\alpha \, , 
  \label{eq:epsxx_alt} \\
  \varepsilon_{yy} &= 1 -  \sum_\alpha\frac{\omega_{\rm p\alpha}^2}{2 k^2 \zeta} \frac{\mu_\alpha^2 \gamma_\alpha}{K_2(\mu_\alpha)}  \mathcal{A}_{yy}^\alpha \,,
  \label{eq:epsyy_alt} \\
  \varepsilon_{xy } &=   \sum_\alpha \frac{\omega_{\rm p\alpha}^2}{k^2 \zeta}\mu_\alpha \beta_\alpha \gamma_\alpha^2
  - \frac{\omega_{\rm p\alpha}^2}{2 k^2} \frac{\mu_\alpha^2 \gamma_\alpha}{K_2(\mu_\alpha)} \mathcal{A}_{xy}^\alpha \,,
  \label{eq:epsxy_alt} 
\end{align}
where the $\mathcal{A}_{lm}^\alpha$ terms are defined as
\begin{align}
  \mathcal{A}_{xx}^\alpha &=	\frac{2}{\mu_\alpha \sqrt{1-\zeta^2}}  \int _{-\infty}^\infty {\rm d}s \,\frac{1}{\chi_\alpha  -s}\nonumber\\ &\times  \left\{ \frac{\gamma_\alpha^2 \beta_\alpha^2}{(s^2+1)^{3/2}}
  + \frac{1}{\mu_\alpha} \left[ \frac{1}{s^2+1}
  + \frac{3\beta_\alpha^2\gamma_\alpha^2}{(s^2+1)^2} \right] \right. +  \left. \frac{1}{\mu_\alpha^2} \left[\frac{1}{(s^2+1)^{3/2}}
  +\frac{3\beta_\alpha^2\gamma_\alpha^2}{(s^2+1)^{5/2}} \right] \right\} e^{-\mu_\alpha \sqrt{s^2+1}} \,,
  \label{eq:int_Axx_alt} \\
  \mathcal{A}_{yy}^\alpha &= \frac{1}{\mu_\alpha \gamma_\alpha^2 \sqrt{1-\zeta^2}}
  \int_{-\infty}^{\infty} {\rm d}s\, \frac{s^2}{\chi_\alpha - s}\nonumber\\ &\times  \left\{ \frac{1}{(s^2+1)^{1/2}} + \frac{\gamma_\alpha^2 \beta_\alpha^2}{(s^2+1)^{3/2}} \right.  \left.+ \frac{1}{\mu_\alpha} \left[ \frac{2}{s^2+1} + \frac{3\gamma_\alpha^2 \beta_\alpha^2}{(s^2+1)^2} \right] \right.+ \left. \frac{1}{\mu_\alpha^2} \left[ \frac{2}{(s^2+1)^{3/2}} + \frac{3\gamma_\alpha^2 \beta_\alpha^2}{(s^2+1)^{5/2}} \right]\right\}  e^{-\mu_\alpha \sqrt{s^2+1}} \,,
  \label{eq:int_Ayy_alt} \\
  \mathcal{A}_{xy}^\alpha &=	\frac{\beta_\alpha}{\mu_\alpha \sqrt{1-\zeta^2}}  \int_{-\infty}^{\infty} {\rm d}s\, \frac{1}{\chi_\alpha  -s} \nonumber\\ &\times \left\{ \frac{1}{(s^2+1)^{1/2}} +\frac{\gamma_\alpha^2 \beta_\alpha^2}{(s^2+1)^{3/2}} \right. \left.+ \frac{3}{\mu_\alpha} \left[ \frac{1}{s^2+1} + \frac{\gamma_\alpha^2 \beta_\alpha^2}{(s^2+1)^2} \right]  \right. \left.+ \frac{3}{\mu_\alpha^2} \left[ \frac{1}{(s^2+1)^{3/2}} + \frac{\gamma_\alpha^2 \beta_\alpha^2}{(s^2+1)^{5/2}}  \right] \right\} e^{-\mu_\alpha \sqrt{s^2+1}} \,,
  \label{eq:int_Axy_alt}
\end{align}
and we have introduced         
\begin{equation}
  \chi_\alpha = \frac{\gamma_\alpha \zeta}{\sqrt{1-\zeta^2}} \,.
\label{eq:chi}
\end{equation}

\section{Low-temperature expression of the kinetic dielectric tensor}
\label{app:low-temp_expansions}

In the following we expand the dielectric tensor elements in the nonrelativistic thermal limit $\mu_\alpha \gg 1$, of particular relevance to the
background plasma particles. Our starting point is Eqs.~\eqref{eq:epsxx_alt}-\eqref{eq:int_Axy_alt}.

Let us first address $\varepsilon_{xx}$, by rewriting \eqref{eq:int_Axx_alt} as
\begin{equation} \label{eq:Axx_alt2}
  \mathcal{A}_{xx}^\alpha = \frac{\beta_\alpha}{\mu_\alpha (1-\zeta^2)}
  \mathcal{B}_{xx}^\alpha(\mu_\alpha,\chi_\alpha)
\end{equation}
where the integral function $\mathcal{B}_{xx}^\alpha(\mu_\alpha,\chi_\alpha)$ is of the form
\begin{equation}
  \mathcal{B}_{xx}^\alpha(\mu_\alpha,\chi_\alpha) = \int_{-\infty}^{\infty} {\rm d}s\,\frac{f_\alpha(s)}{\chi_\alpha -s} e^{-\mu_\alpha \varphi(s)} \,.
\end{equation}
Applying Laplace's method to $\mathcal{B}_{xx}^\alpha (\mu_\alpha,\chi_\alpha)$ \cite{Bender_1978}, one obtains to first order in $1/\mu_\alpha$: 
\begin{align}
  \mathcal{B}_{xx}^\alpha(\mu_\alpha, \chi_\alpha) & \simeq e^{-\mu_\alpha \varphi(0)} \int_{-\infty}^{\infty} {\rm d}s \, \frac{e^{-s^2}}{\sqrt{\frac{\mu_\alpha \chi_\alpha^2 \varphi''(0)}{2}}-s} \nonumber\\ &\times
   \left\{ f_\alpha (0) + \frac{1}{\mu_\alpha} \left[ \frac{f''_\alpha(0)}{\varphi''(0)}s^2 - \left( \frac{\varphi^{(4)}f_\alpha(0)}{6\varphi''(0)^2} + \frac{2\varphi^{(3)}f'_\alpha(0)}{\varphi''(0)^2} \right)s^4
  \right. \right.   \left. \left.+ \frac{\varphi^{(3)}(0)^2f_\alpha(0)}{\varphi''(0)^3}s^6  \right] \right\} \, .
\end{align}
Using $\varphi(0)=1$, $\varphi''(0)=1$, $\varphi^{(3)}(0)=0$ and $\varphi^{(4)}(0)=-3$, as well as the expansions $f_\alpha(0)\simeq \gamma_\alpha^2 \beta_\alpha^2
+ (1+3 \gamma_\alpha^2 \beta_\alpha^2)/\mu_\alpha$ and $f''_\alpha(0) \simeq -3\gamma_\alpha^2 \beta_\alpha^2 $, one gets
\begin{equation} \label{eq:Int_I}
  \mathcal{B}_{xx}^\alpha(\mu_\alpha, \chi_\alpha) \simeq e^{-\mu_\alpha} \int_{-\infty}^{\infty} {\rm d}s \, \frac{e^{-s^2}}{\tilde{\chi}_\alpha - s}  \left\{ \gamma_\alpha^2 \beta_\alpha^2  + \frac{1}{\mu_\alpha} \left[1 + 3\gamma_\alpha^2 \beta_\alpha^2 - 3\gamma_\alpha^2 \beta_\alpha^2 s^2
  + \frac{\gamma_\alpha^2 \beta_\alpha^2}{2} s^4 \right]  \right\} \, ,
\end{equation}
where we have defined 
\begin{equation}
  \tilde{\chi}_\alpha = \sqrt{\frac{\mu_\alpha}{2}} \chi_\alpha = \sqrt{\frac{\mu_\alpha}{2}} \frac{\gamma_\alpha \zeta}{\sqrt{1-\zeta^2}}  \, .
\label{eq:chi_tilde}
\end{equation}
Introducing the well-known plasma dispersion function \cite{NRL_2013}
\begin{equation}
  \mathcal{Z}(\eta) = \frac{1}{\sqrt{\pi}} \int_{-\infty}^{\infty} {\rm d}s\, \frac{e^{-s^2}}{x-\eta} \,,
\end{equation}
and noting that $\mathcal{Z}'(\eta) = -2 \left[1+\eta \mathcal{Z}(\eta)\right]$, Eq.~\eqref{eq:Int_I} can be conveniently expressed as
\begin{equation}
\mathcal{B}_{xx}^\alpha (\mu_\alpha, \chi_\alpha) \simeq - \sqrt{\pi} e^{-\mu_\alpha}
  \left\{ \left[ \gamma_\alpha^2 \beta_\alpha^2 + \frac{1}{\mu_\alpha} (1+3 \gamma_\alpha^2 \beta_\alpha^2) \right] \mathcal{Z}(\tilde{\chi}_\alpha) \right. + \left. \tilde{\chi}_\alpha \frac{\gamma_\alpha^2 \beta_\alpha^2}{\mu_\alpha} \left[ \frac{3}{2} \mathcal{Z}'(\tilde{\chi}_\alpha)
+ \frac{1}{4} \left(1-\tilde{\chi}_\alpha^2 \mathcal{Z}'(\tilde{\chi}_\alpha) \right) \right] \right\}  \, .
\end{equation}  
By combining this equation with Eqs.~\eqref{eq:epsxx_alt} and \eqref{eq:Axx_alt2}, and using the small-argument expansion
$K_2(x) \simeq \sqrt{\frac{\pi}{2x}}e^{-x} (1+\frac{15}{8x})$ \cite{Abramowitz_1972}, one obtains the following low-temperature approximation
of $\varepsilon_{xx}$:
\begin{equation} \label{eq:epsxx_warm}
\varepsilon_{xx} \simeq 1 -  \sum_\alpha\frac{\omega_{\rm p\alpha}^2}{k^2 \zeta^2} \mu_\alpha \gamma_\alpha^2
\left\{ \frac{\beta_\alpha^2}{2} \mathcal{Z}'(\tilde{\chi}_\alpha)  \right. \left. +\frac{1}{8 \mu_\alpha} \left( 9 - \frac{1}{\gamma_\alpha^2} \right)
\left( 1+ \frac{\mathcal{Z}'(\tilde{\chi}_\alpha)}{2} \right) \right. -\left. \frac{\beta_\alpha^2}{\mu_\alpha} \tilde{\chi}_\alpha^2 \left[ \frac{3}{2} \mathcal{Z}'(\tilde{\chi}_\alpha) 
  + \frac{1}{4} \left( 1-\tilde{\chi}_\alpha^2 \mathcal{Z}'(\tilde{\chi}_\alpha ) \right) \right]  \right\}  \, .
\end{equation}
Note that $\tilde{\chi}_\alpha$ is the correct relativistic equivalent of the standard argument ($\eta = \sqrt{\frac{\mu_\alpha}{2}}\zeta$)
of the $\mathcal{Z}(\eta)$ function involved in the nonrelativistic CFI dispersion relation \cite{Davidson_1972}. A similar result
was obtained in Ref.~\cite{Schlickeiser_1997} in the case of electrostatic plasma waves.

Likewise, the integral involved in $\varepsilon_{yy}$ \eqref{eq:epsyy_alt} can be expanded to first order in $1/\mu_\alpha$
\begin{equation}
  \mathcal{A}_{yy}^\alpha \simeq \frac{2 e^{-\mu_\alpha}}{\mu_\alpha^2 \sqrt{1-\zeta^2}} \int_{-\infty}^{\infty} {\rm d}s \, \frac{s^2}{\tilde{\chi}_\alpha-s}
  e^{-s^2} \,,
\end{equation}
so that
\begin{align} \label{eq:epsyy_warm}
&\varepsilon_{yy} \simeq 1 -  \sum_\alpha \frac{\omega_{\rm p\alpha}^2}{2k^2 (1-\zeta^2)} \mu_\alpha \gamma_\alpha^2 \mathcal{Z}'\left( \tilde{\chi}_\alpha \right) \,.
\end{align}

Finally, we expand the integral involved in $\varepsilon_{xy}$ \eqref{eq:epsxy_alt} as
\begin{equation}
 \mathcal{A}_{xy}^\alpha (\mu_\alpha, \chi_\alpha) \simeq \frac{\beta_\alpha e^{-\mu_\alpha}}{\mu_\alpha (1-\zeta^2)}
 \int_{-\infty}^{\infty} {\rm d}s \, \frac{e^{-s^2}}{\tilde{\chi}_\alpha - s}  \left\{ \gamma_\alpha^2 + \frac{1}{\mu_\alpha} \left[ 3\gamma_\alpha^2 + (2-3\gamma_\alpha^2)s^2 + \frac{\gamma_\alpha^2}{2} s^4 \right]  \right\} \,.
\end{equation}
Combining this expression with Eq.~\eqref{eq:epsxy_alt} and identifying the $\mathcal{Z}$ and $\mathcal{Z}'$ functions yields
\begin{equation} \label{eq:epsxy_warm}
  \varepsilon_{xy} \simeq  \sum_\alpha\frac{\omega_{\rm p\alpha}^2}{2k^2 \zeta} \mu_\alpha \beta_\alpha \gamma_\alpha^2
\left\{ \frac{9\tilde{\chi}_\alpha}{4\mu_\alpha} \mathcal{Z}(\tilde{\chi}_\alpha) - \mathcal{Z}'(\tilde{\chi}_\alpha) \right. + \left. \frac{\tilde{\chi}_\alpha^2}{\mu_\alpha} \left[ \left( 3-\frac{2}{\gamma_\alpha^2}\right)\mathcal{Z}'(\tilde{\chi}_\alpha)
  + \frac{1}{2} \left( 1-\tilde{\chi}_\alpha^2 \mathcal{Z}'(\tilde{\chi}_\alpha ) \right) \right] 
  \right\}  \, .
\end{equation}

Expressions \eqref{eq:epsxx_warm}, \eqref{eq:epsyy_warm} and \eqref{eq:epsxy_warm} provide the sought-for low-temperature expansions of the CFI dielectric tensor. The involved $\mathcal{Z}$ and $\mathcal{Z}'$ functions can be readily evaluated in the entire complex 
plane using, \emph{e.g.}, the fast solver developed in Ref.~\cite{Weideman_1994}.

\section{Series expansion of the dielectric tensor}\label{app:series_expansion}

In similar fashion to Ref.~\cite{Schlickeiser_1997}, the dielectric tensor can be expanded in the form of an infinite series, which proves
convenient for deriving approximations in the kinetic regime.

Let us first address $\varepsilon_{xx}$, remarking that when $\Im \chi_\alpha >0$ Eq.~\eqref{eq:int_Axx_alt} can be rewritten as
\begin{align}
  \mathcal{A}_{xx}^\alpha &=
  -\frac{2i}{\mu_\alpha \sqrt{1-\zeta^2}} \int_0^\infty {\rm d}t \, e^{it \chi_\alpha} \int_{-\infty}^\infty {\rm d}s \, e^{-\mu_\alpha \sqrt{s^2+1}-its} \nonumber \\
  &\times \left\{ \frac{\gamma_\alpha^2 \beta_\alpha^2}{(s^2+1)^{3/2}}
  + \frac{1}{\mu_\alpha} \left[ \frac{1}{s^2+1} + \frac{3\beta_\alpha^2 \gamma_\alpha^2}{(s^2+1)^2} \right] \right. \left. +  \frac{1}{\mu_\alpha^2} \left[\frac{1}{(s^2+1)^{3/2}}
  + \frac{3 \beta_\alpha^2 \gamma_\alpha^2}{(s^2+1)^{5/2}} \right] \right\} e^{-\mu_\alpha \sqrt{s^2+1}} \,,
\label{eq:int_Axx_alt2}
\end{align}
where we have exploited the identity
\begin{equation}
  \frac{1}{\chi_\alpha - s} = -i\int_0^{\infty \Im \chi_\alpha} dt\, e^{it(\chi_\alpha-s)} \,,
\end{equation}
Equation~\eqref{eq:int_Axx_alt2} can be put in the form
\begin{equation}
  \mathcal{A}_{xx}^\alpha =
  \frac{4 i}{\mu_\alpha^3 \sqrt{1-\zeta^2}} \int_0^\infty {\rm d}t\, e^{it\chi_\alpha} 
  \left[ \frac{\partial}{\partial b}J(t,\mu_\alpha,1) -2\beta_\alpha^2 \gamma_\alpha^2 \frac{\partial^2}{\partial b^2} J(t,\mu_\alpha,1) \right] \,.
\end{equation}
where we have introduced \cite{Erdelyi_1954}
\begin{equation}
  J(t,\lambda,b) = \int_{-\infty}^{\infty} {\rm d}s\, \frac{e^{-\lambda \sqrt{s^2+b}-its}}{\sqrt{s^2+b}} 
  = 2K_0 \left[\sqrt{b(\lambda^2+t^2)}\right] \,.
  \label{eq:int_J}
\end{equation}
Substituting the above expression into Eq.~\eqref{eq:int_Axx_alt2} yields
\begin{equation}
  \mathcal{A}_{xx}^\alpha =-\frac{4 i}{\mu_\alpha^3 \sqrt{1-\zeta^2}} \int_0^\infty {\rm d}t\, e^{it\chi_\alpha}  \left[ 
  \sqrt{\mu_\alpha^2 + t^2} K_1 \left(\sqrt{t^2+\mu_\alpha^2}\right) \right. \left.+ \beta_\alpha^2 \gamma_\alpha^2 \left( t^2 + \mu_\alpha^2 \right) K_2 \left(\sqrt{t^2+\mu_\alpha^2} \right) \right] \,.
\end{equation}
Expanding $e^{it\chi_\alpha}=\sum_{n=0}^{\infty} (it \chi_\alpha)^n/n!$, and making use of the identities \cite{Gradshteyn_1980}
\begin{align}
  &\int_0^\infty{\rm d}t \, t^n K_0\left(x \sqrt{t^2+\mu^2} \right) = \frac{2^{\frac{n-1}{2}}
  \Gamma\left(\frac{n+1}{2}\right)}{x^{\frac{n+1}{2}}} K_{\frac{n+1}{2}}(x \mu) \,, \\
  & \frac{d^2}{dx^2} K_0(x z) = \frac{z^2}{2} \left[ K_0(x z) + K_2(x z) \right] \,,
\label{eq:int_id1}
\end{align}
we obtain after some algebra, 
\begin{align}
  \mathcal{A}_{xx}^\alpha &= -\frac{2\sqrt{\pi}\gamma_\alpha \zeta}{(1-\zeta^2)}  \Biggl\{ i\sum_{m=0}^\infty \frac{(-1)^m}{\Gamma(m+1)} 
  \left(\frac{\mu_\alpha \chi_\alpha^2}{2} \right)^{m-\frac{1}{2}}   \left[ \frac{1}{\mu_\alpha} K_{m+\frac{3}{2}}(\mu_\alpha) 
  + \beta_\alpha^2 \gamma_\alpha^2 K_{m+\frac{5}{2}}(\mu_\alpha) \right]  \nonumber \\
  & - \sum_{m=0}^\infty \frac{(-1)^m}{\Gamma(m+\frac{3}{2})} 
  \left( \frac{\mu_\alpha \chi_\alpha^2}{2} \right)^m  \left[ \frac{1}{\mu_\alpha} K_{m+2} (\mu_\alpha) 
  + \beta_\alpha^2 \gamma_\alpha^2 K_{m+3}( \mu_\alpha ) \right] \Biggr\} \,.
\label{eq:int_Axx_series}  
\end{align}
The above series can be further simplified using the multiplication theorem of Bessel functions
\cite{Abramowitz_1972}, giving finally
\begin{align}
  \varepsilon_{xx} &= 1 +  \sum_\alpha\frac{\omega_{\rm p\alpha}^2}{\omega^2} \mu_\alpha \beta_\alpha^2 \gamma_\alpha^2 	
  + \frac{\sqrt{\pi}\omega_{\rm p\alpha}^2}{2 k^2(1-\zeta^2)} \frac{\mu_\alpha^2  \gamma_\alpha^2}{K_2(\mu_\alpha)}  \biggl\{ i \sqrt{\frac{2}{\mu_\alpha \chi_\alpha^2}}
  \left[ \frac{K_{3/2} \left( \mu_\alpha \sqrt{1+\chi_\alpha^2} \right)}{\mu_\alpha (1 + \chi_\alpha^2)^{3/4}} 
+\beta_\alpha^2 \gamma_\alpha^2 \frac{K_{5/2} \left( \mu_\alpha \sqrt{1+\chi_\alpha^2} \right)}{(1+\chi_\alpha^2)^{5/4}} \right] 
    \nonumber \\
  & - \sum_{m=0}^\infty \frac{(-1)^m}{\Gamma(m+\frac{3}{2})}  \left( \frac{\mu_\alpha \chi_\alpha^2}{2} \right)^m 
\left[ \frac{1}{\mu_\alpha} K_{m+2}(\mu_\alpha) 
  + \beta_\alpha^2 \gamma_\alpha^2 K_{m+3}(\mu_\alpha) \right] \biggr\} \,.
\label{eq:epsxx_series}
\end{align}

Likewise, the integral involved in $\varepsilon_{yy}$ \eqref{eq:epsyy_alt} can be recast as
\begin{equation}
 \mathcal{A}_{yy}^\alpha = -\frac{i}{\mu_\alpha \gamma_\alpha^2 \sqrt{1-\zeta^2}}
 \int_0^\infty {\rm d}t\, e^{it \chi_\alpha}  \left[ I(t,\mu_\alpha,1)  - \frac{4}{\mu_\alpha^2} \frac{\partial}{\partial b} I(t,\mu_\alpha,1)
 + \frac{4 \beta_\alpha^2 \gamma_\alpha^2}{\mu_\alpha^2} \frac{\partial^2}{\partial b^2} I(t,\mu_\alpha,1) \right]\,.
 \label{eq:int_Ayy_alt2}
\end{equation}
Here, we have defined 
\begin{equation}
  I(t,\lambda,b)= \int_{-\infty}^{\infty} {\rm d}s\, 
  \frac{s^2 e^{-\lambda \sqrt{s^2+b}-its}}{\sqrt{s^2+b}} \,,
\label{eq:int_I}
\end{equation}	
which can be solved in closed form as \cite{Erdelyi_1954}
\begin{equation}
  I(t,\lambda,b)=-\frac{2bt^2}{t^2+\lambda^2}K_0 \left[\sqrt{b(\lambda^2+t^2)}\right] + \frac{2b^{1/2}(\lambda^2-t^2)}{(t^2+\lambda^2)^{3/2}} K_1\left[\sqrt{b(\lambda^2+t^2)}\right] \,.
\label{eq:func_I}
\end{equation}
As in the previous derivation, we substitute Eqs.~\eqref{eq:func_I} into Eq.~\eqref{eq:int_Ayy_alt2}, expand the exponential
factor and exploit \cite{Gradshteyn_1980}
\begin{equation}
  \int_0^{\infty} {\rm d}t \, \frac{t^{2\lambda+1}K_\nu \big[\sqrt{t^2+z^2}\big]}{(t^2+z^2)^{\nu/2}}
  = \frac{2^\lambda \Gamma(\lambda+1)}{z^{\nu-\lambda-1}} K_{\nu-\lambda-1} (z) \,.
\label{eq:int_id2}
\end{equation}
There follows
\begin{align}
  \mathcal{A}_{yy}^\alpha &= -\frac{2\sqrt{\pi}\gamma_\alpha \zeta}{\mu_\alpha(1-\zeta^2)}  \Biggl\{ i \sum_{m=0}^\infty \frac{(-1)^m}{\Gamma(m+1)} 
  \left(\frac{\mu_\alpha \chi_\alpha^2}{2} \right)^{m+\frac{1}{2}}   \left[ K_{m+\frac{1}{2}}(\mu_\alpha)
  + \frac{2}{\mu_\alpha} \left(1+\beta_\alpha^2 \left(m+\frac{1}{2}\right) \right) K_{m+\frac{3}{2}}(\mu_\alpha) \right]  \nonumber \\
  & + \sum_{m=0}^\infty \frac{(-1)^m}{\Gamma(m+\frac{1}{2})} \left(\frac{\mu_\alpha \chi_\alpha^2}{2} \right)^m  \left[ K_m(\mu_\alpha) + \frac{2}{\mu_\alpha} (1+\beta_\alpha^2 m) K_{m+1}(\mu_\alpha) \right] \Biggr\} \,.
\label{eq:int_Ayy_series}  
\end{align}
Applying again the multiplication theorem of Bessel functions, we find the following
expression for $\varepsilon_{yy}$
\begin{align}
  \varepsilon_{yy} &= 1 + \sum_\alpha\frac{\sqrt{\pi}\omega_{\rm p\alpha}^2}{k^2(1-\zeta^2)}  \frac{\mu_\alpha \gamma_\alpha^2}{K_2(\mu_\alpha)}  
\Biggl\{ i \sqrt{\frac{\mu_\alpha \chi_\alpha^2}{2}} 
  \Biggl[ \frac{K_{1/2} \left(\mu_\alpha \sqrt{1+\chi_\alpha^2} \right)}{(1+\chi_\alpha^2)^{1/4}}  +\frac{2}{\mu_\alpha} \left( 1+\frac{\beta_\alpha^2}{2} \right)
  \frac{K_{3/2}\left( \mu_\alpha \sqrt{1+\chi_\alpha^2}\right)}{(1+\chi_\alpha^2)^{3/4}}  \nonumber \\
  & - \beta_\alpha^2 \chi_\alpha^2 \frac{K_{5/2}\left(\mu_\alpha \sqrt{1+\chi_\alpha^2} \right)}{(1+\chi_\alpha^2)^{5/4}} \Biggr]  + \sum_{m=0}^\infty \frac{(-1)^m}{\Gamma(m+\frac{1}{2})} 
  \left( \frac{\mu_\alpha \chi_\alpha^2}{2} \right)^m  \left[ K_m(\mu_\alpha) + \frac{2}{\mu_\alpha} (1+\beta_\alpha^2 m) K_{m+1}(\mu_\alpha) \right] 
\Biggr\} \,.
\label{eq:epsyy_series}
\end{align}

Reiterating the previous procedure, we first express the integral term involved in $\varepsilon_{xy}$ \eqref{eq:epsxy_alt} as
\begin{equation}
  \mathcal{A}_{xy}^\alpha = -\frac{i \beta_\alpha}{\mu_\alpha^3 \sqrt{1-\zeta^2}}  \int_0^\infty {\rm d}t\, e^{it\chi_\alpha} 
  \left[\mu_\alpha^2 J(t,\mu_\alpha,1) - 6\frac{\partial}{\partial b}J(t,\mu_\alpha,1) \right. \left. + 4\beta_\alpha^2 \gamma_\alpha^2  \frac{\partial^2}{\partial b^2} J(t,\mu_\alpha,1) \right] \,.
\label{eq:epsxy_alt2}	
\end{equation}
Evaluating the derivatives of the function $J$ defined by Eq.~\eqref{eq:int_J}, we obtain
\begin{align}
  \mathcal{A}_{xy}^\alpha &= \frac{\sqrt{\pi}\gamma_\alpha \beta_\alpha \zeta}{(1-\zeta^2)}  \left\{ i\sum_{m=0}^\infty \frac{(-1)^m}{\Gamma(m+1)} \left( \frac{\mu_\alpha \chi_\alpha^2}{2} \right)^{m-\frac{1}{2}} \right. \left.  \left[ \frac{2m}{\mu_\alpha} K_{m+\frac{3}{2}}(\mu_\alpha) - \gamma_\alpha^2 K_{m+\frac{5}{2}}(\mu_\alpha) \right] \right. \nonumber \\
  &\left. - \sum_{m=0}^\infty \frac{(-1)^m}{\Gamma(m+\frac{3}{2})}  \left( \frac{\mu_\alpha \chi_\alpha^2}{2} \right)^m \right. \left. \left[ \frac{(2m+1)}{\mu_\alpha} K_{m+2}(\mu_\alpha) - \gamma_\alpha^2 K_{m+3}(\mu_\alpha) \right] \right\} \,.
\label{eq:int_Axy_series}  
\end{align}
Finally, Bessel-function identities allow us to obtain the following alternative expression for $\varepsilon_{xy}$ 
\begin{align}
  &\varepsilon_{xy} = \sum_\alpha\frac{\omega_{\rm p\alpha}^2}{k^2 \zeta}   \mu_\alpha \gamma_\alpha^2 \beta_\alpha
  + \frac{\sqrt{\pi}\omega_{\rm p\alpha}^2\zeta}{k^2(1-\zeta^2)} \frac{\mu_\alpha \gamma_\alpha^2 \beta_\alpha}{K_2(\mu_\alpha)}  \Biggl\{ 
  2i \sqrt{\frac{\mu_\alpha \chi_\alpha^2}{2}}
  \left( 1+\frac{\gamma_\alpha^2}{\chi_\alpha^2}\right)
  \frac{K_{5/2} \left(\mu_\alpha \sqrt{1+\chi_\alpha^2}\right)}{(1+\chi_\alpha^2)^{5/4}}  \nonumber \\
  & + \sum_{m=0}^\infty \frac{(-1)^m}{\Gamma(m+\frac{3}{2})} 
  \left( \frac{\mu_\alpha \chi_\alpha^2}{2} \right)^m  \left[ (2m+1) K_{m+2}(\mu_\alpha)  - \mu_\alpha \gamma_\alpha^2 K_{m+3}(\mu_\alpha) \right] \Biggr\} \,.
\label{eq:epsxy_series}	
\end{align}

\end{widetext}


\bibliographystyle{apsrev4-1}

\bibliography{biblio}

\begin{thebibliography}{78}%
\makeatletter
\providecommand \@ifxundefined [1]{%
 \@ifx{#1\undefined}
}%
\providecommand \@ifnum [1]{%
 \ifnum #1\expandafter \@firstoftwo
 \else \expandafter \@secondoftwo
 \fi
}%
\providecommand \@ifx [1]{%
 \ifx #1\expandafter \@firstoftwo
 \else \expandafter \@secondoftwo
 \fi
}%
\providecommand \natexlab [1]{#1}%
\providecommand \enquote  [1]{``#1''}%
\providecommand \bibnamefont  [1]{#1}%
\providecommand \bibfnamefont [1]{#1}%
\providecommand \citenamefont [1]{#1}%
\providecommand \href@noop [0]{\@secondoftwo}%
\providecommand \href [0]{\begingroup \@sanitize@url \@href}%
\providecommand \@href[1]{\@@startlink{#1}\@@href}%
\providecommand \@@href[1]{\endgroup#1\@@endlink}%
\providecommand \@sanitize@url [0]{\catcode `\\12\catcode `\$12\catcode
  `\&12\catcode `\#12\catcode `\^12\catcode `\_12\catcode `\%12\relax}%
\providecommand \@@startlink[1]{}%
\providecommand \@@endlink[0]{}%
\providecommand \url  [0]{\begingroup\@sanitize@url \@url }%
\providecommand \@url [1]{\endgroup\@href {#1}{\urlprefix }}%
\providecommand \urlprefix  [0]{URL }%
\providecommand \Eprint [0]{\href }%
\providecommand \doibase [0]{http://dx.doi.org/}%
\providecommand \selectlanguage [0]{\@gobble}%
\providecommand \bibinfo  [0]{\@secondoftwo}%
\providecommand \bibfield  [0]{\@secondoftwo}%
\providecommand \translation [1]{[#1]}%
\providecommand \BibitemOpen [0]{}%
\providecommand \bibitemStop [0]{}%
\providecommand \bibitemNoStop [0]{.\EOS\space}%
\providecommand \EOS [0]{\spacefactor3000\relax}%
\providecommand \BibitemShut  [1]{\csname bibitem#1\endcsname}%
\let\auto@bib@innerbib\@empty
\bibitem [{\citenamefont {{Weibel}}(1959)}]{Weibel_1959}%
  \BibitemOpen
  \bibfield  {author} {\bibinfo {author} {\bibfnamefont {E.~S.}\ \bibnamefont
  {{Weibel}}},\ }\href {\doibase 10.1103/PhysRevLett.2.83} {\bibfield
  {journal} {\bibinfo  {journal} {Phys. Rev. Lett.}\ }\textbf {\bibinfo
  {volume} {2}},\ \bibinfo {pages} {83} (\bibinfo {year} {1959})}\BibitemShut
  {NoStop}%
\bibitem [{\citenamefont {{Fried}}(1959)}]{Fried_1959}%
  \BibitemOpen
  \bibfield  {author} {\bibinfo {author} {\bibfnamefont {B.~D.}\ \bibnamefont
  {{Fried}}},\ }\href {\doibase 10.1063/1.1705933} {\bibfield  {journal}
  {\bibinfo  {journal} {Phys. Fluids}\ }\textbf {\bibinfo {volume} {2}},\
  \bibinfo {pages} {337} (\bibinfo {year} {1959})}\BibitemShut {NoStop}%
\bibitem [{\citenamefont {Davidson}\ \emph {et~al.}(1972)\citenamefont
  {Davidson}, \citenamefont {Hammer}, \citenamefont {Haber},\ and\
  \citenamefont {Wagner}}]{Davidson_1972}%
  \BibitemOpen
  \bibfield  {author} {\bibinfo {author} {\bibfnamefont {R.~C.}\ \bibnamefont
  {Davidson}}, \bibinfo {author} {\bibfnamefont {D.~A.}\ \bibnamefont
  {Hammer}}, \bibinfo {author} {\bibfnamefont {I.}~\bibnamefont {Haber}}, \
  and\ \bibinfo {author} {\bibfnamefont {C.~E.}\ \bibnamefont {Wagner}},\
  }\href {\doibase 10.1063/1.1693910} {\bibfield  {journal} {\bibinfo
  {journal} {Phys. Fluids}\ }\textbf {\bibinfo {volume} {15}},\ \bibinfo
  {pages} {317} (\bibinfo {year} {1972})}\BibitemShut {NoStop}%
\bibitem [{\citenamefont {Califano}\ \emph {et~al.}(1997)\citenamefont
  {Califano}, \citenamefont {Pegoraro},\ and\ \citenamefont
  {Bulanov}}]{Califano_1997}%
  \BibitemOpen
  \bibfield  {author} {\bibinfo {author} {\bibfnamefont {F.}~\bibnamefont
  {Califano}}, \bibinfo {author} {\bibfnamefont {F.}~\bibnamefont {Pegoraro}},
  \ and\ \bibinfo {author} {\bibfnamefont {S.~V.}\ \bibnamefont {Bulanov}},\
  }\href {\doibase 10.1103/PhysRevE.56.963} {\bibfield  {journal} {\bibinfo
  {journal} {Phys. Rev. E}\ }\textbf {\bibinfo {volume} {56}},\ \bibinfo
  {pages} {963} (\bibinfo {year} {1997})}\BibitemShut {NoStop}%
\bibitem [{\citenamefont {{Achterberg}}\ and\ \citenamefont
  {{Wiersma}}(2007)}]{Achterberg_2007_I}%
  \BibitemOpen
  \bibfield  {author} {\bibinfo {author} {\bibfnamefont {A.}~\bibnamefont
  {{Achterberg}}}\ and\ \bibinfo {author} {\bibfnamefont {J.}~\bibnamefont
  {{Wiersma}}},\ }\href {\doibase 10.1051/0004-6361:20065365} {\bibfield
  {journal} {\bibinfo  {journal} {Astron. Astrophys.}\ }\textbf {\bibinfo
  {volume} {475}},\ \bibinfo {pages} {1} (\bibinfo {year} {2007})}\BibitemShut
  {NoStop}%
\bibitem [{\citenamefont {{Moiseev}}\ and\ \citenamefont
  {{Sagdeev}}(1963)}]{Moiseev_1963}%
  \BibitemOpen
  \bibfield  {author} {\bibinfo {author} {\bibfnamefont {S.~S.}\ \bibnamefont
  {{Moiseev}}}\ and\ \bibinfo {author} {\bibfnamefont {R.~Z.}\ \bibnamefont
  {{Sagdeev}}},\ }\href@noop {} {\bibfield  {journal} {\bibinfo  {journal}
  {Journal of Nuclear Energy. Part C, Plasma Physics, Accelerators,
  Thermonuclear Research}\ }\textbf {\bibinfo {volume} {5}},\ \bibinfo {pages}
  {43} (\bibinfo {year} {1963})}\BibitemShut {NoStop}%
\bibitem [{\citenamefont {{Medvedev}}\ and\ \citenamefont
  {{Loeb}}(1999)}]{Medvedev_1999}%
  \BibitemOpen
  \bibfield  {author} {\bibinfo {author} {\bibfnamefont {M.~V.}\ \bibnamefont
  {{Medvedev}}}\ and\ \bibinfo {author} {\bibfnamefont {A.}~\bibnamefont
  {{Loeb}}},\ }\href {\doibase 10.1086/308038} {\bibfield  {journal} {\bibinfo
  {journal} {Astrophys. J.}\ }\textbf {\bibinfo {volume} {526}},\ \bibinfo
  {pages} {697} (\bibinfo {year} {1999})}\BibitemShut {NoStop}%
\bibitem [{\citenamefont {{Gruzinov}}\ and\ \citenamefont
  {{Waxman}}(1999)}]{1999ApJ...511..852G}%
  \BibitemOpen
  \bibfield  {author} {\bibinfo {author} {\bibfnamefont {A.}~\bibnamefont
  {{Gruzinov}}}\ and\ \bibinfo {author} {\bibfnamefont {E.}~\bibnamefont
  {{Waxman}}},\ }\href {\doibase 10.1086/306720} {\bibfield  {journal}
  {\bibinfo  {journal} {Astrophys. J.}\ }\textbf {\bibinfo {volume} {511}},\
  \bibinfo {pages} {852} (\bibinfo {year} {1999})},\ \Eprint
  {http://arxiv.org/abs/astro-ph/9807111} {astro-ph/9807111} \BibitemShut
  {NoStop}%
\bibitem [{\citenamefont {Sentoku}\ \emph {et~al.}(2003)\citenamefont
  {Sentoku}, \citenamefont {Mima}, \citenamefont {Kaw},\ and\ \citenamefont
  {Nishikawa}}]{Sentoku_2003}%
  \BibitemOpen
  \bibfield  {author} {\bibinfo {author} {\bibfnamefont {Y.}~\bibnamefont
  {Sentoku}}, \bibinfo {author} {\bibfnamefont {K.}~\bibnamefont {Mima}},
  \bibinfo {author} {\bibfnamefont {P.}~\bibnamefont {Kaw}}, \ and\ \bibinfo
  {author} {\bibfnamefont {K.}~\bibnamefont {Nishikawa}},\ }\href {\doibase
  10.1103/PhysRevLett.90.155001} {\bibfield  {journal} {\bibinfo  {journal}
  {Phys. Rev. Lett.}\ }\textbf {\bibinfo {volume} {90}},\ \bibinfo {pages}
  {155001} (\bibinfo {year} {2003})}\BibitemShut {NoStop}%
\bibitem [{\citenamefont {Adam}\ \emph {et~al.}(2006)\citenamefont {Adam},
  \citenamefont {H\'eron},\ and\ \citenamefont {Laval}}]{Adam_2006}%
  \BibitemOpen
  \bibfield  {author} {\bibinfo {author} {\bibfnamefont {J.~C.}\ \bibnamefont
  {Adam}}, \bibinfo {author} {\bibfnamefont {A.}~\bibnamefont {H\'eron}}, \
  and\ \bibinfo {author} {\bibfnamefont {G.}~\bibnamefont {Laval}},\ }\href
  {\doibase 10.1103/PhysRevLett.97.205006} {\bibfield  {journal} {\bibinfo
  {journal} {Phys. Rev. Lett.}\ }\textbf {\bibinfo {volume} {97}},\ \bibinfo
  {pages} {205006} (\bibinfo {year} {2006})}\BibitemShut {NoStop}%
\bibitem [{\citenamefont {Allen}\ \emph {et~al.}(2012)\citenamefont {Allen},
  \citenamefont {Yakimenko}, \citenamefont {Babzien}, \citenamefont {Fedurin},
  \citenamefont {Kusche},\ and\ \citenamefont {Muggli}}]{Allen_2012}%
  \BibitemOpen
  \bibfield  {author} {\bibinfo {author} {\bibfnamefont {B.}~\bibnamefont
  {Allen}}, \bibinfo {author} {\bibfnamefont {V.}~\bibnamefont {Yakimenko}},
  \bibinfo {author} {\bibfnamefont {M.}~\bibnamefont {Babzien}}, \bibinfo
  {author} {\bibfnamefont {M.}~\bibnamefont {Fedurin}}, \bibinfo {author}
  {\bibfnamefont {K.}~\bibnamefont {Kusche}}, \ and\ \bibinfo {author}
  {\bibfnamefont {P.}~\bibnamefont {Muggli}},\ }\href {\doibase
  10.1103/PhysRevLett.109.185007} {\bibfield  {journal} {\bibinfo  {journal}
  {Phys. Rev. Lett.}\ }\textbf {\bibinfo {volume} {109}},\ \bibinfo {pages}
  {185007} (\bibinfo {year} {2012})}\BibitemShut {NoStop}%
\bibitem [{\citenamefont {Debayle}\ \emph {et~al.}(2010)\citenamefont
  {Debayle}, \citenamefont {Honrubia}, \citenamefont {d'Humi\`eres},\ and\
  \citenamefont {Tikhonchuk}}]{Debayle_2010}%
  \BibitemOpen
  \bibfield  {author} {\bibinfo {author} {\bibfnamefont {A.}~\bibnamefont
  {Debayle}}, \bibinfo {author} {\bibfnamefont {J.~J.}\ \bibnamefont
  {Honrubia}}, \bibinfo {author} {\bibfnamefont {E.}~\bibnamefont
  {d'Humi\`eres}}, \ and\ \bibinfo {author} {\bibfnamefont {V.~T.}\
  \bibnamefont {Tikhonchuk}},\ }\href {\doibase 10.1103/PhysRevE.82.036405}
  {\bibfield  {journal} {\bibinfo  {journal} {Phys. Rev. E}\ }\textbf {\bibinfo
  {volume} {82}},\ \bibinfo {pages} {036405} (\bibinfo {year}
  {2010})}\BibitemShut {NoStop}%
\bibitem [{\citenamefont {{Masson-Laborde}}\ \emph {et~al.}(2010)\citenamefont
  {{Masson-Laborde}}, \citenamefont {Rozmus}, \citenamefont {Peng},
  \citenamefont {Pesme}, \citenamefont {{H\"uller}}, \citenamefont {Casanova},
  \citenamefont {Bychenkov}, \citenamefont {Chapman},\ and\ \citenamefont
  {Loiseau}}]{Masson-Laborde_2010}%
  \BibitemOpen
  \bibfield  {author} {\bibinfo {author} {\bibfnamefont {P.~E.}\ \bibnamefont
  {{Masson-Laborde}}}, \bibinfo {author} {\bibfnamefont {W.}~\bibnamefont
  {Rozmus}}, \bibinfo {author} {\bibfnamefont {Z.}~\bibnamefont {Peng}},
  \bibinfo {author} {\bibfnamefont {D.}~\bibnamefont {Pesme}}, \bibinfo
  {author} {\bibfnamefont {S.}~\bibnamefont {{H\"uller}}}, \bibinfo {author}
  {\bibfnamefont {M.}~\bibnamefont {Casanova}}, \bibinfo {author}
  {\bibfnamefont {V.~Y.}\ \bibnamefont {Bychenkov}}, \bibinfo {author}
  {\bibfnamefont {T.}~\bibnamefont {Chapman}}, \ and\ \bibinfo {author}
  {\bibfnamefont {P.}~\bibnamefont {Loiseau}},\ }\href {\doibase
  10.1063/1.3474619} {\bibfield  {journal} {\bibinfo  {journal} {Phys.
  Plasmas}\ }\textbf {\bibinfo {volume} {17}},\ \bibinfo {pages} {092704}
  (\bibinfo {year} {2010})}\BibitemShut {NoStop}%
\bibitem [{\citenamefont {{Mondal}}\ \emph {et~al.}(2012)\citenamefont
  {{Mondal}}, \citenamefont {{Narayanan}}, \citenamefont {{Ding}},
  \citenamefont {{Lad}}, \citenamefont {{Hao}}, \citenamefont {{Ahmad}},
  \citenamefont {{Wang}}, \citenamefont {{Sheng}}, \citenamefont {{Sengupta}},
  \citenamefont {{Kaw}}, \citenamefont {{Das}},\ and\ \citenamefont
  {{Kumar}}}]{Mondal_2012}%
  \BibitemOpen
  \bibfield  {author} {\bibinfo {author} {\bibfnamefont {S.}~\bibnamefont
  {{Mondal}}}, \bibinfo {author} {\bibfnamefont {V.}~\bibnamefont
  {{Narayanan}}}, \bibinfo {author} {\bibfnamefont {W.~J.}\ \bibnamefont
  {{Ding}}}, \bibinfo {author} {\bibfnamefont {A.~D.}\ \bibnamefont {{Lad}}},
  \bibinfo {author} {\bibfnamefont {B.}~\bibnamefont {{Hao}}}, \bibinfo
  {author} {\bibfnamefont {S.}~\bibnamefont {{Ahmad}}}, \bibinfo {author}
  {\bibfnamefont {W.~M.}\ \bibnamefont {{Wang}}}, \bibinfo {author}
  {\bibfnamefont {Z.~M.}\ \bibnamefont {{Sheng}}}, \bibinfo {author}
  {\bibfnamefont {S.}~\bibnamefont {{Sengupta}}}, \bibinfo {author}
  {\bibfnamefont {P.}~\bibnamefont {{Kaw}}}, \bibinfo {author} {\bibfnamefont
  {A.}~\bibnamefont {{Das}}}, \ and\ \bibinfo {author} {\bibfnamefont {G.~R.}\
  \bibnamefont {{Kumar}}},\ }\href {\doibase 10.1073/pnas.1200753109}
  {\bibfield  {journal} {\bibinfo  {journal} {Proc. Natl. Acad. Sci. USA}\
  }\textbf {\bibinfo {volume} {109}},\ \bibinfo {pages} {8011} (\bibinfo {year}
  {2012})}\BibitemShut {NoStop}%
\bibitem [{\citenamefont {Quinn}\ \emph {et~al.}(2012)\citenamefont {Quinn},
  \citenamefont {Romagnani}, \citenamefont {Ramakrishna}, \citenamefont
  {Sarri}, \citenamefont {Dieckmann}, \citenamefont {Wilson}, \citenamefont
  {Fuchs}, \citenamefont {Lancia}, \citenamefont {Pipahl}, \citenamefont
  {Toncian}, \citenamefont {Willi}, \citenamefont {Clarke}, \citenamefont
  {Notley}, \citenamefont {Macchi},\ and\ \citenamefont
  {Borghesi}}]{Quinn_2012}%
  \BibitemOpen
  \bibfield  {author} {\bibinfo {author} {\bibfnamefont {K.}~\bibnamefont
  {Quinn}}, \bibinfo {author} {\bibfnamefont {L.}~\bibnamefont {Romagnani}},
  \bibinfo {author} {\bibfnamefont {B.}~\bibnamefont {Ramakrishna}}, \bibinfo
  {author} {\bibfnamefont {G.}~\bibnamefont {Sarri}}, \bibinfo {author}
  {\bibfnamefont {M.~E.}\ \bibnamefont {Dieckmann}}, \bibinfo {author}
  {\bibfnamefont {P.~A.}\ \bibnamefont {Wilson}}, \bibinfo {author}
  {\bibfnamefont {J.}~\bibnamefont {Fuchs}}, \bibinfo {author} {\bibfnamefont
  {L.}~\bibnamefont {Lancia}}, \bibinfo {author} {\bibfnamefont
  {A.}~\bibnamefont {Pipahl}}, \bibinfo {author} {\bibfnamefont
  {T.}~\bibnamefont {Toncian}}, \bibinfo {author} {\bibfnamefont
  {O.}~\bibnamefont {Willi}}, \bibinfo {author} {\bibfnamefont {R.~J.}\
  \bibnamefont {Clarke}}, \bibinfo {author} {\bibfnamefont {M.}~\bibnamefont
  {Notley}}, \bibinfo {author} {\bibfnamefont {A.}~\bibnamefont {Macchi}}, \
  and\ \bibinfo {author} {\bibfnamefont {M.}~\bibnamefont {Borghesi}},\ }\href
  {\doibase 10.1103/PhysRevLett.108.135001} {\bibfield  {journal} {\bibinfo
  {journal} {Phys. Rev. Lett.}\ }\textbf {\bibinfo {volume} {108}},\ \bibinfo
  {pages} {135001} (\bibinfo {year} {2012})}\BibitemShut {NoStop}%
\bibitem [{\citenamefont {Fiuza}\ \emph {et~al.}(2012)\citenamefont {Fiuza},
  \citenamefont {Fonseca}, \citenamefont {Tonge}, \citenamefont {Mori},\ and\
  \citenamefont {Silva}}]{Fiuza_2012}%
  \BibitemOpen
  \bibfield  {author} {\bibinfo {author} {\bibfnamefont {F.}~\bibnamefont
  {Fiuza}}, \bibinfo {author} {\bibfnamefont {R.~A.}\ \bibnamefont {Fonseca}},
  \bibinfo {author} {\bibfnamefont {J.}~\bibnamefont {Tonge}}, \bibinfo
  {author} {\bibfnamefont {W.~B.}\ \bibnamefont {Mori}}, \ and\ \bibinfo
  {author} {\bibfnamefont {L.~O.}\ \bibnamefont {Silva}},\ }\href {\doibase
  10.1103/PhysRevLett.108.235004} {\bibfield  {journal} {\bibinfo  {journal}
  {Phys. Rev. Lett.}\ }\textbf {\bibinfo {volume} {108}},\ \bibinfo {pages}
  {235004} (\bibinfo {year} {2012})}\BibitemShut {NoStop}%
\bibitem [{\citenamefont {Ruyer}\ \emph
  {et~al.}(2015{\natexlab{a}})\citenamefont {Ruyer}, \citenamefont
  {Gremillet},\ and\ \citenamefont {Bonnaud}}]{Ruyer_2015b}%
  \BibitemOpen
  \bibfield  {author} {\bibinfo {author} {\bibfnamefont {C.}~\bibnamefont
  {Ruyer}}, \bibinfo {author} {\bibfnamefont {L.}~\bibnamefont {Gremillet}}, \
  and\ \bibinfo {author} {\bibfnamefont {G.}~\bibnamefont {Bonnaud}},\ }\href
  {\doibase 10.1063/1.4928096} {\bibfield  {journal} {\bibinfo  {journal}
  {Phys. Plasmas}\ }\textbf {\bibinfo {volume} {22}},\ \bibinfo {pages}
  {082107} (\bibinfo {year} {2015}{\natexlab{a}})}\BibitemShut {NoStop}%
\bibitem [{\citenamefont {Fox}\ \emph {et~al.}(2013)\citenamefont {Fox},
  \citenamefont {Fiksel}, \citenamefont {Bhattacharjee}, \citenamefont {Chang},
  \citenamefont {Germaschewski}, \citenamefont {Hu},\ and\ \citenamefont
  {Nilson}}]{Fox_2013}%
  \BibitemOpen
  \bibfield  {author} {\bibinfo {author} {\bibfnamefont {W.}~\bibnamefont
  {Fox}}, \bibinfo {author} {\bibfnamefont {G.}~\bibnamefont {Fiksel}},
  \bibinfo {author} {\bibfnamefont {A.}~\bibnamefont {Bhattacharjee}}, \bibinfo
  {author} {\bibfnamefont {P.-Y.}\ \bibnamefont {Chang}}, \bibinfo {author}
  {\bibfnamefont {K.}~\bibnamefont {Germaschewski}}, \bibinfo {author}
  {\bibfnamefont {S.~X.}\ \bibnamefont {Hu}}, \ and\ \bibinfo {author}
  {\bibfnamefont {P.~M.}\ \bibnamefont {Nilson}},\ }\href {\doibase
  10.1103/PhysRevLett.111.225002} {\bibfield  {journal} {\bibinfo  {journal}
  {Phys. Rev. Lett.}\ }\textbf {\bibinfo {volume} {111}},\ \bibinfo {pages}
  {225002} (\bibinfo {year} {2013})}\BibitemShut {NoStop}%
\bibitem [{\citenamefont {{Huntington}}\ \emph {et~al.}(2015)\citenamefont
  {{Huntington}}, \citenamefont {{Fiuza}}, \citenamefont {{Ross}},
  \citenamefont {{Zylstra}}, \citenamefont {{Drake}}, \citenamefont {{Froula}},
  \citenamefont {{Gregori}}, \citenamefont {{Kugland}}, \citenamefont
  {{Kuranz}}, \citenamefont {{Levy}}, \citenamefont {{Li}}, \citenamefont
  {{Meinecke}}, \citenamefont {{Morita}}, \citenamefont {{Petrasso}},
  \citenamefont {{Plechaty}}, \citenamefont {{Remington}}, \citenamefont
  {{Ryutov}}, \citenamefont {{Sakawa}}, \citenamefont {{Spitkovsky}},
  \citenamefont {{Takabe}},\ and\ \citenamefont {{Park}}}]{Huntington_2015}%
  \BibitemOpen
  \bibfield  {author} {\bibinfo {author} {\bibfnamefont {C.~M.}\ \bibnamefont
  {{Huntington}}}, \bibinfo {author} {\bibfnamefont {F.}~\bibnamefont
  {{Fiuza}}}, \bibinfo {author} {\bibfnamefont {J.~S.}\ \bibnamefont {{Ross}}},
  \bibinfo {author} {\bibfnamefont {A.~B.}\ \bibnamefont {{Zylstra}}}, \bibinfo
  {author} {\bibfnamefont {R.~P.}\ \bibnamefont {{Drake}}}, \bibinfo {author}
  {\bibfnamefont {D.~H.}\ \bibnamefont {{Froula}}}, \bibinfo {author}
  {\bibfnamefont {G.}~\bibnamefont {{Gregori}}}, \bibinfo {author}
  {\bibfnamefont {N.~L.}\ \bibnamefont {{Kugland}}}, \bibinfo {author}
  {\bibfnamefont {C.~C.}\ \bibnamefont {{Kuranz}}}, \bibinfo {author}
  {\bibfnamefont {M.~C.}\ \bibnamefont {{Levy}}}, \bibinfo {author}
  {\bibfnamefont {C.~K.}\ \bibnamefont {{Li}}}, \bibinfo {author}
  {\bibfnamefont {J.}~\bibnamefont {{Meinecke}}}, \bibinfo {author}
  {\bibfnamefont {T.}~\bibnamefont {{Morita}}}, \bibinfo {author}
  {\bibfnamefont {R.}~\bibnamefont {{Petrasso}}}, \bibinfo {author}
  {\bibfnamefont {C.}~\bibnamefont {{Plechaty}}}, \bibinfo {author}
  {\bibfnamefont {B.~A.}\ \bibnamefont {{Remington}}}, \bibinfo {author}
  {\bibfnamefont {D.~D.}\ \bibnamefont {{Ryutov}}}, \bibinfo {author}
  {\bibfnamefont {Y.}~\bibnamefont {{Sakawa}}}, \bibinfo {author}
  {\bibfnamefont {A.}~\bibnamefont {{Spitkovsky}}}, \bibinfo {author}
  {\bibfnamefont {H.}~\bibnamefont {{Takabe}}}, \ and\ \bibinfo {author}
  {\bibfnamefont {H.-S.}\ \bibnamefont {{Park}}},\ }\href {\doibase
  10.1038/nphys3178} {\bibfield  {journal} {\bibinfo  {journal} {Nat. Phys.}\
  }\textbf {\bibinfo {volume} {11}},\ \bibinfo {pages} {173} (\bibinfo {year}
  {2015})}\BibitemShut {NoStop}%
\bibitem [{\citenamefont {{Drake}}\ and\ \citenamefont
  {{Gregori}}(2012)}]{Drake_2012}%
  \BibitemOpen
  \bibfield  {author} {\bibinfo {author} {\bibfnamefont {R.~P.}\ \bibnamefont
  {{Drake}}}\ and\ \bibinfo {author} {\bibfnamefont {G.}~\bibnamefont
  {{Gregori}}},\ }\href {\doibase 10.1088/0004-637X/749/2/171} {\bibfield
  {journal} {\bibinfo  {journal} {Astrophys. J.}\ }\textbf {\bibinfo {volume}
  {749}},\ \bibinfo {pages} {171} (\bibinfo {year} {2012})}\BibitemShut
  {NoStop}%
\bibitem [{\citenamefont {{Chen}}\ \emph {et~al.}(2015)\citenamefont {{Chen}},
  \citenamefont {{Fiuza}}, \citenamefont {{Link}}, \citenamefont {{Hazi}},
  \citenamefont {{Hill}}, \citenamefont {{Hoarty}}, \citenamefont {{James}},
  \citenamefont {{Kerr}}, \citenamefont {{Meyerhofer}}, \citenamefont
  {{Myatt}}, \citenamefont {{Park}}, \citenamefont {{Sentoku}},\ and\
  \citenamefont {{Williams}}}]{Chen_PRL_114_215001_2015}%
  \BibitemOpen
  \bibfield  {author} {\bibinfo {author} {\bibfnamefont {H.}~\bibnamefont
  {{Chen}}}, \bibinfo {author} {\bibfnamefont {F.}~\bibnamefont {{Fiuza}}},
  \bibinfo {author} {\bibfnamefont {A.}~\bibnamefont {{Link}}}, \bibinfo
  {author} {\bibfnamefont {A.}~\bibnamefont {{Hazi}}}, \bibinfo {author}
  {\bibfnamefont {M.}~\bibnamefont {{Hill}}}, \bibinfo {author} {\bibfnamefont
  {D.}~\bibnamefont {{Hoarty}}}, \bibinfo {author} {\bibfnamefont
  {S.}~\bibnamefont {{James}}}, \bibinfo {author} {\bibfnamefont
  {S.}~\bibnamefont {{Kerr}}}, \bibinfo {author} {\bibfnamefont {D.~D.}\
  \bibnamefont {{Meyerhofer}}}, \bibinfo {author} {\bibfnamefont
  {J.}~\bibnamefont {{Myatt}}}, \bibinfo {author} {\bibfnamefont
  {J.}~\bibnamefont {{Park}}}, \bibinfo {author} {\bibfnamefont
  {Y.}~\bibnamefont {{Sentoku}}}, \ and\ \bibinfo {author} {\bibfnamefont
  {G.~J.}\ \bibnamefont {{Williams}}},\ }\href {\doibase
  10.1103/PhysRevLett.114.215001} {\bibfield  {journal} {\bibinfo  {journal}
  {Phys. Rev. Lett.}\ }\textbf {\bibinfo {volume} {114}},\ \bibinfo {eid}
  {215001} (\bibinfo {year} {2015})}\BibitemShut {NoStop}%
\bibitem [{\citenamefont {Lobet}\ \emph {et~al.}(2015)\citenamefont {Lobet},
  \citenamefont {Ruyer}, \citenamefont {Debayle}, \citenamefont {d'Humi\`eres},
  \citenamefont {Grech}, \citenamefont {Lemoine},\ and\ \citenamefont
  {Gremillet}}]{Lobet_2015}%
  \BibitemOpen
  \bibfield  {author} {\bibinfo {author} {\bibfnamefont {M.}~\bibnamefont
  {Lobet}}, \bibinfo {author} {\bibfnamefont {C.}~\bibnamefont {Ruyer}},
  \bibinfo {author} {\bibfnamefont {A.}~\bibnamefont {Debayle}}, \bibinfo
  {author} {\bibfnamefont {E.}~\bibnamefont {d'Humi\`eres}}, \bibinfo {author}
  {\bibfnamefont {M.}~\bibnamefont {Grech}}, \bibinfo {author} {\bibfnamefont
  {M.}~\bibnamefont {Lemoine}}, \ and\ \bibinfo {author} {\bibfnamefont
  {L.}~\bibnamefont {Gremillet}},\ }\href {\doibase
  10.1103/PhysRevLett.115.215003} {\bibfield  {journal} {\bibinfo  {journal}
  {Phys. Rev. Lett.}\ }\textbf {\bibinfo {volume} {115}},\ \bibinfo {pages}
  {215003} (\bibinfo {year} {2015})}\BibitemShut {NoStop}%
\bibitem [{\citenamefont {Ruyer}\ \emph {et~al.}(2016)\citenamefont {Ruyer},
  \citenamefont {Gremillet}, \citenamefont {Bonnaud},\ and\ \citenamefont
  {Riconda}}]{Ruyer_2016}%
  \BibitemOpen
  \bibfield  {author} {\bibinfo {author} {\bibfnamefont {C.}~\bibnamefont
  {Ruyer}}, \bibinfo {author} {\bibfnamefont {L.}~\bibnamefont {Gremillet}},
  \bibinfo {author} {\bibfnamefont {G.}~\bibnamefont {Bonnaud}}, \ and\
  \bibinfo {author} {\bibfnamefont {C.}~\bibnamefont {Riconda}},\ }\href
  {\doibase 10.1103/PhysRevLett.117.065001} {\bibfield  {journal} {\bibinfo
  {journal} {Phys. Rev. Lett.}\ }\textbf {\bibinfo {volume} {117}},\ \bibinfo
  {pages} {065001} (\bibinfo {year} {2016})}\BibitemShut {NoStop}%
\bibitem [{\citenamefont {Ruyer}\ \emph {et~al.}(2017)\citenamefont {Ruyer},
  \citenamefont {Gremillet}, \citenamefont {Bonnaud},\ and\ \citenamefont
  {Riconda}}]{Ruyer_2017}%
  \BibitemOpen
  \bibfield  {author} {\bibinfo {author} {\bibfnamefont {C.}~\bibnamefont
  {Ruyer}}, \bibinfo {author} {\bibfnamefont {L.}~\bibnamefont {Gremillet}},
  \bibinfo {author} {\bibfnamefont {G.}~\bibnamefont {Bonnaud}}, \ and\
  \bibinfo {author} {\bibfnamefont {C.}~\bibnamefont {Riconda}},\ }\href
  {\doibase 10.1063/1.4979187} {\bibfield  {journal} {\bibinfo  {journal}
  {Phys. Plasmas}\ }\textbf {\bibinfo {volume} {24}},\ \bibinfo {pages}
  {041409} (\bibinfo {year} {2017})}\BibitemShut {NoStop}%
\bibitem [{\citenamefont {{Wiersma}}\ and\ \citenamefont
  {{Achterberg}}(2004)}]{Wiersma_2004}%
  \BibitemOpen
  \bibfield  {author} {\bibinfo {author} {\bibfnamefont {J.}~\bibnamefont
  {{Wiersma}}}\ and\ \bibinfo {author} {\bibfnamefont {A.}~\bibnamefont
  {{Achterberg}}},\ }\href {\doibase 10.1051/0004-6361:20041882} {\bibfield
  {journal} {\bibinfo  {journal} {Astron. Astrophys.}\ }\textbf {\bibinfo
  {volume} {428}},\ \bibinfo {pages} {365} (\bibinfo {year}
  {2004})}\BibitemShut {NoStop}%
\bibitem [{\citenamefont {Lyubarsky}\ and\ \citenamefont
  {Eichler}(2006)}]{Lyubarsky_2006a}%
  \BibitemOpen
  \bibfield  {author} {\bibinfo {author} {\bibfnamefont {Y.}~\bibnamefont
  {Lyubarsky}}\ and\ \bibinfo {author} {\bibfnamefont {D.}~\bibnamefont
  {Eichler}},\ }\href {\doibase 10.1086/505523} {\bibfield  {journal} {\bibinfo
   {journal} {Astrophy. J.}\ }\textbf {\bibinfo {volume} {647}},\ \bibinfo
  {pages} {1250} (\bibinfo {year} {2006})}\BibitemShut {NoStop}%
\bibitem [{\citenamefont {{Medvedev}}\ \emph {et~al.}(2004)\citenamefont
  {{Medvedev}}, \citenamefont {{Fiore}}, \citenamefont {{Fonseca}},
  \citenamefont {{Silva}},\ and\ \citenamefont {{Mori}}}]{Medvedev_2005}%
  \BibitemOpen
  \bibfield  {author} {\bibinfo {author} {\bibfnamefont {M.~V.}\ \bibnamefont
  {{Medvedev}}}, \bibinfo {author} {\bibfnamefont {M.}~\bibnamefont {{Fiore}}},
  \bibinfo {author} {\bibfnamefont {R.~A.}\ \bibnamefont {{Fonseca}}}, \bibinfo
  {author} {\bibfnamefont {L.~O.}\ \bibnamefont {{Silva}}}, \ and\ \bibinfo
  {author} {\bibfnamefont {W.~B.}\ \bibnamefont {{Mori}}},\ }\href {\doibase
  10.1086/427921} {\bibfield  {journal} {\bibinfo  {journal} {Astrophys. J.}\
  }\textbf {\bibinfo {volume} {618}},\ \bibinfo {pages} {L75} (\bibinfo {year}
  {2004})}\BibitemShut {NoStop}%
\bibitem [{\citenamefont {{Milosavljevic}}\ and\ \citenamefont
  {{Nakar}}(2006)}]{Milosavljevic_2006a}%
  \BibitemOpen
  \bibfield  {author} {\bibinfo {author} {\bibfnamefont {M.}~\bibnamefont
  {{Milosavljevic}}}\ and\ \bibinfo {author} {\bibfnamefont {E.}~\bibnamefont
  {{Nakar}}},\ }\href {\doibase 10.1086/500654} {\bibfield  {journal} {\bibinfo
   {journal} {Astrophys. J.}\ }\textbf {\bibinfo {volume} {641}},\ \bibinfo
  {pages} {978} (\bibinfo {year} {2006})}\BibitemShut {NoStop}%
\bibitem [{\citenamefont {{Achterberg}}\ \emph {et~al.}(2007)\citenamefont
  {{Achterberg}}, \citenamefont {{Wiersma}},\ and\ \citenamefont
  {{Norman}}}]{Achterberg_2007_II}%
  \BibitemOpen
  \bibfield  {author} {\bibinfo {author} {\bibfnamefont {A.}~\bibnamefont
  {{Achterberg}}}, \bibinfo {author} {\bibfnamefont {J.}~\bibnamefont
  {{Wiersma}}}, \ and\ \bibinfo {author} {\bibfnamefont {C.~A.}\ \bibnamefont
  {{Norman}}},\ }\href {\doibase 10.1051/0004-6361:20065366} {\bibfield
  {journal} {\bibinfo  {journal} {Astron. Astrophys.}\ }\textbf {\bibinfo
  {volume} {475}},\ \bibinfo {pages} {19} (\bibinfo {year} {2007})}\BibitemShut
  {NoStop}%
\bibitem [{\citenamefont {{Bret}}\ \emph {et~al.}(2013)\citenamefont {{Bret}},
  \citenamefont {{Stockem}}, \citenamefont {{Fiuza}}, \citenamefont {{Ruyer}},
  \citenamefont {{Gremillet}}, \citenamefont {{Narayan}},\ and\ \citenamefont
  {{Silva}}}]{Bret_2013}%
  \BibitemOpen
  \bibfield  {author} {\bibinfo {author} {\bibfnamefont {A.}~\bibnamefont
  {{Bret}}}, \bibinfo {author} {\bibfnamefont {A.}~\bibnamefont {{Stockem}}},
  \bibinfo {author} {\bibfnamefont {D.}~\bibnamefont {{Fiuza}}}, \bibinfo
  {author} {\bibfnamefont {C.}~\bibnamefont {{Ruyer}}}, \bibinfo {author}
  {\bibfnamefont {L.}~\bibnamefont {{Gremillet}}}, \bibinfo {author}
  {\bibfnamefont {R.}~\bibnamefont {{Narayan}}}, \ and\ \bibinfo {author}
  {\bibfnamefont {L.~O.}\ \bibnamefont {{Silva}}},\ }\href {\doibase
  10.1063/1.4798541} {\bibfield  {journal} {\bibinfo  {journal} {Phys.
  Plasmas}\ }\textbf {\bibinfo {volume} {20}},\ \bibinfo {pages} {042102}
  (\bibinfo {year} {2013})}\BibitemShut {NoStop}%
\bibitem [{\citenamefont {Ruyer}\ \emph
  {et~al.}(2015{\natexlab{b}})\citenamefont {Ruyer}, \citenamefont {Gremillet},
  \citenamefont {Debayle},\ and\ \citenamefont {Bonnaud}}]{Ruyer_2015a}%
  \BibitemOpen
  \bibfield  {author} {\bibinfo {author} {\bibfnamefont {C.}~\bibnamefont
  {Ruyer}}, \bibinfo {author} {\bibfnamefont {L.}~\bibnamefont {Gremillet}},
  \bibinfo {author} {\bibfnamefont {A.}~\bibnamefont {Debayle}}, \ and\
  \bibinfo {author} {\bibfnamefont {G.}~\bibnamefont {Bonnaud}},\ }\href
  {\doibase 10.1063/1.4913651} {\bibfield  {journal} {\bibinfo  {journal}
  {Phys. Plasmas}\ }\textbf {\bibinfo {volume} {22}},\ \bibinfo {pages}
  {032102} (\bibinfo {year} {2015}{\natexlab{b}})}\BibitemShut {NoStop}%
\bibitem [{\citenamefont {{Vanthieghem}}\ \emph {et~al.}(2018)\citenamefont
  {{Vanthieghem}}, \citenamefont {{Lemoine}},\ and\ \citenamefont
  {{Gremillet}}}]{Vanthieghem_2018}%
  \BibitemOpen
  \bibfield  {author} {\bibinfo {author} {\bibfnamefont {A.}~\bibnamefont
  {{Vanthieghem}}}, \bibinfo {author} {\bibfnamefont {M.}~\bibnamefont
  {{Lemoine}}}, \ and\ \bibinfo {author} {\bibfnamefont {L.}~\bibnamefont
  {{Gremillet}}},\ }\href {\doibase 10.1063/1.5033562} {\bibfield  {journal}
  {\bibinfo  {journal} {Phys. Plasmas}\ }\textbf {\bibinfo {volume} {25}},\
  \bibinfo {eid} {072115} (\bibinfo {year} {2018})}\BibitemShut {NoStop}%
\bibitem [{\citenamefont {Lee}\ and\ \citenamefont {Lampe}(1973)}]{Lee_1973}%
  \BibitemOpen
  \bibfield  {author} {\bibinfo {author} {\bibfnamefont {R.}~\bibnamefont
  {Lee}}\ and\ \bibinfo {author} {\bibfnamefont {M.}~\bibnamefont {Lampe}},\
  }\href {\doibase 10.1103/PhysRevLett.31.1390} {\bibfield  {journal} {\bibinfo
   {journal} {Phys. Rev. Lett.}\ }\textbf {\bibinfo {volume} {31}},\ \bibinfo
  {pages} {1390} (\bibinfo {year} {1973})}\BibitemShut {NoStop}%
\bibitem [{\citenamefont {{Silva}}\ \emph {et~al.}(2003)\citenamefont
  {{Silva}}, \citenamefont {{Fonseca}}, \citenamefont {{Tonge}}, \citenamefont
  {{Dawson}}, \citenamefont {{Mori}},\ and\ \citenamefont
  {{Medvedev}}}]{Silva_2003}%
  \BibitemOpen
  \bibfield  {author} {\bibinfo {author} {\bibfnamefont {L.~O.}\ \bibnamefont
  {{Silva}}}, \bibinfo {author} {\bibfnamefont {R.~A.}\ \bibnamefont
  {{Fonseca}}}, \bibinfo {author} {\bibfnamefont {J.~W.}\ \bibnamefont
  {{Tonge}}}, \bibinfo {author} {\bibfnamefont {J.~M.}\ \bibnamefont
  {{Dawson}}}, \bibinfo {author} {\bibfnamefont {W.~B.}\ \bibnamefont
  {{Mori}}}, \ and\ \bibinfo {author} {\bibfnamefont {M.~V.}\ \bibnamefont
  {{Medvedev}}},\ }\href {\doibase 10.1086/379156} {\bibfield  {journal}
  {\bibinfo  {journal} {Astrophys. J. Lett.}\ }\textbf {\bibinfo {volume}
  {596}},\ \bibinfo {pages} {L121} (\bibinfo {year} {2003})}\BibitemShut
  {NoStop}%
\bibitem [{\citenamefont {{Frederiksen}}\ \emph {et~al.}(2004)\citenamefont
  {{Frederiksen}}, \citenamefont {{Hededal}}, \citenamefont {{Haugb{\o}lle}},\
  and\ \citenamefont {{Nordlund}}}]{Frederiksen_2004}%
  \BibitemOpen
  \bibfield  {author} {\bibinfo {author} {\bibfnamefont {J.~T.}\ \bibnamefont
  {{Frederiksen}}}, \bibinfo {author} {\bibfnamefont {C.~B.}\ \bibnamefont
  {{Hededal}}}, \bibinfo {author} {\bibfnamefont {T.}~\bibnamefont
  {{Haugb{\o}lle}}}, \ and\ \bibinfo {author} {\bibfnamefont
  {{\AA}.}~\bibnamefont {{Nordlund}}},\ }\href {\doibase 10.1086/421262}
  {\bibfield  {journal} {\bibinfo  {journal} {Astrophys. J. Lett.}\ }\textbf
  {\bibinfo {volume} {608}},\ \bibinfo {pages} {L13} (\bibinfo {year}
  {2004})}\BibitemShut {NoStop}%
\bibitem [{\citenamefont {{Jaroschek}}\ \emph {et~al.}(2005)\citenamefont
  {{Jaroschek}}, \citenamefont {{Lesch}},\ and\ \citenamefont
  {{Treumann}}}]{Jaroschek_2005}%
  \BibitemOpen
  \bibfield  {author} {\bibinfo {author} {\bibfnamefont {C.~H.}\ \bibnamefont
  {{Jaroschek}}}, \bibinfo {author} {\bibfnamefont {H.}~\bibnamefont
  {{Lesch}}}, \ and\ \bibinfo {author} {\bibfnamefont {R.~A.}\ \bibnamefont
  {{Treumann}}},\ }\href {\doibase 10.1086/426066} {\bibfield  {journal}
  {\bibinfo  {journal} {Astrophys. J.}\ }\textbf {\bibinfo {volume} {618}},\
  \bibinfo {pages} {822} (\bibinfo {year} {2005})}\BibitemShut {NoStop}%
\bibitem [{\citenamefont {{Nishikawa}}\ \emph {et~al.}(2009)\citenamefont
  {{Nishikawa}}, \citenamefont {{Niemiec}}, \citenamefont {{Hardee}},
  \citenamefont {{Medvedev}}, \citenamefont {{Sol}}, \citenamefont {{Mizuno}},
  \citenamefont {{Zhang}}, \citenamefont {{Pohl}}, \citenamefont {{Oka}},\ and\
  \citenamefont {{Hartmann}}}]{Nishikawa_2009}%
  \BibitemOpen
  \bibfield  {author} {\bibinfo {author} {\bibfnamefont {K.-I.}\ \bibnamefont
  {{Nishikawa}}}, \bibinfo {author} {\bibfnamefont {J.}~\bibnamefont
  {{Niemiec}}}, \bibinfo {author} {\bibfnamefont {P.~E.}\ \bibnamefont
  {{Hardee}}}, \bibinfo {author} {\bibfnamefont {M.}~\bibnamefont
  {{Medvedev}}}, \bibinfo {author} {\bibfnamefont {H.}~\bibnamefont {{Sol}}},
  \bibinfo {author} {\bibfnamefont {Y.}~\bibnamefont {{Mizuno}}}, \bibinfo
  {author} {\bibfnamefont {B.}~\bibnamefont {{Zhang}}}, \bibinfo {author}
  {\bibfnamefont {M.}~\bibnamefont {{Pohl}}}, \bibinfo {author} {\bibfnamefont
  {M.}~\bibnamefont {{Oka}}}, \ and\ \bibinfo {author} {\bibfnamefont {D.~H.}\
  \bibnamefont {{Hartmann}}},\ }\href {\doibase 10.1088/0004-637X/698/1/L10}
  {\bibfield  {journal} {\bibinfo  {journal} {Astrophys. J. Lett.}\ }\textbf
  {\bibinfo {volume} {698}},\ \bibinfo {pages} {L10} (\bibinfo {year}
  {2009})}\BibitemShut {NoStop}%
\bibitem [{\citenamefont {{Shvets}}\ \emph {et~al.}(2009)\citenamefont
  {{Shvets}}, \citenamefont {{Polomarov}}, \citenamefont {{Khudik}},
  \citenamefont {{Siemon}},\ and\ \citenamefont {{Kaganovich}}}]{Shvets_2009}%
  \BibitemOpen
  \bibfield  {author} {\bibinfo {author} {\bibfnamefont {G.}~\bibnamefont
  {{Shvets}}}, \bibinfo {author} {\bibfnamefont {O.}~\bibnamefont
  {{Polomarov}}}, \bibinfo {author} {\bibfnamefont {V.}~\bibnamefont
  {{Khudik}}}, \bibinfo {author} {\bibfnamefont {C.}~\bibnamefont {{Siemon}}},
  \ and\ \bibinfo {author} {\bibfnamefont {I.}~\bibnamefont {{Kaganovich}}},\
  }\href {\doibase 10.1063/1.3093477} {\bibfield  {journal} {\bibinfo
  {journal} {Phys. Plasmas}\ }\textbf {\bibinfo {volume} {16}},\ \bibinfo
  {pages} {056303} (\bibinfo {year} {2009})}\BibitemShut {NoStop}%
\bibitem [{\citenamefont {{Bret}}\ \emph
  {et~al.}(2010{\natexlab{a}})\citenamefont {{Bret}}, \citenamefont
  {{Gremillet}},\ and\ \citenamefont {{Dieckman}}}]{Bret_2010a}%
  \BibitemOpen
  \bibfield  {author} {\bibinfo {author} {\bibfnamefont {A.}~\bibnamefont
  {{Bret}}}, \bibinfo {author} {\bibfnamefont {L.}~\bibnamefont {{Gremillet}}},
  \ and\ \bibinfo {author} {\bibfnamefont {M.~E.}\ \bibnamefont {{Dieckman}}},\
  }\href {\doibase 10.1063/1.3514586} {\bibfield  {journal} {\bibinfo
  {journal} {Phys. Plasmas}\ }\textbf {\bibinfo {volume} {17}},\ \bibinfo
  {pages} {120501} (\bibinfo {year} {2010}{\natexlab{a}})}\BibitemShut
  {NoStop}%
\bibitem [{\citenamefont {{Kato}}(2007)}]{Kato_2007}%
  \BibitemOpen
  \bibfield  {author} {\bibinfo {author} {\bibfnamefont {T.~N.}\ \bibnamefont
  {{Kato}}},\ }\href {\doibase 10.1086/521297} {\bibfield  {journal} {\bibinfo
  {journal} {Astrophys. J. Lett.}\ }\textbf {\bibinfo {volume} {668}},\
  \bibinfo {pages} {974} (\bibinfo {year} {2007})}\BibitemShut {NoStop}%
\bibitem [{\citenamefont {{Spitkovsky}}(2008)}]{Spitkovsky_2008a}%
  \BibitemOpen
  \bibfield  {author} {\bibinfo {author} {\bibfnamefont {A.}~\bibnamefont
  {{Spitkovsky}}},\ }\href {\doibase 10.1086/590248} {\bibfield  {journal}
  {\bibinfo  {journal} {Astrophys. J. Lett.}\ }\textbf {\bibinfo {volume}
  {682}},\ \bibinfo {eid} {L5} (\bibinfo {year} {2008})}\BibitemShut {NoStop}%
\bibitem [{\citenamefont {{Martins}}\ \emph {et~al.}(2009)\citenamefont
  {{Martins}}, \citenamefont {{Fonseca}}, \citenamefont {{Silva}},\ and\
  \citenamefont {{Mori}}}]{Martins_2009}%
  \BibitemOpen
  \bibfield  {author} {\bibinfo {author} {\bibfnamefont {S.~F.}\ \bibnamefont
  {{Martins}}}, \bibinfo {author} {\bibfnamefont {R.~A.}\ \bibnamefont
  {{Fonseca}}}, \bibinfo {author} {\bibfnamefont {L.~O.}\ \bibnamefont
  {{Silva}}}, \ and\ \bibinfo {author} {\bibfnamefont {W.~B.}\ \bibnamefont
  {{Mori}}},\ }\href {\doibase 10.1088/0004-637X/695/2/L189} {\bibfield
  {journal} {\bibinfo  {journal} {Astrophys. J. Lett.}\ }\textbf {\bibinfo
  {volume} {695}},\ \bibinfo {pages} {L189} (\bibinfo {year}
  {2009})}\BibitemShut {NoStop}%
\bibitem [{\citenamefont {{Keshet}}\ \emph {et~al.}(2009)\citenamefont
  {{Keshet}}, \citenamefont {{Katz}}, \citenamefont {{Spitkovsky}},\ and\
  \citenamefont {{Waxman}}}]{Keshet_2009}%
  \BibitemOpen
  \bibfield  {author} {\bibinfo {author} {\bibfnamefont {U.}~\bibnamefont
  {{Keshet}}}, \bibinfo {author} {\bibfnamefont {B.}~\bibnamefont {{Katz}}},
  \bibinfo {author} {\bibfnamefont {A.}~\bibnamefont {{Spitkovsky}}}, \ and\
  \bibinfo {author} {\bibfnamefont {E.}~\bibnamefont {{Waxman}}},\ }\href
  {\doibase 10.1088/0004-637X/693/2/L127} {\bibfield  {journal} {\bibinfo
  {journal} {Astrophys. J.}\ }\textbf {\bibinfo {volume} {693}},\ \bibinfo
  {pages} {L127} (\bibinfo {year} {2009})}\BibitemShut {NoStop}%
\bibitem [{\citenamefont {Sironi}\ and\ \citenamefont
  {Spitkovsky}(2009)}]{Sironi_2009b}%
  \BibitemOpen
  \bibfield  {author} {\bibinfo {author} {\bibfnamefont {L.}~\bibnamefont
  {Sironi}}\ and\ \bibinfo {author} {\bibfnamefont {A.}~\bibnamefont
  {Spitkovsky}},\ }\href {\doibase 10.1088/0004-637X/698/2/1523} {\bibfield
  {journal} {\bibinfo  {journal} {Astrophys. J.}\ }\textbf {\bibinfo {volume}
  {698}},\ \bibinfo {pages} {1523} (\bibinfo {year} {2009})}\BibitemShut
  {NoStop}%
\bibitem [{\citenamefont {{Sironi}}\ \emph {et~al.}(2013)\citenamefont
  {{Sironi}}, \citenamefont {{Spitkovsky}},\ and\ \citenamefont
  {{Arons}}}]{Sironi_2013}%
  \BibitemOpen
  \bibfield  {author} {\bibinfo {author} {\bibfnamefont {L.}~\bibnamefont
  {{Sironi}}}, \bibinfo {author} {\bibfnamefont {A.}~\bibnamefont
  {{Spitkovsky}}}, \ and\ \bibinfo {author} {\bibfnamefont {J.}~\bibnamefont
  {{Arons}}},\ }\href {\doibase 10.1088/0004-637X/771/1/54} {\bibfield
  {journal} {\bibinfo  {journal} {Astrophys. J.}\ }\textbf {\bibinfo {volume}
  {771}},\ \bibinfo {eid} {54} (\bibinfo {year} {2013})}\BibitemShut {NoStop}%
\bibitem [{\citenamefont {{Haugb{\o}lle}}(2011)}]{Haugbolle_2011}%
  \BibitemOpen
  \bibfield  {author} {\bibinfo {author} {\bibfnamefont {T.}~\bibnamefont
  {{Haugb{\o}lle}}},\ }\href {\doibase 10.1088/2041-8205/739/2/L42} {\bibfield
  {journal} {\bibinfo  {journal} {Astrophys. J. Lett.}\ }\textbf {\bibinfo
  {volume} {739}},\ \bibinfo {pages} {42} (\bibinfo {year} {2011})}\BibitemShut
  {NoStop}%
\bibitem [{\citenamefont {{Kumar}}\ \emph {et~al.}(2015)\citenamefont
  {{Kumar}}, \citenamefont {{Eichler}},\ and\ \citenamefont
  {{Gedalin}}}]{Kumar_2015}%
  \BibitemOpen
  \bibfield  {author} {\bibinfo {author} {\bibfnamefont {R.}~\bibnamefont
  {{Kumar}}}, \bibinfo {author} {\bibfnamefont {D.}~\bibnamefont {{Eichler}}},
  \ and\ \bibinfo {author} {\bibfnamefont {M.}~\bibnamefont {{Gedalin}}},\
  }\href {\doibase 10.1088/0004-637X/806/2/165} {\bibfield  {journal} {\bibinfo
   {journal} {Astrophys. J.}\ }\textbf {\bibinfo {volume} {806}},\ \bibinfo
  {eid} {165} (\bibinfo {year} {2015})}\BibitemShut {NoStop}%
\bibitem [{\citenamefont {{Lemoine}}\ and\ \citenamefont
  {{Pelletier}}(2010)}]{Lemoine_2010}%
  \BibitemOpen
  \bibfield  {author} {\bibinfo {author} {\bibfnamefont {M.}~\bibnamefont
  {{Lemoine}}}\ and\ \bibinfo {author} {\bibfnamefont {G.}~\bibnamefont
  {{Pelletier}}},\ }\href {\doibase 10.1111/j.1365-2966.2009.15869.x}
  {\bibfield  {journal} {\bibinfo  {journal} {Mon. Not. Roy. Astron. Soc.}\
  }\textbf {\bibinfo {volume} {402}},\ \bibinfo {pages} {321} (\bibinfo {year}
  {2010})}\BibitemShut {NoStop}%
\bibitem [{\citenamefont {{Rabinak}}\ \emph {et~al.}(2011)\citenamefont
  {{Rabinak}}, \citenamefont {{Katz}},\ and\ \citenamefont
  {{Waxman}}}]{Rabinak_2011}%
  \BibitemOpen
  \bibfield  {author} {\bibinfo {author} {\bibfnamefont {I.}~\bibnamefont
  {{Rabinak}}}, \bibinfo {author} {\bibfnamefont {B.}~\bibnamefont {{Katz}}}, \
  and\ \bibinfo {author} {\bibfnamefont {E.}~\bibnamefont {{Waxman}}},\ }\href
  {\doibase 10.1088/0004-637X/736/2/157} {\bibfield  {journal} {\bibinfo
  {journal} {Astrophys. J.}\ }\textbf {\bibinfo {volume} {736}},\ \bibinfo
  {eid} {157} (\bibinfo {year} {2011})}\BibitemShut {NoStop}%
\bibitem [{\citenamefont {{Lemoine}}\ and\ \citenamefont
  {{Pelletier}}(2011)}]{Lemoine_2011}%
  \BibitemOpen
  \bibfield  {author} {\bibinfo {author} {\bibfnamefont {M.}~\bibnamefont
  {{Lemoine}}}\ and\ \bibinfo {author} {\bibfnamefont {G.}~\bibnamefont
  {{Pelletier}}},\ }\href {\doibase 10.1111/j.1365-2966.2011.19331.x}
  {\bibfield  {journal} {\bibinfo  {journal} {Mon. Not. Roy. Astron. Soc.}\
  }\textbf {\bibinfo {volume} {417}},\ \bibinfo {pages} {1148} (\bibinfo {year}
  {2011})}\BibitemShut {NoStop}%
\bibitem [{\citenamefont {{Shaisultanov}}\ \emph {et~al.}(2012)\citenamefont
  {{Shaisultanov}}, \citenamefont {{Lyubarsky}},\ and\ \citenamefont
  {{Eichler}}}]{Shaisultanov_2012}%
  \BibitemOpen
  \bibfield  {author} {\bibinfo {author} {\bibfnamefont {R.}~\bibnamefont
  {{Shaisultanov}}}, \bibinfo {author} {\bibfnamefont {Y.}~\bibnamefont
  {{Lyubarsky}}}, \ and\ \bibinfo {author} {\bibfnamefont {D.}~\bibnamefont
  {{Eichler}}},\ }\href {\doibase 10.1088/0004-637X/744/2/182} {\bibfield
  {journal} {\bibinfo  {journal} {Astrophys. J.}\ }\textbf {\bibinfo {volume}
  {744}},\ \bibinfo {eid} {182} (\bibinfo {year} {2012})}\BibitemShut {NoStop}%
\bibitem [{\citenamefont {{Lemoine}}\ \emph
  {et~al.}(2019{\natexlab{a}})\citenamefont {{Lemoine}}, \citenamefont
  {{Gremillet}}, \citenamefont {{Pelletier}},\ and\ \citenamefont
  {{Vanthieghem}}}]{L1}%
  \BibitemOpen
  \bibfield  {author} {\bibinfo {author} {\bibfnamefont {M.}~\bibnamefont
  {{Lemoine}}}, \bibinfo {author} {\bibfnamefont {L.}~\bibnamefont
  {{Gremillet}}}, \bibinfo {author} {\bibfnamefont {G.}~\bibnamefont
  {{Pelletier}}}, \ and\ \bibinfo {author} {\bibfnamefont {A.}~\bibnamefont
  {{Vanthieghem}}},\ }\href {\doibase 10.1103/PhysRevLett.123.035101}
  {\bibfield  {journal} {\bibinfo  {journal} {Phys. Rev. Lett.}\ }\textbf
  {\bibinfo {volume} {123}},\ \bibinfo {pages} {035101} (\bibinfo {year}
  {2019}{\natexlab{a}})},\ \Eprint {http://arxiv.org/abs/1907.07595}
  {arXiv:1907.07595} \BibitemShut {NoStop}%
\bibitem [{\citenamefont {{Lemoine}}\ \emph
  {et~al.}(2019{\natexlab{b}})\citenamefont {{Lemoine}}, \citenamefont
  {{Vanthieghem}}, \citenamefont {{Pelletier}},\ and\ \citenamefont
  {{Gremillet}}}]{pap2}%
  \BibitemOpen
  \bibfield  {author} {\bibinfo {author} {\bibfnamefont {M.}~\bibnamefont
  {{Lemoine}}}, \bibinfo {author} {\bibfnamefont {A.}~\bibnamefont
  {{Vanthieghem}}}, \bibinfo {author} {\bibfnamefont {G.}~\bibnamefont
  {{Pelletier}}}, \ and\ \bibinfo {author} {\bibfnamefont {L.}~\bibnamefont
  {{Gremillet}}},\ }\href@noop {} {\bibfield  {journal} {\bibinfo  {journal}
  {Phys. Rev. E, submitted (Pap. II)}\ } (\bibinfo {year}
  {2019}{\natexlab{b}})},\ \Eprint {http://arxiv.org/abs/1907.08219}
  {arXiv:1907.08219 [astro-ph.HE]} \BibitemShut {NoStop}%
\bibitem [{\citenamefont {{Lemoine}}\ \emph
  {et~al.}(2019{\natexlab{c}})\citenamefont {{Lemoine}}, \citenamefont
  {{Pelletier}}, \citenamefont {{Vanthieghem}},\ and\ \citenamefont
  {{Gremillet}}}]{pap3}%
  \BibitemOpen
  \bibfield  {author} {\bibinfo {author} {\bibfnamefont {M.}~\bibnamefont
  {{Lemoine}}}, \bibinfo {author} {\bibfnamefont {G.}~\bibnamefont
  {{Pelletier}}}, \bibinfo {author} {\bibfnamefont {A.}~\bibnamefont
  {{Vanthieghem}}}, \ and\ \bibinfo {author} {\bibfnamefont {L.}~\bibnamefont
  {{Gremillet}}},\ }\href@noop {} {\bibfield  {journal} {\bibinfo  {journal}
  {Phys. Rev. E, submitted (Pap. III)}\ } (\bibinfo {year}
  {2019}{\natexlab{c}})},\ \Eprint {http://arxiv.org/abs/1907.10294}
  {arXiv:1907.10294 [astro-ph.HE]} \BibitemShut {NoStop}%
\bibitem [{\citenamefont {{Vanthieghem}}\ \emph {et~al.}(2019)\citenamefont
  {{Vanthieghem}}, \citenamefont {{Lemoine}}, \citenamefont {{Gremillet}},\
  and\ \citenamefont {{Pelletier}}}]{pap4}%
  \BibitemOpen
  \bibfield  {author} {\bibinfo {author} {\bibfnamefont {A.}~\bibnamefont
  {{Vanthieghem}}}, \bibinfo {author} {\bibfnamefont {M.}~\bibnamefont
  {{Lemoine}}}, \bibinfo {author} {\bibfnamefont {L.}~\bibnamefont
  {{Gremillet}}}, \ and\ \bibinfo {author} {\bibfnamefont {G.}~\bibnamefont
  {{Pelletier}}},\ }\href@noop {} {\bibfield  {journal} {\bibinfo  {journal}
  {Phys. Rev. E, in prep. (Pap. IV)}\ } (\bibinfo {year} {2019})}\BibitemShut
  {NoStop}%
\bibitem [{\citenamefont {{Blandford}}\ and\ \citenamefont
  {{McKee}}(1976)}]{1976PhFl...19.1130B}%
  \BibitemOpen
  \bibfield  {author} {\bibinfo {author} {\bibfnamefont {R.~D.}\ \bibnamefont
  {{Blandford}}}\ and\ \bibinfo {author} {\bibfnamefont {C.~F.}\ \bibnamefont
  {{McKee}}},\ }\href {\doibase 10.1063/1.861619} {\bibfield  {journal}
  {\bibinfo  {journal} {Phys. Fluids}\ }\textbf {\bibinfo {volume} {19}},\
  \bibinfo {pages} {1130} (\bibinfo {year} {1976})}\BibitemShut {NoStop}%
\bibitem [{\citenamefont {Silva}\ \emph {et~al.}(2002)\citenamefont {Silva},
  \citenamefont {Fonseca}, \citenamefont {Tonge}, \citenamefont {Mori},\ and\
  \citenamefont {Dawson}}]{Silva_2002}%
  \BibitemOpen
  \bibfield  {author} {\bibinfo {author} {\bibfnamefont {L.~O.}\ \bibnamefont
  {Silva}}, \bibinfo {author} {\bibfnamefont {R.~A.}\ \bibnamefont {Fonseca}},
  \bibinfo {author} {\bibfnamefont {J.~W.}\ \bibnamefont {Tonge}}, \bibinfo
  {author} {\bibfnamefont {W.~B.}\ \bibnamefont {Mori}}, \ and\ \bibinfo
  {author} {\bibfnamefont {J.~M.}\ \bibnamefont {Dawson}},\ }\href {\doibase
  10.1063/1.1476004} {\bibfield  {journal} {\bibinfo  {journal} {Phys.
  Plasmas}\ }\textbf {\bibinfo {volume} {9}},\ \bibinfo {pages} {2458}
  (\bibinfo {year} {2002})}\BibitemShut {NoStop}%
\bibitem [{\citenamefont {Ichimaru}(1973)}]{Ichimaru_1973}%
  \BibitemOpen
  \bibfield  {author} {\bibinfo {author} {\bibfnamefont {S.}~\bibnamefont
  {Ichimaru}},\ }\href {https://books.google.fr/books?id=YiRRAAAAMAAJ} {\emph
  {\bibinfo {title} {{Basic Principles of Plasma Physics: A Statistical
  Approach}}}},\ A lecture note and reprint series\ (\bibinfo  {publisher} {W.
  A. Benjamin},\ \bibinfo {address} {Reading, MA (USA)},\ \bibinfo {year}
  {1973})\BibitemShut {NoStop}%
\bibitem [{\citenamefont {Molvig}(1975)}]{Molvig_1975}%
  \BibitemOpen
  \bibfield  {author} {\bibinfo {author} {\bibfnamefont {K.}~\bibnamefont
  {Molvig}},\ }\href {\doibase 10.1103/PhysRevLett.35.1504} {\bibfield
  {journal} {\bibinfo  {journal} {Phys. Rev. Lett.}\ }\textbf {\bibinfo
  {volume} {35}},\ \bibinfo {pages} {1504} (\bibinfo {year}
  {1975})}\BibitemShut {NoStop}%
\bibitem [{\citenamefont {{Cary}}\ \emph {et~al.}(1981)\citenamefont {{Cary}},
  \citenamefont {{Thode}}, \citenamefont {{Lemons}}, \citenamefont {{Jones}},\
  and\ \citenamefont {{Mostrom}}}]{Cary_1981}%
  \BibitemOpen
  \bibfield  {author} {\bibinfo {author} {\bibfnamefont {J.~R.}\ \bibnamefont
  {{Cary}}}, \bibinfo {author} {\bibfnamefont {L.~E.}\ \bibnamefont {{Thode}}},
  \bibinfo {author} {\bibfnamefont {D.~S.}\ \bibnamefont {{Lemons}}}, \bibinfo
  {author} {\bibfnamefont {M.~E.}\ \bibnamefont {{Jones}}}, \ and\ \bibinfo
  {author} {\bibfnamefont {M.~A.}\ \bibnamefont {{Mostrom}}},\ }\href {\doibase
  10.1063/1.863262} {\bibfield  {journal} {\bibinfo  {journal} {Phys. Fluids}\
  }\textbf {\bibinfo {volume} {24}},\ \bibinfo {pages} {1818} (\bibinfo {year}
  {1981})}\BibitemShut {NoStop}%
\bibitem [{\citenamefont {{Okada}}\ and\ \citenamefont
  {{Niu}}(1980)}]{Okada_1980a}%
  \BibitemOpen
  \bibfield  {author} {\bibinfo {author} {\bibfnamefont {T.}~\bibnamefont
  {{Okada}}}\ and\ \bibinfo {author} {\bibfnamefont {K.}~\bibnamefont
  {{Niu}}},\ }\href {\doibase 10.1017/S0022377800022431} {\bibfield  {journal}
  {\bibinfo  {journal} {J. Plasma Phys.}\ }\textbf {\bibinfo {volume} {23}},\
  \bibinfo {pages} {423} (\bibinfo {year} {1980})}\BibitemShut {NoStop}%
\bibitem [{\citenamefont {{Hill}}\ \emph {et~al.}(2005)\citenamefont {{Hill}},
  \citenamefont {{Key}}, \citenamefont {{Hatchett}},\ and\ \citenamefont
  {{Freeman}}}]{Hill_2005}%
  \BibitemOpen
  \bibfield  {author} {\bibinfo {author} {\bibfnamefont {J.~M.}\ \bibnamefont
  {{Hill}}}, \bibinfo {author} {\bibfnamefont {M.~H.}\ \bibnamefont {{Key}}},
  \bibinfo {author} {\bibfnamefont {S.~P.}\ \bibnamefont {{Hatchett}}}, \ and\
  \bibinfo {author} {\bibfnamefont {R.~R.}\ \bibnamefont {{Freeman}}},\ }\href
  {\doibase 10.1063/1.1986988} {\bibfield  {journal} {\bibinfo  {journal}
  {Phys. Plasmas}\ }\textbf {\bibinfo {volume} {12}},\ \bibinfo {eid} {082304}
  (\bibinfo {year} {2005})}\BibitemShut {NoStop}%
\bibitem [{\citenamefont {{Bret}}\ \emph {et~al.}(2004)\citenamefont {{Bret}},
  \citenamefont {Firpo},\ and\ \citenamefont {Deutsch}}]{Bret_2004}%
  \BibitemOpen
  \bibfield  {author} {\bibinfo {author} {\bibfnamefont {A.}~\bibnamefont
  {{Bret}}}, \bibinfo {author} {\bibfnamefont {M.-C.}\ \bibnamefont {Firpo}}, \
  and\ \bibinfo {author} {\bibfnamefont {C.}~\bibnamefont {Deutsch}},\ }\href
  {\doibase 10.1103/PhysRevE.70.046401} {\bibfield  {journal} {\bibinfo
  {journal} {Phys. Rev. E}\ }\textbf {\bibinfo {volume} {70}},\ \bibinfo
  {pages} {046401} (\bibinfo {year} {2004})}\BibitemShut {NoStop}%
\bibitem [{\citenamefont {Bret}\ \emph {et~al.}(2007)\citenamefont {Bret},
  \citenamefont {Gremillet},\ and\ \citenamefont {Bellido}}]{Bret_2007}%
  \BibitemOpen
  \bibfield  {author} {\bibinfo {author} {\bibfnamefont {A.}~\bibnamefont
  {Bret}}, \bibinfo {author} {\bibfnamefont {L.}~\bibnamefont {Gremillet}}, \
  and\ \bibinfo {author} {\bibfnamefont {J.~C.}\ \bibnamefont {Bellido}},\
  }\href {\doibase 10.1063/1.2710810} {\bibfield  {journal} {\bibinfo
  {journal} {Phys. Plasmas}\ }\textbf {\bibinfo {volume} {14}},\ \bibinfo
  {pages} {032103} (\bibinfo {year} {2007})}\BibitemShut {NoStop}%
\bibitem [{\citenamefont {{J{\"u}ttner}}(1911)}]{Juttner_1911}%
  \BibitemOpen
  \bibfield  {author} {\bibinfo {author} {\bibfnamefont {F.}~\bibnamefont
  {{J{\"u}ttner}}},\ }\href {\doibase 10.1002/andp.19113390503} {\bibfield
  {journal} {\bibinfo  {journal} {Annalen der Physik}\ }\textbf {\bibinfo
  {volume} {339}},\ \bibinfo {pages} {856} (\bibinfo {year}
  {1911})}\BibitemShut {NoStop}%
\bibitem [{\citenamefont {{Wright}}\ and\ \citenamefont
  {{Hadley}}(1975)}]{Wright_1975}%
  \BibitemOpen
  \bibfield  {author} {\bibinfo {author} {\bibfnamefont {T.~P.}\ \bibnamefont
  {{Wright}}}\ and\ \bibinfo {author} {\bibfnamefont {G.~R.}\ \bibnamefont
  {{Hadley}}},\ }\href {\doibase 10.1103/PhysRevA.12.686} {\bibfield  {journal}
  {\bibinfo  {journal} {Phys. Rev. A}\ }\textbf {\bibinfo {volume} {12}},\
  \bibinfo {pages} {686} (\bibinfo {year} {1975})}\BibitemShut {NoStop}%
\bibitem [{\citenamefont {Huba}(2013)}]{NRL_2013}%
  \BibitemOpen
  \bibfield  {author} {\bibinfo {author} {\bibfnamefont {J.~D.}\ \bibnamefont
  {Huba}},\ }\href {http://wwwppd.nrl.navy.mil/nrlformulary/} {\emph {\bibinfo
  {title} {Plasma Physics}}}\ (\bibinfo  {publisher} {Naval Research
  Laboratory},\ \bibinfo {address} {Washington, DC},\ \bibinfo {year} {2013})\
  pp.\ \bibinfo {pages} {1--71}\BibitemShut {NoStop}%
\bibitem [{\citenamefont {{Lefebvre}}\ \emph {et~al.}(2003)\citenamefont
  {{Lefebvre}}, \citenamefont {{Cochet}}, \citenamefont {{Fritzler}},
  \citenamefont {{Malka}}, \citenamefont {{Al{\'e}onard}}, \citenamefont
  {{Chemin}}, \citenamefont {{Darbon}}, \citenamefont {{Disdier}},
  \citenamefont {{Faure}}, \citenamefont {{Fedotoff}}, \citenamefont
  {{Landoas}}, \citenamefont {{Malka}}, \citenamefont {{M{\'e}ot}},
  \citenamefont {{Morel}}, \citenamefont {{Rabec LeGloahec}}, \citenamefont
  {{Rouyer}}, \citenamefont {{Rubbelynck}}, \citenamefont {{Tikhonchuk}},
  \citenamefont {{Wrobel}}, \citenamefont {{Audebert}},\ and\ \citenamefont
  {{Rousseaux}}}]{Lefebvre_2003}%
  \BibitemOpen
  \bibfield  {author} {\bibinfo {author} {\bibfnamefont {E.}~\bibnamefont
  {{Lefebvre}}}, \bibinfo {author} {\bibfnamefont {N.}~\bibnamefont
  {{Cochet}}}, \bibinfo {author} {\bibfnamefont {S.}~\bibnamefont
  {{Fritzler}}}, \bibinfo {author} {\bibfnamefont {V.}~\bibnamefont {{Malka}}},
  \bibinfo {author} {\bibfnamefont {M.-M.}\ \bibnamefont {{Al{\'e}onard}}},
  \bibinfo {author} {\bibfnamefont {J.-F.}\ \bibnamefont {{Chemin}}}, \bibinfo
  {author} {\bibfnamefont {S.}~\bibnamefont {{Darbon}}}, \bibinfo {author}
  {\bibfnamefont {L.}~\bibnamefont {{Disdier}}}, \bibinfo {author}
  {\bibfnamefont {J.}~\bibnamefont {{Faure}}}, \bibinfo {author} {\bibfnamefont
  {A.}~\bibnamefont {{Fedotoff}}}, \bibinfo {author} {\bibfnamefont
  {O.}~\bibnamefont {{Landoas}}}, \bibinfo {author} {\bibfnamefont
  {G.}~\bibnamefont {{Malka}}}, \bibinfo {author} {\bibfnamefont
  {V.}~\bibnamefont {{M{\'e}ot}}}, \bibinfo {author} {\bibfnamefont
  {P.}~\bibnamefont {{Morel}}}, \bibinfo {author} {\bibfnamefont
  {M.}~\bibnamefont {{Rabec LeGloahec}}}, \bibinfo {author} {\bibfnamefont
  {A.}~\bibnamefont {{Rouyer}}}, \bibinfo {author} {\bibfnamefont
  {C.}~\bibnamefont {{Rubbelynck}}}, \bibinfo {author} {\bibfnamefont
  {V.}~\bibnamefont {{Tikhonchuk}}}, \bibinfo {author} {\bibfnamefont
  {R.}~\bibnamefont {{Wrobel}}}, \bibinfo {author} {\bibfnamefont
  {P.}~\bibnamefont {{Audebert}}}, \ and\ \bibinfo {author} {\bibfnamefont
  {C.}~\bibnamefont {{Rousseaux}}},\ }\href {\doibase
  10.1088/0029-5515/43/7/317} {\bibfield  {journal} {\bibinfo  {journal} {Nucl.
  Fus.}\ }\textbf {\bibinfo {volume} {43}},\ \bibinfo {pages} {629} (\bibinfo
  {year} {2003})}\BibitemShut {NoStop}%
\bibitem [{\citenamefont {{Godfrey}}\ and\ \citenamefont
  {{Vay}}(2014)}]{Godfrey_2014}%
  \BibitemOpen
  \bibfield  {author} {\bibinfo {author} {\bibfnamefont {B.~B.}\ \bibnamefont
  {{Godfrey}}}\ and\ \bibinfo {author} {\bibfnamefont {J.-L.}\ \bibnamefont
  {{Vay}}},\ }\href {\doibase 10.1016/j.jcp.2014.02.022} {\bibfield  {journal}
  {\bibinfo  {journal} {J. Comput. Phys.}\ }\textbf {\bibinfo {volume} {267}},\
  \bibinfo {pages} {1} (\bibinfo {year} {2014})}\BibitemShut {NoStop}%
\bibitem [{\citenamefont {{Bret}}\ \emph
  {et~al.}(2010{\natexlab{b}})\citenamefont {{Bret}}, \citenamefont
  {{Gremillet}},\ and\ \citenamefont {{B\'enisti}}}]{Bret_2010b}%
  \BibitemOpen
  \bibfield  {author} {\bibinfo {author} {\bibfnamefont {A.}~\bibnamefont
  {{Bret}}}, \bibinfo {author} {\bibfnamefont {L.}~\bibnamefont {{Gremillet}}},
  \ and\ \bibinfo {author} {\bibfnamefont {D.}~\bibnamefont {{B\'enisti}}},\
  }\href {\doibase 10.1103/PhysRevE.81.036402} {\bibfield  {journal} {\bibinfo
  {journal} {Phys. Rev. E}\ }\textbf {\bibinfo {volume} {81}},\ \bibinfo
  {pages} {036402} (\bibinfo {year} {2010}{\natexlab{b}})}\BibitemShut
  {NoStop}%
\bibitem [{\citenamefont {Abramowitz}\ and\ \citenamefont
  {Stegun}(1972)}]{Abramowitz_1972}%
  \BibitemOpen
  \bibfield  {author} {\bibinfo {author} {\bibfnamefont {M.}~\bibnamefont
  {Abramowitz}}\ and\ \bibinfo {author} {\bibfnamefont {I.~A.}\ \bibnamefont
  {Stegun}},\ }\href@noop {} {\emph {\bibinfo {title} {Handbook of Mathematical
  Functions}}}\ (\bibinfo  {publisher} {Dover Publications},\ \bibinfo
  {address} {New York},\ \bibinfo {year} {1972})\BibitemShut {NoStop}%
\bibitem [{\citenamefont {Gradshteyn}\ and\ \citenamefont
  {Rizhik}(1980)}]{Gradshteyn_1980}%
  \BibitemOpen
  \bibfield  {author} {\bibinfo {author} {\bibfnamefont {I.~S.}\ \bibnamefont
  {Gradshteyn}}\ and\ \bibinfo {author} {\bibfnamefont {I.~M.}\ \bibnamefont
  {Rizhik}},\ }\href@noop {} {\emph {\bibinfo {title} {Tables of Integrals,
  Series and Products}}}\ (\bibinfo  {publisher} {Academic Press},\ \bibinfo
  {address} {New York},\ \bibinfo {year} {1980})\BibitemShut {NoStop}%
\bibitem [{\citenamefont {{Bret}}\ \emph {et~al.}(2008)\citenamefont {{Bret}},
  \citenamefont {{Gremillet}}, \citenamefont {{B{\'e}nisti}},\ and\
  \citenamefont {{Lefebvre}}}]{Bret_2008}%
  \BibitemOpen
  \bibfield  {author} {\bibinfo {author} {\bibfnamefont {A.}~\bibnamefont
  {{Bret}}}, \bibinfo {author} {\bibfnamefont {L.}~\bibnamefont {{Gremillet}}},
  \bibinfo {author} {\bibfnamefont {D.}~\bibnamefont {{B{\'e}nisti}}}, \ and\
  \bibinfo {author} {\bibfnamefont {E.}~\bibnamefont {{Lefebvre}}},\ }\href
  {\doibase 10.1103/PhysRevLett.100.205008} {\bibfield  {journal} {\bibinfo
  {journal} {Phys. Rev. Lett.}\ }\textbf {\bibinfo {volume} {100}},\ \bibinfo
  {eid} {205008} (\bibinfo {year} {2008})}\BibitemShut {NoStop}%
\bibitem [{\citenamefont {{Mikhailovskii}}(1981)}]{Mikhailovskii_1981}%
  \BibitemOpen
  \bibfield  {author} {\bibinfo {author} {\bibfnamefont {A.~B.}\ \bibnamefont
  {{Mikhailovskii}}},\ }\href {\doibase 10.1088/0032-1028/23/5/003} {\bibfield
  {journal} {\bibinfo  {journal} {Plasma Phys.}\ }\textbf {\bibinfo {volume}
  {23}},\ \bibinfo {pages} {413} (\bibinfo {year} {1981})}\BibitemShut
  {NoStop}%
\bibitem [{\citenamefont {{Bender}}\ and\ \citenamefont
  {{Orszag}}(1978)}]{Bender_1978}%
  \BibitemOpen
  \bibfield  {author} {\bibinfo {author} {\bibfnamefont {C.~M.}\ \bibnamefont
  {{Bender}}}\ and\ \bibinfo {author} {\bibfnamefont {S.~A.}\ \bibnamefont
  {{Orszag}}},\ }\href@noop {} {\emph {\bibinfo {title} {{Advanced Mathematical
  Methods for Scientists and Engineers}}}}\ (\bibinfo  {publisher}
  {McGraw-Hill},\ \bibinfo {address} {New York},\ \bibinfo {year}
  {1978})\BibitemShut {NoStop}%
\bibitem [{\citenamefont {{Schlickeiser}}\ and\ \citenamefont
  {{Kneller}}(1997)}]{Schlickeiser_1997}%
  \BibitemOpen
  \bibfield  {author} {\bibinfo {author} {\bibfnamefont {R.}~\bibnamefont
  {{Schlickeiser}}}\ and\ \bibinfo {author} {\bibfnamefont {M.}~\bibnamefont
  {{Kneller}}},\ }\href {\doibase 10.1017/S0022377897005485} {\bibfield
  {journal} {\bibinfo  {journal} {J. Plasma Phys.}\ }\textbf {\bibinfo {volume}
  {57}},\ \bibinfo {pages} {709} (\bibinfo {year} {1997})}\BibitemShut
  {NoStop}%
\bibitem [{\citenamefont {Weideman}(1994)}]{Weideman_1994}%
  \BibitemOpen
  \bibfield  {author} {\bibinfo {author} {\bibfnamefont {J.~A.~C.}\
  \bibnamefont {Weideman}},\ }\href {\doibase 10.1137/0731077} {\bibfield
  {journal} {\bibinfo  {journal} {{SIAM J. Numer. Anal.}}\ }\textbf {\bibinfo
  {volume} {31}},\ \bibinfo {pages} {1497} (\bibinfo {year}
  {1994})}\BibitemShut {NoStop}%
\bibitem [{\citenamefont {Erdelyi}\ \emph {et~al.}(1954)\citenamefont
  {Erdelyi}, \citenamefont {Magnus}, \citenamefont {Oberhettinger},\ and\
  \citenamefont {Triconi}}]{Erdelyi_1954}%
  \BibitemOpen
  \bibfield  {author} {\bibinfo {author} {\bibfnamefont {A.}~\bibnamefont
  {Erdelyi}}, \bibinfo {author} {\bibfnamefont {W.}~\bibnamefont {Magnus}},
  \bibinfo {author} {\bibfnamefont {F.}~\bibnamefont {Oberhettinger}}, \ and\
  \bibinfo {author} {\bibfnamefont {F.~G.}\ \bibnamefont {Triconi}},\
  }\href@noop {} {\emph {\bibinfo {title} {Tables of Integral Transforms}}},\
  Vol.~\bibinfo {volume} {1}\ (\bibinfo  {publisher} {McGraw-Hill},\ \bibinfo
  {address} {New York},\ \bibinfo {year} {1954})\BibitemShut {NoStop}%
\end{thebibliography}%

\end{document}